\documentclass[useAMS,usenatbib,usegraphicx]{mn2e}
\usepackage{lscape}
\usepackage{rotating}
\usepackage{times}
\usepackage{hyperref}

\def\swift{{\sl Swift}}

\def\astrosat{{\sl Astrosat}}

\def\rxte{{\sl RXTE}}

\def\vlt{{VLT}}
\def\p{$\pm$}
\def\ltsim{\mathrel{\hbox{\rlap{\hbox{\lower4pt\hbox{$\sim$}}}\hbox{$<$}}}}
\def\gtsim{\mathrel{\hbox{\rlap{\hbox{\lower4pt\hbox{$\sim$}}}\hbox{$>$}}}}

\def\Msun{M$_{\odot}$}

\def\nh{$N_{\rm H}$}
\def\iraf{{\sc iraf}}
\def\ftools{{\sc ftools}}
\def\heasoft{{\sc heasoft}}
\def\xspec{{\sc xspec}}
\def\diskbb{{\sc diskbb}}
\def\diskir{{\sc diskir}}
\def\xronos{{\sc xronos}}
\def\pexrav{{\sc pexrav}}
\def\idlextract{{\sc idl\_extract}}

\def\ha{H$\alpha$}
\def\hb{H$\beta$}
\def\hg{H$\gamma$}
\def\hd{H$\delta$}

\def\l{$\lambda$}

\def\heii{He~{\sc ii}}
\def\hei{He~{\sc i}}
\def\nai{Na~{\sc i}}

\def\gravrad{$R_{\rm G}$}

\def\ledd{$L_{\rm Edd}$}
\def\lx{$L_{\rm X}$}

\def\gx339{GX~339--4}
\def\swiftj1753{SWIFT~J1753.5--0127}
\def\xtej1118{XTE~J1118+480}
\def\av{$A_{\rm V}$}
\def\ebv{$E_{\rm B-V}$}
\def\numax{$\nu_{\rm max}$}

\title[Optical and X-ray timing of GX~339--4]{Rapid optical and X-ray timing observations of GX~339--4: multi-component optical variability in the low/hard state}

\author[P. Gandhi et al.]{P. Gandhi,$^{1,2}$ V.S. Dhillon,$^3$ M. Durant,$^{4,5}$ A.C. Fabian,$^6$ A. Kubota,$^7$ K. Makishima,$^{8,2}$\newauthor  J. Malzac,$^9$ T.R. Marsh,$^{10}$ J.M. Miller,$^{11}$ T. Shahbaz,$^4$ H.C. Spruit$^{12}$ \& P. Casella$^{13}$\\
$^{1}$Institute of Space and Astronautical Science (ISAS), Japan Aerospace Exploration Agency, 3-1-1 Yoshinodai, chuo-ku, Sagamihara, Kanagawa 229-8510, Japan\\
$^{2}$RIKEN Cosmic Radiation Lab, 2-1 Hirosawa, Wakoshi, Saitama 351-0198, Japan\\
$^{3}$Department of Physics and Astronomy, University of Sheffield, Sheffield S3 7RH\\
$^{4}$Instituto de Astrof\'{i}sica de Canarias, La Laguna, E38205 Tenerife, Spain\\
$^{5}$University of Florida, Bryant Space Center, Gainesville, FL 32611-2055, USA\\
$^{6}$Institute of Astronomy, Madingley Road, Cambridge, CB3 0HA\\
$^{7}$Department of Electronic Information Systems, Shibaura Institute of Technology, 307 Fukasaku, Minuma-ku, Saitama-shi, Saitama, 337-8570, Japan\\
$^{8}$Department of Physics, University of Tokyo, 7-3-1 Hongo, Bunkyo-ku, Tokyo 113-0033, Japan\\
$^{9}$CESR, Universit\'e de Toulouse (UPS), CNRS (UMR 5187), 9 Avenue du colonel Roche BP44346, 31028 Toulouse Cedex 4, France\\
$^{10}$Department of Physics, University of Warwick, Coventry, CV4 7AL\\
$^{11}$Department of Astronomy, University of Michigan, 500 Church Street, Ann Arbor, MI 48109\\
$^{12}$Max-Planck-Institut f\"{u}r Astrophysik, Postfach 1317, 85741 Garching bei M\"{u}nchen, Germany\\
$^{13}$School of Physics and Astronomy, University of Southampton, Southampton, Hampshire, SO17 1BJ
}

\begin{document}

\date{Accepted May 24 2010.
      Received Jan 07 2010.}

\pagerange{\pageref{firstpage}--\pageref{lastpage}} 
\pubyear{2010}

\maketitle
\label{firstpage}

\begin{abstract}
A rapid timing analysis of \vlt/ULTRACAM (optical) and \rxte\ (X-ray) observations of the Galactic black hole binary GX~339--4 in the low/hard, post-outburst state of June 2007 is presented. The optical light curves in the $r'$, $g'$ and $u'$ filters show slow ($\sim$ 20 s) quasi-periodic variability. Upon this is superposed fast flaring activity on times approaching the best time resolution probed ($\sim$ 50 ms in $r'$ and $g'$) and with maximum strengths of more than twice the local mean. Power spectral analysis over $\sim$ 0.004--10 Hz is presented, and shows that although the average optical variability amplitude is lower than that in X-rays, the peak variability power emerges at a higher Fourier frequency in the optical. 
Energetically, we measure a large optical vs. X-ray flux ratio, higher than that seen on previous occasions when the source was fully jet-dominated. Such a large ratio cannot be easily explained with a disc alone. 
Studying the optical:X-ray cross-spectrum in Fourier space shows a markedly different behaviour above and below $\sim$0.2 Hz. The peak of the coherence function above this threshold is associated with a short optical time lag with respect to X-rays, also seen as the dominant feature in the time-domain cross-correlation at $\approx$150 ms. The rms energy spectrum of these fast variations is best described by distinct physical components over the optical and X-ray regimes, and also suggests a maximal irradiated disc fraction of 20 per cent around 5000 \AA. If the constant time delay is due to propagation of fluctuations to (or within) the jet, this is the clearest optical evidence to date of the location of this component. The low-frequency quasi-periodic oscillation is seen in the optical but not in X-rays, and is associated with a low coherence. 
Evidence of reprocessing emerges at the lowest Fourier frequencies, with optical lags at $\sim$10 s and strong coherence in the blue $u'$ filter. Consistent with this, simultaneous optical spectroscopy also shows the Bowen fluorescence blend, though its emission location is unclear. But canonical disc reprocessing cannot dominate the optical power easily, nor explain the fast variability.
\end{abstract}
\begin{keywords}
accretion: stars -- individual: GX339--4 -- stars: X-rays: binaries.
\end{keywords}

\section{Introduction}

The radiative output of active X-ray binaries (XRBs) peaks at X-ray wavelengths, hence their name. A characteristic property of this radiation is its strong and aperiodic variability over a range of time scales \citep[e.g. ][]{vanderklis89}. This is a manifestation of rich underlying structure in the accretion environments of these sources on a range of physical scales. X-ray timing studies can thus provide key information for our understanding of volatile XRB environments, and X-ray timing is now a mature field in this respect \citep[e.g. ][]{hasingervanderklis89,vanderklis95,remillardmcclintock06}. 

At lower energies, e.g. in the optical and infrared, the picture is thought to be complex. For many low mass XRBs, the donor star contribution can be ignored during active accretion states. Optical emission from the accretion disc arises predominantly from its outer, cool portions with temperatures much below $\sim$10$^6$ K. Consequently, large variations of this component should occur only on a relatively-slow viscous timescale. Any faster variations may be the result of high-energy photons being reprocessed on the outer disc. In fact, such X-ray heating is thought to be the primary engine generating the bulk of the ultraviolet and optical emission seen from XRBs \citep{vanparadijsmcclintock94}. Weaker, but highly significant reprocessing on the companion star is also an important effect which can be used to determine the physical parameters of the binary system \citep[e.g. ][]{horne85, marsh88, obrien02}.

Observations are now also beginning to provide evidence of fast variability (on $\sim$1~s timescales or less) at low energies which is not associated with reprocessing, and recent works attribute much of this to non-thermal contributions from the jet and corona. Some observations of fast low energy variations in several sources have actually existed for many years \citep[e.g.][]{motch82, uemura02_v404cyg, fender97}. In several cases, simultaneous low and high energy observations argue against a reprocessing origin. This was best demonstrated in \xtej1118, for which rapid optical and X-ray timing observations revealed complex flux variations, including an anti-correlation between optical and X-rays -- exactly the opposite of that predicted by reprocessing scenarios \citep[][]{kanbach01, spruitkanbach02}. Many works have successfully modelled the optical power and broad-band spectrum as a result of (cyclo-)synchrotron emission, though the location and characteristics of the emitting plasma remain a matter of debate \citep[e.g. ][ and more]{merloni00, esin01, markoff05, yuan05}. Self-consistently incorporating the timing properties in this framework has proven to be trickier. An energy reservoir model feeding a common jet and corona system developed by \citet{malzac04} can reproduce both the shape and the strength of the observed timing correlations very well. But physical modeling of the reservoir itself and of the impulses of energy release governing the observed variability remains somewhat ad hoc, being modelled simply as a linear superposition of fluctuating stochastic shots. \citet{g09_rmsflux} discusses a simple modification of this phenomenological model to reproduce another observable (the \lq rms--flux\rq\ relation), by making the light curves a non-linear coupling of shots, rather than an additive superposition.

Other sources with intriguing multi-wavelength timing correlations include GRS~1915+105, studied by \citet{eikenberry98} and \citet{arai09}, among others. And in a recent study, \citet[][ hereafter Paper I]{g08} and \citet{durant08} observed two other Galactic BH candidates in the X-ray low/hard state, \gx339\ and \swiftj1753, simultaneously in X-rays with the {\sl Rossi X-ray Timing Explorer} (\rxte) and in the optical with ULTRACAM mounted on the Very Large Telescope (VLT). Cross-correlating the light curves in the two wavebands revealed complex variability patterns in both systems. \citet{durant09} have discussed the second source in detail, suggesting that optical cyclotron anti-correlates with an X-ray (presumably inverse Compton dominated) response (e.g. \citealt{fabian82}). In the present paper\footnote{Based on observations obtained during ESO programs 079.D-0535 and 279.D-5021.}, we present the full spectral and timing properties of \gx339, supplementing the cross-correlation function presented in Paper I. 

GX~339--4 (4U~1658--48) is one of the best-studied Galactic black hole (BH) candidates over a wide range of wavelengths. The source is a classic X-ray transient. It is usually found in the active, low/hard flux state, but is known to undergo dramatic changes on month--year timespans, showing a broad range of X-ray states from \lq quiescence\rq, to the \lq very high\rq\ state \citep[e.g. ][]{zdziarski04, dunn08}. Such changes are accompanied by strong modulations in its radio \citep{corbel00}, and optical \citep{makishima86} properties. The source has proven to be key for the discovery and elucidation of several important aspects of XRB behaviour. It was one of the first to show clear evidence of the existence of an inner accretion disc in the X-ray low/hard state \citep{miller06}. It is the prototype source for defining an important relationship between the radio and X-ray fluxes of XRBs \citep[e.g. ][]{corbel00, gallo03}. 

It was also one of the first sources found to show very fast optical flickering \citep{motch82}. Follow-up rapid simultaneous optical and X-ray studies uncovered evidence of a slow anti-correlated response between the two bands, but such studies remained scarce \citep{motch83, makishima86}. In Paper I, we presented a new correlation feature in the faint low/hard state, with a sharp, sub-second optical response in a red ($r'$) filter, delayed with respect to X-rays. As was discussed in that paper, and elaborated upon herein, simple linear reprocessing transfer functions cannot explain this feature. We now present our full multi-filter variability dataset, and relate this with the simultaneous X-ray timing properties over a range of timescales. The optical and X-ray power spectra and correlations in time and Fourier space are analysed. Supplementary optical spectroscopy with the FORS2/\vlt\ instrument is also presented. The observations are discussed in the context of multiple varying components with some (as yet not fully understood) underlying interactions. Our broad-band analysis provides first simultaneous constraints on the physical characteristics, emitted powers {\em and} relative locations of these components within the accretion environment. 

The donor star of this low mass XRB system is very faint \citep{shahbaz01}, which complicates measurement of physical parameters of the binary, but means that the optical flux is completely dominated by the accretion flow, which is our main interest herein. We assume a compact object mass of 6~\Msun, consistent with the best determination of the mass function of 5.8\p0.5 \Msun \citep{hynes03_gx339}, and a distance of 8 kpc (a range of 6--15 kpc has been suggested in the literature; e.g. \citealt{hynes04, zdziarski04}). The most-widely accepted value of the binary orbital period is 1.75 days, based on outburst radial velocity variations of the optical Bowen emission blend reprocessed on the secondary star \citep{hynes03_gx339}, and we adopt this value here.

\section{Observations and data analysis}

\subsection{Optical timing (ULTRACAM)}
\label{sec:ultracamobs}

Our optical observations were carried out with ULTRACAM, which is a fast imaging camera employing frame-transfer CCDs \citep{ultracam}. Although not a \lq photon-counting\rq\ instrument, ULTRACAM satisfies two critical requirements for astronomical studies with CCD detectors on fast (sub-second) timescales -- 1) a low readout noise and, 2) a small dead time between exposures. Unlike proportional counters, CCD dead time does not depend on the incident count rate. Instead, it is just the time taken to read out, or clock, the exposed CCD pixels, during which any incident photons are not accumulated. Frame-transfer CCDs are constructed with an extra storage area identical to their exposed pixel area, and clocking quickly transfers the photo-generated charge to this storage area from where it is subsequently read out and digitised. The maximum dead time in ULTRACAM is only 24 ms. But this holds true for full CCD read out and signal digitisation. The dead time can be much reduced by reading out only small portions of the CCD which are of interest.

\begin{figure}
  \begin{center}
    \includegraphics[angle=90,width=8.5cm]{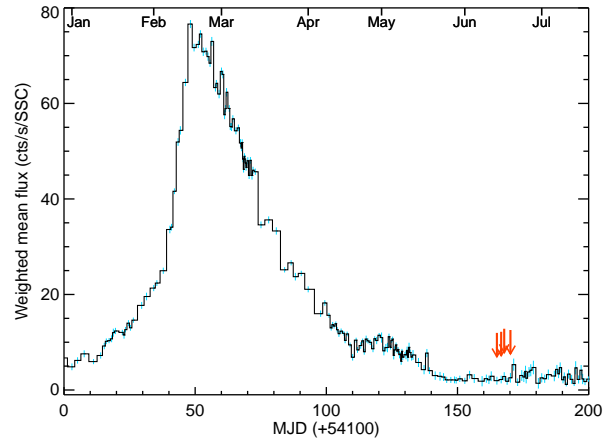}
    \caption{Long-term \rxte\ All Sky Monitor light curve of \gx339, obtained and rebinned from the quick-look results provided by the ASM/\rxte\ team (http://xte.mit.edu/asmlc/). The arrows are the nights of our observations, denoted Nights 4, 3, 2 and 1 from left to right, in terms of improving weather. The first day of several months in 2007 are marked at the top. As is apparent, our observations were carried out soon after the source had returned to a low flux (but active) state following its 2007 outburst.
\label{fig:asmlc}}
  \end{center}
\end{figure}

ULTRACAM was mounted on the Nasmyth focus of VLT/UT3 as a visitor instrument during 2007 Jun 9--24, and our observational time window was governed by this mounting period. We happened to catch \gx339\ in its typical X-ray low/hard state, soon after it had emerged from an outburst that peaked in mid Feb \citep[][]{kalemci07}. Fig.~\ref{fig:asmlc} shows the long-term light curve of \gx339; our days of observations are denoted by arrows. We chose to monitor the source for four hours of integration, split equally across four nights. No observing was possible on the night of UT Jun 12 (Chilean night starting Jun 11) because of clouds. Weather steadily improved thereafter, with changeable conditions on Jun 14 and 16, and very stable conditions on Jun 18. These nights are hereafter referred to as Nights 4, 3, 2 and 1, respectively, ranked according to the weather. 

ULTRACAM has a 2.6$\times$2.6 arcminute field at the VLT. In order to achieve a reasonable signal:noise, the timing resolution of our observations was decided interactively at the beginning of each sequence of exposures based on real-time weather conditions. The size of the CCD window to be read out was adjusted in order to achieve this time resolution, resulting in small window sizes on all nights (not larger than 100$\times$100 pixels centred on the target). In addition, to achieve the fastest frame rates, we used the \lq drift\rq\ observing mode. Instead of clocking (i.e. charge transferring) the entire exposed area of the CCD to the unexposed area, drift mode clocks only the pixel rows corresponding to the window region. As clocking is typically much faster than signal digitisation, the dead time (equal to the clocking time) is very small. The interested reader should refer to \citet{ultracam} for details. For our observations, the frame rates ranged from just under 50 ms to about 136 ms, and dead times were approximately 1.6, 2.3 and 2.3 ms on Nights 1, 2 and 3, respectively. These correspond to about 3.2, 1.7 and 1.7 per cent of the cycle time on each night. Table~\ref{tab:obslog} lists the observation log. Note that the cycle and frame times stated therein are average values and the standard deviation of these is $\approx$ 5 $\mu$s, which means that the light curves are not strictly regular at the sub-millisecond level. But this is not a problem because of the accurate time sampling.

Source flux was monitored by observing a bright comparison star (USNO-B1.0 0412-0564199) within the field of view, centred on a second CCD window identical in size to that used for the target. ULTRACAM is equipped with three separate, but identical, detectors, and with the use of beam-splitters, fully simultaneous observations in three separate optical filters can be carried out. We used the $r'$, $g'$ and $u'$ SDSS filter set. Note that low atmospheric (and inter-stellar) transparency in $u'$ necessitates co-addition of many frames for source detection, with the result that cycle times of faster than a few seconds were not probed for this filter.

\begin{figure}
  \begin{center}
    \includegraphics[angle=90,width=8.5cm]{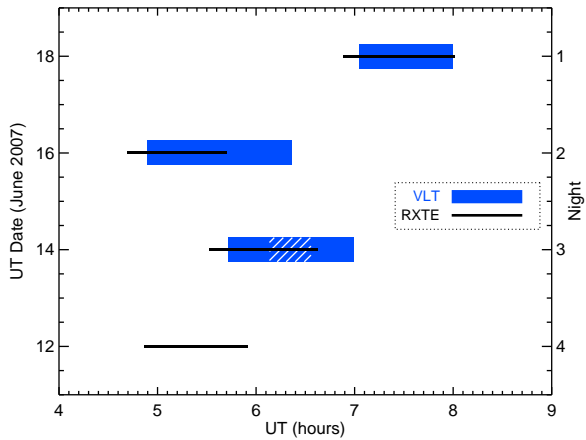}
    \caption{Times of the optical and X-ray observations. For both telescopes, the marked intervals denote the effective \lq good\rq\ time intervals only. The ULTRACAM observations have small (few second gaps) which are excluded on Nights 1 and 2 from the analysis. The worst period of highly variable transparency during Night 3 is a hatched 25 m segment starting at UT06:08, which was also excluded. 
} \label{fig:finaltimes}
  \end{center}
\end{figure}

The ULTRACAM pipeline\footnote{http://deneb.astro.warwick.ac.uk/phsaap/software/ultracam/html/} was used for data cube reduction, including bias subtraction, flat-fielding and photometry, for each of the three filters separately. Aperture photometry was carried out by profile fitting and optimal extraction of target and comparison counts separately, and annular background subtraction. All source light curves were divided by the corresponding comparison star light curves normalised to their mean. The photometric solution is very stable on Night 1, but degrades progressively on the other two nights, with the source being lost on several frames during our hour-long monitoring on Night 3. The pipeline provides some control by allowing variable aperture sizes and centering options, but the worst of these bad weather periods -- a single contiguous 25 min segment on Night 3 (see Fig.~\ref{fig:finaltimes}) -- has been avoided. Relative photometry was thus important for Nights 2 and 3, especially for the long-term ($\gtsim$ minute timespan) variations. Additionally, there are intermittent source detection drops for the lower sensitivity $u'$ data throughout both these nights, and hence these data are not analysed. For Night 1, 50 $u'$ co-adds were used, resulting in best time bin of $\approx$2.5 s for this filter.

In short, we use the full $r'$ and $g'$ datasets on all three nights for the timing analysis presented in this paper, except for the bad transparency segment on Night 3. For the $u'$ filter, only data from Night 1 are used. 

\subsection{Optical spectroscopy (FORS)}

We were also awarded director's discretionary time for short optical spectroscopic observations with the FORS2 instrument \citep{fors}, simultaneous with ULTRACAM. Changeable weather conditions meant that only two spectra were obtained simultaneously, with the data from the best night (Night 1) presented herein. The GRIS~300V grism with the GG375 order sorting filter\footnote{FORS2 User's Manual (http://www.eso.org/sci/facilities/paranal/instru ments/fors/doc/)} was used for a total integration time of 200 s, starting at UT07:09. The effective wavelength coverage was over the $\sim$ 4000--7500 \AA\ regime, with a resolving power $R$$\approx$700 at 5000 \AA. Standard data reduction, wavelength calibration and spectrum extraction were carried out in \iraf. Flux calibration was carried out against a spectro-photometric standard star (LTT~7987; \citealt{hamuy94}) observation from the same night.

\subsection{X-rays (RXTE)}

In X-rays, we observed \gx339\ for four hours (Observation ID prefix 93119-01) with \rxte\ \citep{rxte}, coordinated to be closely simultaneous to the optical observations over the four nights. In practice, \rxte\ orbital constraints resulted in the exact length of overlap being slightly shorter than the desired 60 mins on each night. Table~\ref{tab:obslog} lists the \rxte\ observing log and Fig.~\ref{fig:finaltimes} shows the final overlap period of the optical and X-ray observations in graphical form.

\rxte\ has two narrow-field instruments. The one with a low energy ($\sim$2--60 keV) response is called PCA (the Proportional Counter Array; \citealt{rxtepca}), and that sensitive at higher energies ($\sim$15-250 keV) is HEXTE (the High Energy X-ray Timing Experiment; \citealt{rxtehexte}). HEXTE employs two detection clusters, each originally capable of rocking between the source and a background position. At the time of our observations, one of the two (cluster A) could no longer operate in the rocking mode and hence real-time background subtraction was not possible for this. Furthermore, the count rate of \gx339\ in HEXTE was very low and dominated by the background; the net rate was only 3 ct s$^{-1}$. Hence we do not analyse the HEXTE timing data further herein. But the net counts from cluster B accumulated over the full exposure duration were sufficient for extraction of a spectrum with useful signal.

The PCA has five proportional counter units (PCUs), and two or three of these were operational during our observations, the number changing from night to night. For data reduction, we followed standard procedures described in the \rxte\ cookbook and other material provided by the \rxte\ Guest Observer Facility, using \ftools\footnote{http://heasarc.gsfc.nasa.gov/ftools/} \citep{ftools} and \heasoft\ v.6.8. The detected photons are analysed and delivered in several formats. In the event mode recommended for weak sources, {\tt GoodXenon} mode, every event is telemetered to the ground (with intrinsic time resolution of $\sim$1 $\mu$s) and this was the mode used for our timing analysis. Spectra were extracted from both instruments. For the PCA, we used the binned {\tt Standard2} mode, with a time resolution of 16 s. Basic good time intervals were selected by filtering on a minimum elevation angle of 10 degrees from the Earth's limb and a maximum pointing offset of 0.02. Each PCU has three Xenon detector layers, all of which were summed. Background files were generated by running the {\tt pcabackest} utility, which estimates the background using {\tt Standard2} data files and synthetic background models. For HEXTE, data and background were extracted from the good channels of cluster B. Standard \ftools\ routines were used for response matrix generation and dead-time correction. Spectral modelling was carried out over $\approx$3--30 keV (PCA) and 20--200 keV (HEXTE). Systematic uncertainties of 0.5 per cent\footnote{http://www.universe.nasa.gov/xrays/programs/rxte/pca/doc/rmf/pcarmf-11.7} and 1 per cent\footnote{http://heasarc.gsfc.nasa.gov/docs/xte/ftools/xtefaq\_answers.html} were included for the two instruments, as recommended.

\subsection{Barycentering of light curves}

Several of the timing results that we will describe below depend on accurate calibration of any time lag between the two light curves. Hence, transformation of the light curves to a common time frame is an important part of the data reduction process. In X-rays, this was accomplished with the {\tt faxbary} routine to convert the \rxte\ data file times from mission time to a Barycentric frame in Terrestrial Time (TT). In the optical, a custom-built routine called {\tt tcorr} from the ULTRACAM pipeline was used, which carries out absolute time conversions using the SLA C library for absolute timing (Pat Wallace, priv. comm.), with inputs being the source coordinates (we used J2000 epoch coordinates as RA=17:02:49.5 and Dec=--48:47:23), the name of the telescope (whose latitude, longitude and altitude are internally pre-defined for the VLT as 70$^\circ$24$'$9$''$.9W, 24$^\circ$37$'$30$''$.3S and 2635m, respectively) and the observing time in UT for all time bins. Absolute timing in ULTRACAM is achieved by GPS time stamps from multiple satellites, and is good to at least 1 ms \citep{ultracam}. In the fast drift observing mode, there are additional corrections required for time stamping, because an exposed CCD window is not read out (i.e. digitised) immediately upon completion of its exposure. Instead, with each exposure, it is clocked by the number of rows it contains and is read out only when it reaches the CCD edge. But it is a simple matter for the pipeline to determine the time stamp based on the (very accurately known) clocking and digitisation times per pixel. In X-rays, the absolute accuracy of the raw times are better than 100 $\mu$s\footnote{http://heasarc.gsfc.nasa.gov/docs/xte/abc/time.html}. The accuracy of the Barycentre transformation codes has been compared against pulsar timing, and found to be accurate to better than 100 $\mu$s. Similar transformation applied to the data for another source -- \swiftj1753, described in \citet{durant08} -- gave an optical vs. X-ray cross-correlation with a sharp change exactly at zero delay, giving extra confidence in the timing calibrations.

\begin{table*}
  \begin{center}
    \begin{tabular}{lccccccr}
      \hline
      Night  &    Date   & \multicolumn{2}{c}{\underline{\hspace{1.2cm}\rxte\hspace{1.2cm}}}     &  \multicolumn{3}{c}{\underline{\hspace{1.5cm}ULTRACAM\hspace{1.5cm}}} & Overlap\\
             & 2007 Jun  & Times                 & Active PCUs                                   & Times                 & Exp time  &   Cycle time                      &  \\
             &    UT     & UT                    &  IDs                                          & UT                    &   ms      &     ms                            &   min\\
      \hline
      4      & 12        & 04:51--05:54          &  024                                          & \multicolumn{4}{c}{--}                     \\
      3      & 14        & 05:31--06:37          &  02                                           & 05:43--06:59          &  134.0      &    136.3                            &  49 \\
      2      & 16        & 04:40--05:50          &  024                                          & 04:54--06:22          &  131.0      &    133.3                            &  47 \\
      1      & 18        & 06:55--08:03          &  02                                           & 07:03--08:00          &   48.0      &     49.6                            &  56 \\
      \hline
    \end{tabular}
    \caption{Observation log. The first column states our night designation, based on weather conditions. The third column lists the \rxte\ good time intervals. Night 4 was completely clouded out, and is not discussed in this paper. The \lq Exp time\rq\ and \lq Cycle time\rq\ columns list the individual frame exposure times, and time difference between successive frames, respectively, rounded off to tenths of a milli-second. For the $u'$ filter, only the data from Night 1 are analysed, and these required 50 frame co-adds, resulting in correspondingly longer best time resolution. The last column states the length of the overlapping optical and X-ray observational interval, rounded off in minutes. There are a handful of short gaps in the optical and X-ray data on different nights (and a longer period of bad transparency on Night 3; see Fig.~\ref{fig:finaltimes}), which were removed from the analysis. Only contiguous sections of specific length (typically 256~s) were used for power spectra and coherence computations, as described in the text, leading to further slight shortening of the overlap time in these cases. 
\label{tab:obslog}}
  \end{center}
\end{table*}

\section{Results: Spectroscopy}

\subsection{Optical}
\label{sec:optspec}

\begin{figure*}
  \begin{center}
    \includegraphics[width=8.5cm]{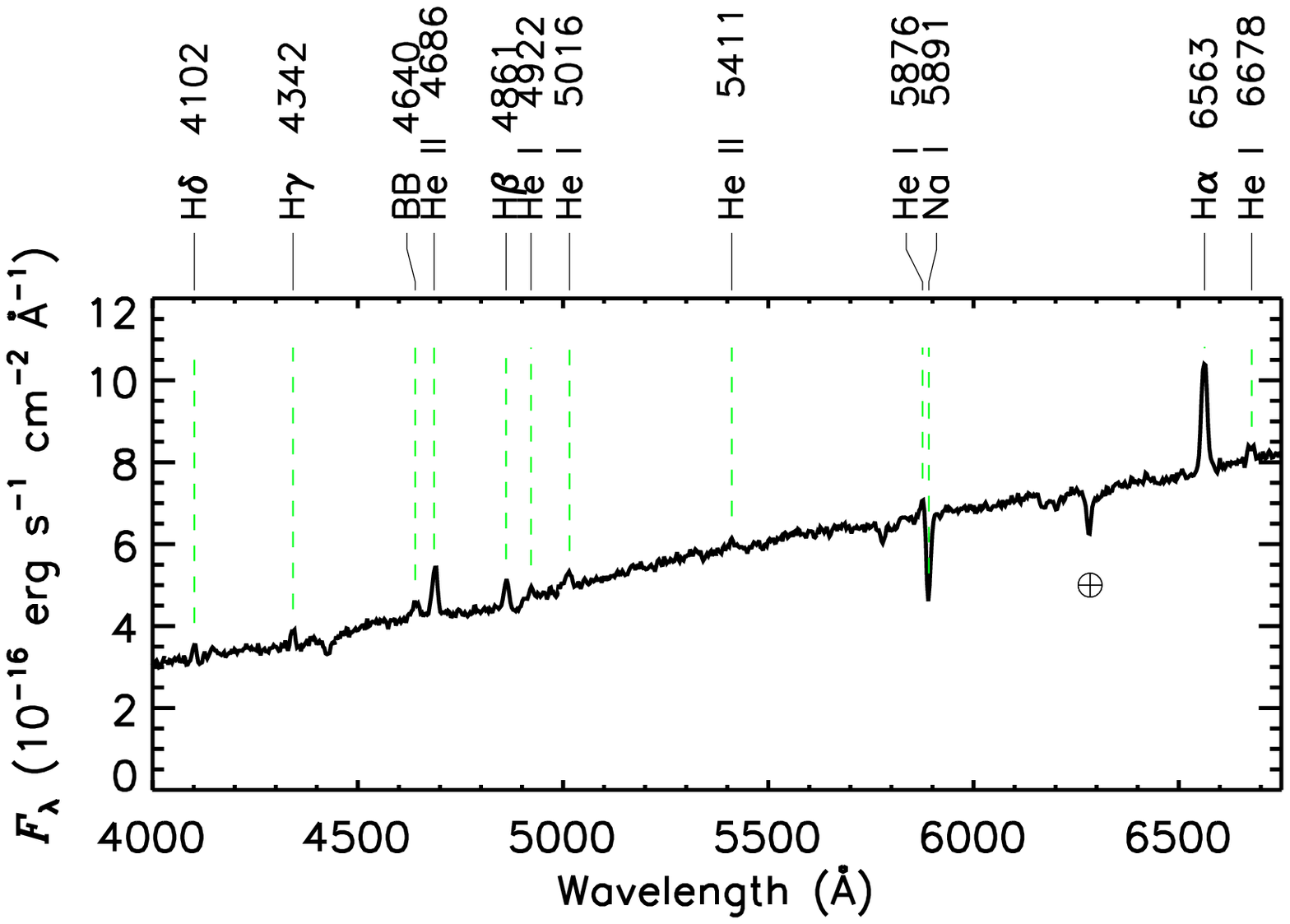}
    \includegraphics[width=8.5cm]{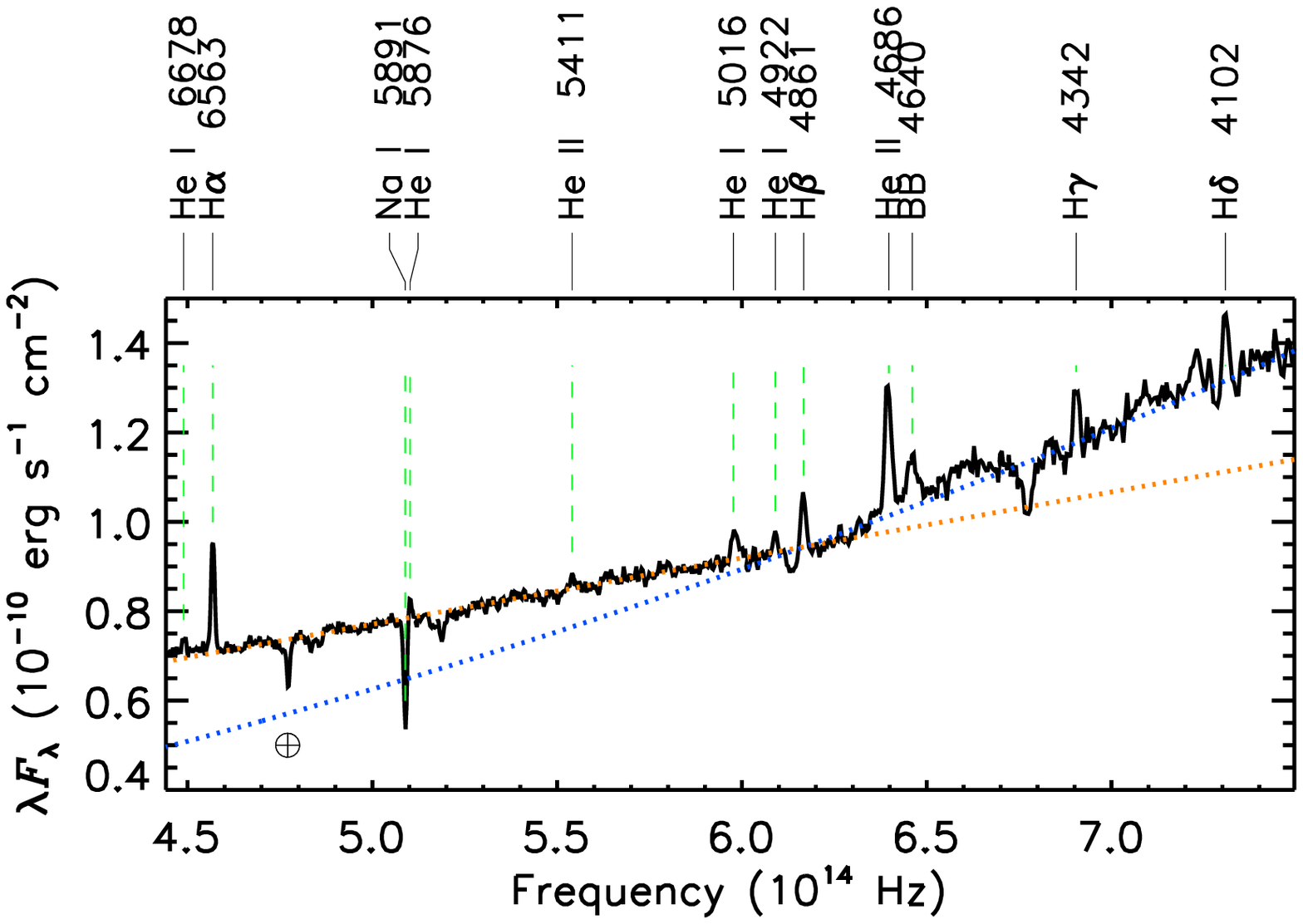}
\caption{Optical spectrum of \gx339 obtained with \vlt/FORS2 on Night 1. {\em (Left)} The observed flux-calibrated spectrum, with main line identifications labelled. The circled cross denotes telluric absorption. {\em (Right)} The dereddened spectrum (assuming \av=3.5) plotted in $\lambda F_{\rm \lambda}$ units and against frequency for better comparison with the broad-band spectral energy distribution discussed later. Note the lower limit of the y-axis is above zero. The dotted lines indicate the continuum power law fits below (blue) and above (orange) 5000 \AA, extrapolated over the full frequency range. 
 \label{fig:optspec}}
  \end{center}
\end{figure*}

The observed optical spectrum is shown in Fig.~\ref{fig:optspec}. \gx339\ lies behind large interstellar reddening, whose exact value has been debated in the literature (see, e.g., extensive discussion in \citealt{zdziarski98}). Among the latest works, \citet{zdziarski98} report \ebv=1.2\p0.1, while \citet{buxton03} quote a value of $\sim$1.1\p0.2. Assuming the standard Galactic value of the total:selective extinction $R_{\rm V}$=3.1 gives \av$\approx$3.5 mags, which was adopted for dereddening corrections based on the Galactic extinction law of \citet{cardelli89}. The error on the reddening translates into a relatively-large uncertainty on the corresponding extinction $\Delta$\av\ of at least 0.5 mag, and hence a dereddened $V$--band flux uncertainty of 1.6, which is likely to be a lower-limit because $R_{\rm V}$ is also known to vary along different sightlines. The final dereddened and flux-calibrated spectrum is also shown in the figure. 

The observed spectrum possesses a continuum that rises towards red wavelengths faster than a flat spectrum source, overlaid with several emission lines. The dereddened continuum, on the other hand, rises towards blue wavelengths, with a simple power-law parametrisation having a form $F_\lambda$$\sim$$\lambda^{-2.4}$, in units of flux density per unit wavelength. A systematic uncertainty of about 0.7 in slope is possible for the full range of reddening correction uncertainty above. But, as can be discerned from the figure, a single power-law is not a good fit, with the slope below $\approx$5000 \AA\ being bluer ($F_\lambda$$\sim$$\lambda^{-3.0}$), and that above being redder ($F_\lambda$$\sim$$\lambda^{-2.0}$) than expected from a single power law. Such a \lq convex-shaped\rq, broken power-law continuum is not a result of over-correcting for the reddening; we confirmed this by using \av\ values at the lower end of expected range. We also note that the flux calibration fit resulted in an excellent rms of 0.01 mags over the range of 4000--7500 \AA, and a broken slope appeared on the other night of data as well (not shown). 

The observed spectral energy density normalisation is $F_{\lambda}$(5500 \AA)= 6$\times 10^{-16}$ erg s$^{-1}$ cm$^{-2}$ \AA$^{-1}$, which is equivalent to $V_{\rm Vega}\approx 17$. This matches contemporaneous photometric monitoring of the source \citep{buxtonbailyn07}. Our average extinction correction and related systematic uncertainties then yield $M_{\rm V}$=--1.02${\pm 0.5}$ or $\lambda L_\lambda$(5500 \AA) $\approx$ 6$_{-2}^{+4}$ $\times$$10^{35}$ erg s$^{-1}$, for the assumed distance of 8 kpc. 
 
The strongest observed emission lines include the Balmer series from \ha\ to \hd, the Bowen blend around 4640 \AA\ and several \hei\ and \heii\ lines. These lines are typical for \gx339\ in the low/hard state \citep{soria99, buxton03}. See Table~\ref{tab:optspeclines} for the line strength measurements. Most lines, especially the strong ones (e.g. \ha\ and \hei~\l 4686) appear single-peaked within the limits of our modest spectral resolution. The Na I interstellar absorption line is also visible for which we measure an equivalent width of 3.8~\AA, though no correction for optical depth effects has been made. This lies within the range of line strengths tabulated by \citet{buxton03}. 

\begin{table}
\begin{center}
  \begin{tabular}{lr}
    \hline
    Line              &   Eq. Width  \\
                      &   \AA  \\
    \hline
    \hd\ 4102         & 1.00\p0.26\\ 
    \hg\ 4342         & 1.09\p0.24\\ 
    Bowen blend (BB)  & 2.12\p0.39\\ 
    \heii\ 4686       & 4.61\p0.52 \\ 
    \hb\ 4861         & 1.86\p0.21\\ 
    \hei\ 4922        & 0.56\p0.08\\ 
    \hei\ 5016        & 1.23\p0.24\\ 
    \heii\ 5411       & 0.70\p0.35\\ 
    \hei\ 5876        & 0.90\p0.22\\ 
    \nai\ 5891        & --3.84\p0.24 \\ 
    \ha\ 6563         & 6.17\p0.17\\ 
    \hei\ 6678        & 0.75\p0.17\\ 
    \hline
  \end{tabular}
  \caption{Line features detected in the FORS optical spectrum of \gx339\ obtained on Night 1. Eq. Width is the observed equivalent width measured from a single symmetric gaussian fit, with negative values denoting absorption. Error propagation on the line width, normalisation and continuum level was used to determine the stated 1$\sigma$ uncertainties.\label{tab:optspeclines}}
\end{center}
\end{table}

\subsection{X-ray}
\label{sec:xspec}

X-ray spectroscopy quasi-simultaneous with our observations has been discussed in detail by \citet{tomsick08} -- their \lq spectrum 2\rq\ was last observed on 2007 Jun 14, corresponding to our Night 3. They used data from the \swift\ satellite and its X-Ray Telescope (XRT) instrument, in combination with \rxte. The XRT employs a CCD detector and good energy resolution below 10 keV, making it superior to \rxte/PCA in terms of isolating sources and for modelling the low energy (and Fe K$\alpha$) spectrum regions. So we present only minimal details of our spectroscopic analysis here. 

A simple absorbed power law with a photon index $\Gamma$=1.63\p0.01 (uncertainty for $\Delta\chi^2$=1) absorbed by a fixed Galactic \nh=5.3$\times$10$^{21}$ cm$^{-2}$ \citep{dickeylongman90} gives a best fit statistic of $\chi^2$/dof=99.5/96 (dof being the number of degrees of freedom) over 3--200 keV, and no significant cross-calibration offset between the two instruments. Adopting a reflection model (\pexrav) instead, results in moderate improvement at $\chi^2$/dof=96.1/95, for $\Gamma$=1.66\p0.04, reflection fraction $\Omega$/2$\pi$=0.15\p0.1 but with unconstrained power-law cut-off energy. This is in good agreement with \citet{tomsick08}, and we refer the reader to that paper for further discussion on the spectral modelling. Here, we simply measure the source power from our data. 

The observed source 3--10 keV flux from the power law model is $F_{3-10}$=1.39(\p0.01)$\times$10$^{-10}$ erg s$^{-1}$ cm$^{-2}$, where the error is for 90 per cent confidence, estimated using the {\tt cflux} model parameter in \xspec\ using the PCA data alone. Extrapolating the model energy range and correcting for absorption gives a 2--10 keV flux of 1.75(\p0.02)$\times$10$^{-10}$ erg s$^{-1}$ cm$^{-2}$ and a luminosity of $L_{2-10}$=1.3$\times$10$^{36}$ erg s$^{-1}$ at 8 kpc. For comparison with the Eddington luminosity (\ledd), a broader band is more appropriate; our model gives $L_{1-100}$=5.3$\times$10$^{36}$ erg s$^{-1}$, corresponding to 0.007$\times$\ledd. No important luminosity variations are seen between the different nights.

\subsection{Broad-band spectral energy distribution}
\label{sec:sed}

\begin{figure}
  \begin{center}
    \includegraphics[width=8.5cm,angle=0]{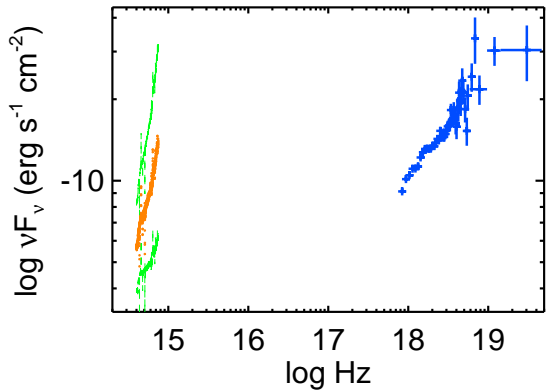}
\caption{Broad-band SED of GX~339--4, with simultaneous intrinsic optical (orange) and unfolded X-ray model (blue) spectral fluxes from Night 1. The green optical spectra represent the systematic uncertainty of the reddening correction, computed for $\Delta$\av=\p 0.5.
 \label{fig:sed}}
  \end{center}
\end{figure}

Fig.~\ref{fig:sed} shows the broad-band spectral energy distribution (SED) of \gx339\ including the simultaneous X-ray and optical data from the previous section. Night 1 measurements are used here, as this had photometric weather conditions. We note that the optical spectral fluxes are close to those obtained by conversion of the mean count rates in ULTRACAM monitoring to $BV$ filter fluxes, assuming standard zero-points and filter transformations applicable for the ULTRACAM Sloan filters \citep{jester05}. The X-ray data shows the unfolded PCA+HEXTE power-law model. 

The optical flux of the source is a significant fraction of its X-ray power, with monochromatic flux ratios of $\lambda L_\lambda$(5500 \AA)/$\lambda L_\lambda$(2 keV)=0.50. In terms of integrated fluxes, $\lambda L_\lambda$(5500 \AA)/$L_{\rm 2-10\ keV}$=0.46, or $L_{\rm 400-700\ nm}$/$L_{\rm 2-10\ keV}$=0.30 and $L_{\rm 400-700\ nm}$/$L_{\rm 1-100\ keV}$=0.07, respectively. Dereddening introduces systematic errors of a factor of $\approx$1.6 around 5500~\AA\ (\S~\ref{sec:optspec}). 

\section{Results: Timing}
\label{sec:timing}

\subsection{Light curves}
\label{sec:opticaltiming}

\begin{figure*}
  \begin{center}
    \includegraphics[width=18cm]{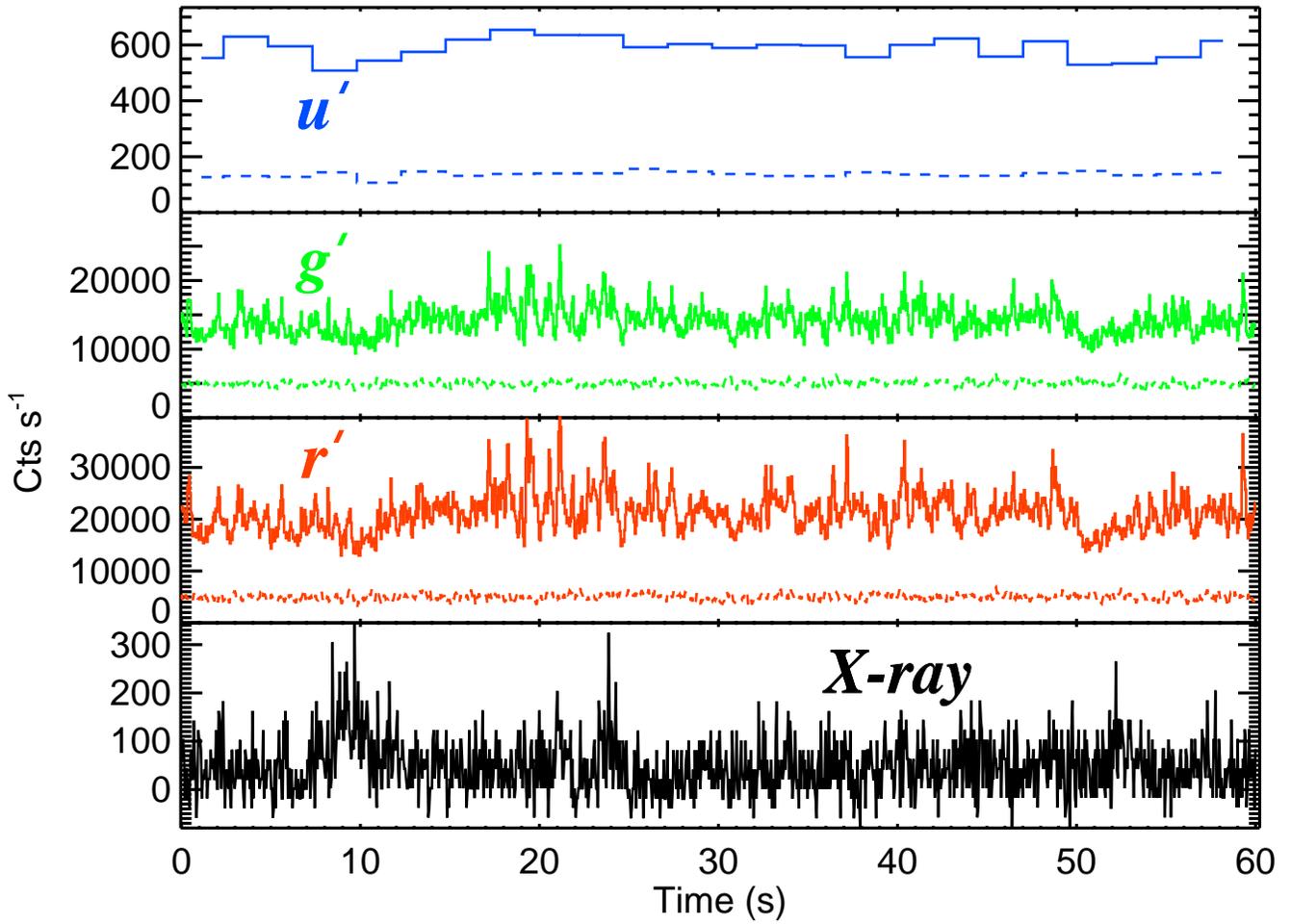}
\caption{Simultaneous section of extracted net barycentered lightcurves from Night 1. The ULTRACAM $r'$ and $g'$ data have a time resolution of 50 ms, and the X-ray PCA full band data (summed over two active PCUs on this night) is matched to this for comparison. The $u'$ data is 50 times slower. The lower dashed curves in all optical filters are for the comparison star and normalised to the mean flux of GX~339--4 in each filter (with an offset downwards for clarity), so any atmospheric scintillation will appear as equal amplitude modulations for both target and comparison.
 \label{fig:lcsection}}
  \end{center}
\end{figure*}

Short 60~s sections of the net barycentered light curves from the best night (Night 1) are shown in Fig.~\ref{fig:lcsection}. These are simultaneous for the three ULTRACAM filters, and the full band \rxte\ PCA {\tt GoodXenon} events. The X-ray time resolution for this display is matched to the $\approx$50 ms binning of the $r'$ and $g'$ filters, while the 50 co-adds for each $u'$ data point result in a time bin of $\approx$2.5 s. Both the $r'$ and $g'$ data show a mean flux that is slowly variable on timescales of several seconds or longer. Superposed on this are rapid flaring events. All this variability is statistically very significant, as can be easily deduced by eye from the high labelled net count rates, as well as the flat comparison star light curve, shown for the $r'$ filter. This comparison light curve is shown normalised to the median flux of GX~339-4 (and offset) for clarity; with this scaling, atmospheric scintillation equally affecting both the target and the standard star should cause variations of identical amplitude in both. It is seen from the comparison star light curve that the effect of the atmosphere is minute, relative to the changes seen in the target, which must be intrinsic. 

\begin{figure}
  \begin{center}
    \includegraphics[width=8.5cm]{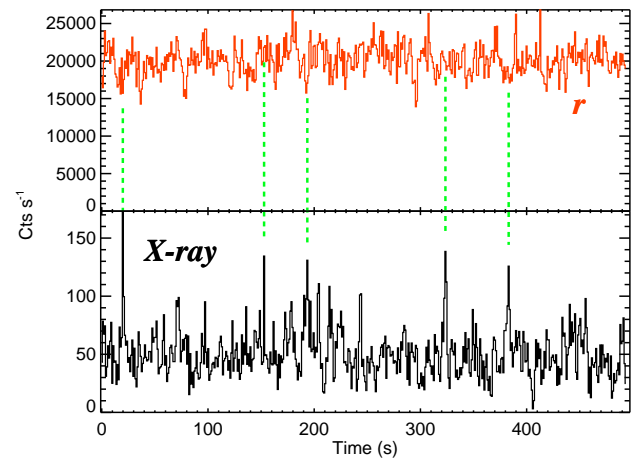}
\caption{Simultaneous 500 s section of extracted net barycentered lightcurves from Night 1 in ULTRACAM $r'$ and X-ray PCA full band data, binned to 1 s time resolution. X-rays flares with $f$$\ge$2.4 are marked.
 \label{fig:lcsection1s}}
  \end{center}
\end{figure}

The novelty of our work is the fast time ($\ltsim$ 1 s) variability of the source at multiple wavelengths. Variability on slower timescales has already been investigated in the literature \citep[e.g. ][]{motch82, steiman-cameron90, homan05}, but in order to ease comparison of results, we present in Fig.~\ref{fig:lcsection1s}) a longer 500 s section of the simultaneous X-ray and $r'$ light curves binned to 1 s resolution (the variations in the other optical filters are similar). This section was chosen randomly with the only criterion that it contain several prominent X-ray flares, the strongest of which are marked according to the superposed flare decryption that we will next describe. The figure shows that there is no significant evolution of the general light curve characteristics on this timescale. 

\begin{figure*}
  \begin{center}
    \includegraphics[width=8cm]{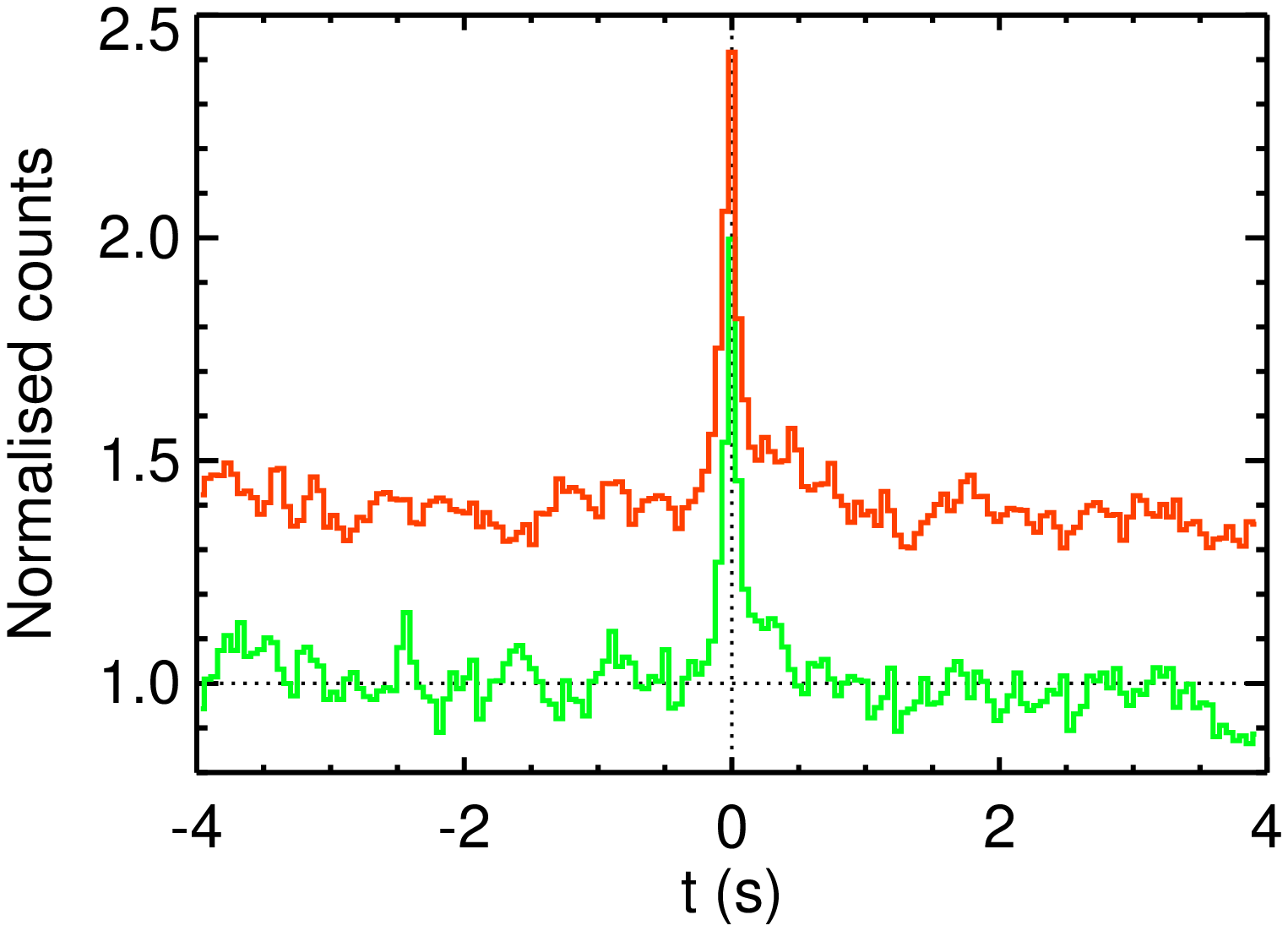}
    \includegraphics[width=8cm]{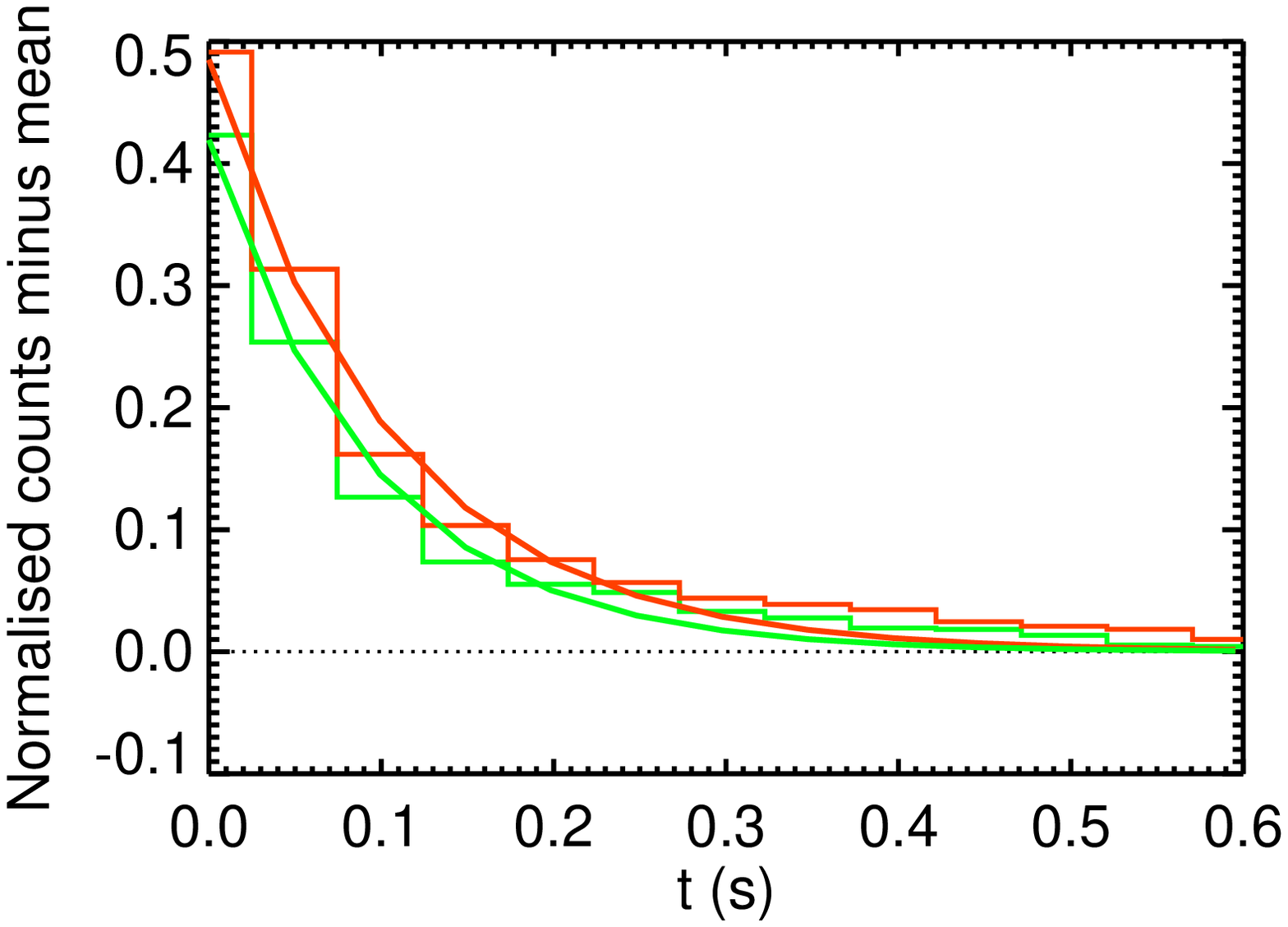}
\caption{Peak-aligned optical flares properties, normalised to the local mean flux level. ({\em Left}) The strongest flares, selected with $f$$\ge$2. The lower (green) histogram is the average for $g'$ and the top (red) histogram is for $r'$ (offset by +0.4 for clarity). In both cases, the peak has a value of about twice the local mean level. ({\em Right}) The superposition of more typical flares selected with $f$$\ge$1.2 is shown, overplotted with best-fit exponential decay profiles with a decay time of 0.11 and 0.09 s in $r'$ (red) and $g'$ (green) respectively. 
 \label{fig:optflares}}
  \end{center}
\end{figure*}

What is the maximum flare strength associated with the fast optical variations? This can be answered by means of a superposed shots analysis (cf. \citealt{negoro01}, \citealt{malzac03}, Paper I). We selected flares with a peak strength of $f$ times above the local mean in a running $t_m$=32 s long section, and also required the flares to be maxima within a contiguous segment of $\pm t_p$=4 s. This assures that only distinct, significant flares are selected. The sections were all continuum-normalised, peak-aligned and averaged. The results for flares with $f$$\ge$2 are shown in the left panel of Fig.~\ref{fig:optflares}. The averaged flares have a peak value of close to a factor of 2 above the local mean, which means that strongest flares in both $r'$ and $g'$ are associated with flux increases of this magnitude. 

The plot shows the flares to be very narrow with widths of at most a few time bins. In order to quantify the typical time scales associated with the optical flaring, we selected more common lower-amplitude flares with $f$$\ge$1.2 to which we fitted a simple exponential time decay model of form $Ae^{-t/\tau}$. This is shown in the right hand panel of Fig.~\ref{fig:optflares}, where the profile of the average from flare superposition in both filters is plotted after removal of the normalised continuum of value 1. Overlaid are decay models fitted at times of $<$1 s from the average flare peak. We find $\tau$=0.11\p0.01 s and 0.09\p0.01 s in $r'$ and $g'$, respectively, implying that the fastest rise and fall timescales of optical flaring that we probe are only $\approx$100 ms.

\begin{figure}
  \begin{center}
    \includegraphics[width=8cm]{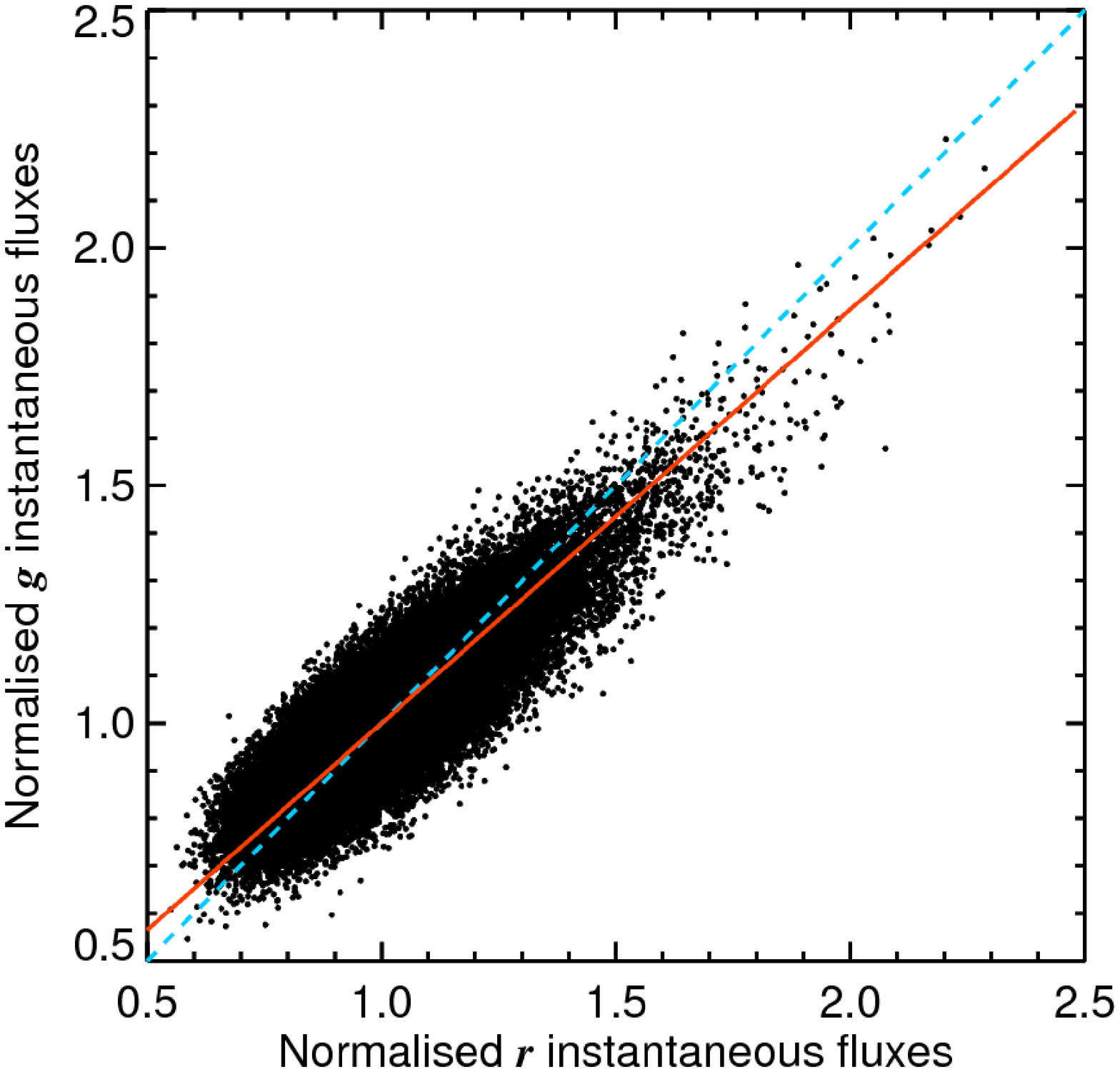}
\caption{Instantaneous light curve count rate measurements for all 68317 time bins from 56 mins of data in $r'$ vs. $g'$ on Night 1, normalized to the mean flux in each filter. The 1:1 line is the dashed light blue one, and the red continuous line is the best-fit to the linear model of $y=mx+c$ obtained by the ordinary least square bisector method of \citet{isobe90}, with $m$=0.871\p0.002 and $c$=0.129\p0.002. The gradient is flatter than the 1:1 line, which means that flaring is stronger in $r'$. 
 \label{fig:rg}}
  \end{center}
\end{figure}

The variations in the two filters match each other well, and the peak to peak variability across the entire $r'$ and $g'$ lightcurves is a factor of 4. This can be discerned, for instance, in Fig.~\ref{fig:rg}, where we plot every simultaneous flux measurement from the light curve of Night 1 in the two filters. The range of variations on both axes encompass a factor of 4. These lightcurves, when binned to the slower $u'$ data, also result in a good match to the $u'$ variations. 

Table~\ref{tab:avgrates} shows the average count rates in different bands. Also listed are the fractional rms variability amplitudes. These are measured from the entire lightcurves and denote the excess variation above that due to Poisson uncertainties \citep[cf. ][]{vaughan03}:

\begin{equation}
{\rm rms}=\frac{\sqrt{var-\overline{{\sigma_{\rm err}^2}}}}{\overline{l}}
\label{eq:rms}
\end{equation}

\noindent
where $\overline{l}$ denotes the mean count rate of the light curve $l$ in any band, $var$ is its raw variance, $\sigma_{\rm err}$ the individual errors on the count rate measurements and $\overline{\sigma_{\rm err}^2}$ the mean square error.

The X-ray light curves were extracted with a time bin of 2$^{-8}$~s from the full PCA band. In addition, light curves in two other bands were extracted for comparison: $\approx$2--5 keV (PCA channels 5--11) and $\approx$5--20 keV (PCA channels 12-47). The X-ray count rate is much lower, and fractional variability higher, as compared to the optical. The strongest optical variability occurs in the $r'$ band, at about 15 per cent rms across the full light curve. This red rms variability is perceptibly stronger than in $g'$, as can also be seen from the slope of the fit to the distribution of instantaneous fluxes in Fig.~\ref{fig:rg}. 

\begin{table}
 \begin{center}
  \begin{tabular}{lcccccr}
    \hline
      Band               &    \multicolumn{2}{c}{\underline{        Night 1            }}&    \multicolumn{2}{c}{\underline{        Night 2            }}&    \multicolumn{2}{c}{\underline{        Night 3            }}\\   
                         &     $R$     &  rms                        &     $R$     &  rms                        &     $R$     &  rms                        \\   
    \hline                                                                                                       
     X-ray PCA full      &      24     &  0.46                       &      22     &  0.42                       &      23     &  0.41                       \\
     X-ray 2--5 keV  &       8     &  0.43                       &       7     &  0.41                       &       7     &  0.38                       \\
     X-ray 5--20 keV &      15     &  0.47                       &      14     &  0.42                       &      14     &  0.42                       \\
       $r'$              &    20190    &  0.15                       &     8080    &  0.15                       &     3030   &  0.16                       \\
       $g'$              &    13790    &  0.12                       &     5430    &  0.13                       &     2040   &  0.12                       \\
       $u'$              &      610    &  0.06                       &      --     &   --                        &       --    &   --                        \\
    \hline
  \end{tabular}
  \caption{Light curve properties. $R$ represents the net counts per second (rounded off to the nearest 10 ct s$^{-1}$ in the optical). In X-rays, this is stated per active PCU in the PCA instrument. rms is the fractional variability amplitude above Poisson uncertainties (Eq.~\ref{eq:rms}). Typical 1$\sigma$ error on this value is $\ltsim$0.02. Light curves with time resolutions matching the best optical time resolution on each night were used for these measurements in both optical and X-rays; only the $u'$ observations are 50 times slower (see Table~\ref{tab:obslog}). Note that non-photometric weather causes the mean optical flux to decrease significantly on Nights 2 and 3.
\label{tab:avgrates}}
 \end{center}
\end{table}

\subsection{Power spectra}
\label{sec:psds}

The power spectral densities (PSDs) are computed as the modulus-square of the discrete Fourier transforms of the light curves as

\begin{equation}
PSD(f)=A\left| \sum_{i=1}^N{(l_i-\bar{l})e^{2\pi{\rm i}ft_i}}\right|^2,
\label{eq:psd}
\end{equation}

\noindent
with the frequencies themselves being discretised at $N/2$ evenly-spaced values from 1/$N\Delta T$ to the Nyquist frequency 1/2$\Delta T$, where $\Delta T$ is the time bin (cf. \citealt{vaughan03}). 

The X-ray PSDs were computed from net lightcurves with time resolution of 2$^{-8}$ s over the full PCA energy range. The data from each night was split into segments of 256 s, and the power spectra were computed using standard \xronos\ procedures, and cross-checked with a custom routine coding Eq.~\ref{eq:psd} and also used for the optical PSDs below. In this case, $\bar{l}$ refers to the mean count rate in any given segment under consideration. The PSDs were averaged across the segments and rebinned in logarithmic frequency bins by a factor of 1.1 in order to improve signal:noise, especially at high Fourier frequencies. The optical PSDs were computed using an identical procedure, on the fastest time resolutions available for each filter and each night. A rebinning factor of 1.05 was sufficient for all data on Nights 1 and 2. On Night 3, no acceptable solution was found for the PSD fits that we will describe shortly, likely as a result of bad weather systematics. For this night, we rebinned by a factor of 1.1. The constant $A$ in Eq.~\ref{eq:psd} was determined following usual rms-squared standardisation $A$=2$\Delta T$/$\bar{l}^2$$N$ \citep[e.g. ][]{vanderklis97,bellonihasinger90,vaughan03}, which normalises the PSD to the fractional light curve variance per unit frequency interval (above Poisson uncertainties if white noise is subtracted, as described below). The PSDs are shown in Fig.~\ref{fig:psds} for each night, in both optical and X-rays.

\subsubsection{Noise levels}

Poisson uncertainties introduce a constant white noise level, which becomes increasingly important at high Fourier frequency. For individual (mean) count rate errors $\sigma_{\rm err}$ ($\overline{\sigma_{\rm err}}$), the noise for rms-squared normalised PSDs given by

\begin{equation}
 n=\frac{2 \Delta T \overline{\sigma^2_{\rm err}}}{\bar{l}^2}.
 \label{eq:psdnoise}
\end{equation}

\noindent
This proved to be an excellent measure of the constant noise level in the X-ray PSDs, as is readily apparent in Fig.~\ref{fig:psds}. It also showed PCA dead time corrections to be unimportant in determining this noise level for the low count rate relevant here; this was confirmed by using Eq.~4 of \citet{nowak99_cygx1_ii} to determine the correction factor. Poisson uncertainties on the gross (source+background) count rates are the only important contribution to the noise in X-rays. In the optical, additional contributions to the variance are CCD readout noise and dead time. The read noise contribution is automatically accounted for in the photometric errors returned by the pipeline. The presence of dead time means that any fast flux variations that occur within the dead time interval are not sampled and cannot be distinguished from source variations on (longer) sampled times. Effectively, some power is \lq aliased\rq\ down to frequencies below the Nyquist sampling frequency of $1/2T_{\rm cycle}$ \citep[e.g. ][]{vanderklis89}. The high duty cycle of our observations (Table 1) means that aliasing will be small and can be approximated by increasing the noise level by the ratio of $T_{\rm cycle}/T_{\rm exp}$; see Eqn. (A3) of \citealt{vaughan03}. The resultant corrections increase the noise power by $\approx$ 2--3 per cent between the nights (cf. \S~\ref{sec:ultracamobs}) and have been applied. Finally, photometric errors from the comparison star against which relative photometry is performed are included in the final noise level via Gaussian propagation of the errors to $\overline{\sigma^2_{\rm err}}$ in Eq.~\ref{eq:psdnoise}. The noise levels are indicated on Fig.~\ref{fig:psds} and show all optical data to lie above the white noise level up to the Nyquist frequency. Several tests of the photometry and errors were carried out to confirm this, and the effect of any possible systematic misestimate in noise levels is discussed alongside other results presented below and in the following sections. In any case, the optical noise levels are generally low because of the high optical count rates (Table~\ref{tab:avgrates}); consequently, their effect on the mean PSD levels is much smaller than in X-rays over the full frequency range probed. The increasing optical noise level on Nights 2 and 3 can be explained by a significant decrease in detected source count rates through high and strongly variable transparency and seeing conditions. This also degraded the quality of the slower $u'$ data, for which only the Night 1 data are presented. 

The results after noise subtraction are displayed in the bottom row of Fig.~\ref{fig:psds} for each night of data separately. The X-ray variability power is larger than that in the optical on all nights. Noise dominates the X-ray PSDs at Fourier frequencies of $\gtsim$10~Hz, with the result that using light curves with longer binning (e.g. 50 ms matched to the optical) produce identical PSDs to those made from the above finer time binning. The optical power spectra also show band-limited noise continua up to several Hz at least. Above $\sim$0.1 Hz, they exhibit strong curvature to high Fourier frequencies. The limiting Nyquist frequency decreases from Night 1 to Night 3 as a result of the increasing time bins with worsening weather. Around $\sim$0.05 Hz, there is a local excess on all nights, whereas at the lowest frequencies, the continuum slope seems to exhibit a change. Despite the changeable weather, the overall PSD structures on all three nights match. 

\subsubsection{Model fits}
\label{sec:psdfits}

In order to parametrise the PSD shapes, we fit them with Lorentzian profiles. Given the low source X-ray count rate, the relatively short total observing times, and the fact that the PSDs appeared similar on all nights, we averaged all three nights in order to characterise the functional form. This was done separately for the full PCA band, as well as the 2--5 keV and 5--20 keV bands. Three Lorentzians were found to produce a good fit in each case, as shown in Fig.~\ref{fig:xpsdfits}. These are all broad (or zero-centred) Lorentzians describing the broad peaks at characteristic frequencies (\numax) of around 0.02, 0.1 and 2 Hz, consistent across the three energy bands. The measurements are listed in Table~\ref{tab:xpsdfits}. No quasi-periodic oscillation (QPO; defined as having a quality factor of $Q>2$) is found, with the narrowest Lorentzian having $Q$$\approx$0.7. The Lorentzian parameters -- $r$, $\nu_0$, $\Delta$, \numax\ and $Q$ -- are defined identically to \citet{belloni02}. 

In the optical, the PSDs were similarly fit with multiple Lorentzians. All PSDs clearly contain a low-frequency QPO and the PSDs covering fast timescales (i.e. the $r'$ and $g'$ data) also demand extra curvature at high Fourier frequencies (Fig.~\ref{fig:psds}). This is especially clear for the Night 1 data which provide the best dataset in terms of probing the broadest range of timescales, and observed under the best weather conditions. A minimum of four Lorentzian components are required to describe this -- one for the QPO and three for the broad-band continuum humps (Fig.~\ref{fig:opsdfitsn1}; Table~\ref{tab:opsdfits}). We also tried different models, including an exponential cut-off power-law, as well as the bending power-law of \citet{mchardy04}, for the PSD fall-off above a Fourier frequency of $\sim$1 Hz. These can roughly describe the curvature over restricted frequency ranges, but require further modification such as additional Lorentzians or multiple bending sub-components, for a reasonable fit. Given our relatively short dataset, we chose to retain the multiple Lorentzian component fits described above as a simple parametrisation of the PSD, and not necessarily as an accurate physical description of the underlying variable components, though as we will describe in the following sections, this parametrisation does provide interesting insight into the X-ray vs. optical coherence and time lags.

Note that we have not included the lowest and highest Fourier frequency bins in the optical fits presented, because the presence of new components are indicated at both ends which cannot be properly modelled. We have kept the fits to below 8 Hz. Including the bins up to 10 Hz changes $\chi^2$ significantly; e.g. $\Delta\chi^2$=+20 for an extra five degree of freedom for the $r'$ fit. The presence of variability noise above our Nyquist sampling frequency is important, and should be probed in future faster observations. A lower limit on the strength of the high frequency component may be obtained by extending the fits shown in Fig.~\ref{fig:opsdfitsn1} and measuring the residuals, which gives an rms of at least 0.5 per cent, over only 8--10 Hz. We also note that the fit is robust to small systematic changes in the white noise level. We tested this by subtracting off a higher noise obtained from fitting a constant (per unit Hz) to the PSD at the highest Fourier frequency bins; the relative normalisation of the individual components does vary, but not the overall fit. It is also worth noting that in the X-ray PSD (Fig.~\ref{fig:xpsdfits}, top panel), there is an apparent excess above the three Lorentzian fit at the high frequency end (centred at $\approx$9.4 Hz). But fitting an extra Lorentzian to this results in a $\Delta\chi^2$ of only --2 for 3 extra degrees of freedom; i.e. this excess is not significant in the present data.

We carried out similar multi-component Lorentzian fits to the $r'$ and $g'$ data for Nights 2 and 3, which cover a narrower range of Fourier frequencies. The PSDs for these two nights are presented in Figs.~\ref{fig:opsdfitsn2} and ~\ref{fig:opsdfitsn3}, and fit parameters listed in Table~\ref{tab:opsdfits}. Three Lorentzian components were sufficient for the data on Night 3 (which cover the narrowest frequency range, and have been rebinned by a factor of 1.1 as mentioned before). The characteristic frequencies of the QPO and the component covering the peak around 1 Hz show very mild variations across the nights, with an apparent increase on Nights 2 and 3. This could simply be a result of the fact that we do not probe the high Fourier frequency Lorentzians well on these nights, which forces the fit peaks to shift slightly in this direction to compensate. In any case, there is consistency between these fit values at the 4 $\sigma$ level across the nights. At low Fourier frequencies below $\approx$0.02 Hz, both Nights 2 and 3 suggest the presence of some small excess power above the Night 1 variability power levels. This may be indicative of incomplete photometric correction when dividing by the comparison star, because any residual transparency variations are likely to be more important on long timescales (low Fourier frequencies). Additional systematic residual effects are apparent on the worst Night 3, where the QPO appears to be broader ($\Delta$$\approx$0.011 in $r'$, compared to a value of 0.004 on the other two nights), and is not well described by a single Lorentzian. 

Within the limits imposed by the above minor caveats, the PSDs may be considered to be stable across the nights, and for the physical interpretation presented in the remainder of the paper, we will mainly refer to the Lorentzian fits for the best Night 1. A more detailed analysis of the PSDs will also be presented in future work.

\subsubsection{The QPO}
\label{sec:results_qpo}

The QPO lies at a Fourier frequency of $\approx$0.05 Hz (on all nights) and has a $Q$ value of 5--6 on the best Nights 1 and 2; Table~\ref{tab:opsdfits}). The $u'$ data covers only low Fourier frequencies, but a simple fit to this clearly shows the QPO to have similar properties in all three filters (Fig.~\ref{fig:opsdfitsn1}). The QPO is not an instrumental artifact, otherwise it would have introduced oscillations in the comparison star light curve as well. It is clear from Fig.~\ref{fig:lcsection} that this is not the case. We also confirmed this by extracting optical PSDs for comparison star. Additionally, there is no known mechanical or electronic cycle either in the instrument, or in the telescope which occurs at these frequencies. 

We also checked the detectability of any low frequency QPO in X-rays by adding a model Lorentzian to the PSD fit of Fig.~\ref{fig:xpsdfits}, with a fixed $\nu_{\rm max}$ and $Q$ identical to that measured in the optical, and obtained a 90\% ($\Delta\chi^2$=2.71) upper limit on its integrated fractional rms of $r_{\rm X-ray}^{\rm QPO}\approx$0.075. This is about 2.5 times larger than $r_{\rm Optical}^{\rm QPO}\approx$0.03 (see Table~\ref{tab:opsdfits}), so we cannot rule out an X-ray QPO identical to that seen in the optical.

\subsubsection{Previous PSD measurements}
\label{sec:previouspsds}

How do these PSDs compare with previous measurements? In X-rays, there have been several studies. For example, our Lorentzian fit is similar to that found by \citet{nowak99_gx339} for their observation 5, which had the faintest source fluxes ($F_{3-9}$=2.5$\times$10$^{-10}$ erg s$^{-1}$ cm$^{-2}$ [\citealt{wilms99}], only about twice as bright as in our observation). In particular, the characteristic frequencies of the two strong broad band features match closely. In the optical, previous measurements have been published below a Fourier frequency of 1 Hz by \citet{motch82, motch83} during a 1981 low/hard state (bright in both optical and X-rays), and by \citet{motch85} during a 1982 X-ray off state. The hard state observation showed a 0.05 Hz (20 s) QPO feature, very similar to our observation. The QPO was found at a higher frequency (0.14 Hz, or 7 s) in the off state. As these data were obtained with a completely different observational setup, this also rules out any instrumental artifice for our QPO detection. The high frequency power decreased rapidly in all these observations (e.g. with a slope of --1.6 in 1981), whereas we find a gentler initial fall-off; fitting the 0.1--1 Hz PSD with a single power-law for comparison with \citet{motch82} gives a slope of $\sim$--0.5. We note that the bright hard state observation of 1981 showed a strong QPO in X-rays as well \citep{motch85}, which is absent in our (fainter) case. Other studies include very fast white light observations by \citet{imamura90} and \citet{steiman-cameron97}, though with no simultaneous X-ray data. The last authors detected strong continuum curvature around 1 Hz and a 0.06 Hz (16 s) QPO during 1996, suggesting some reasonable agreement with our $r'$ and $g'$ PSDs.

\begin{figure*}
  \begin{center}
    \includegraphics[angle=0,width=5.8cm]{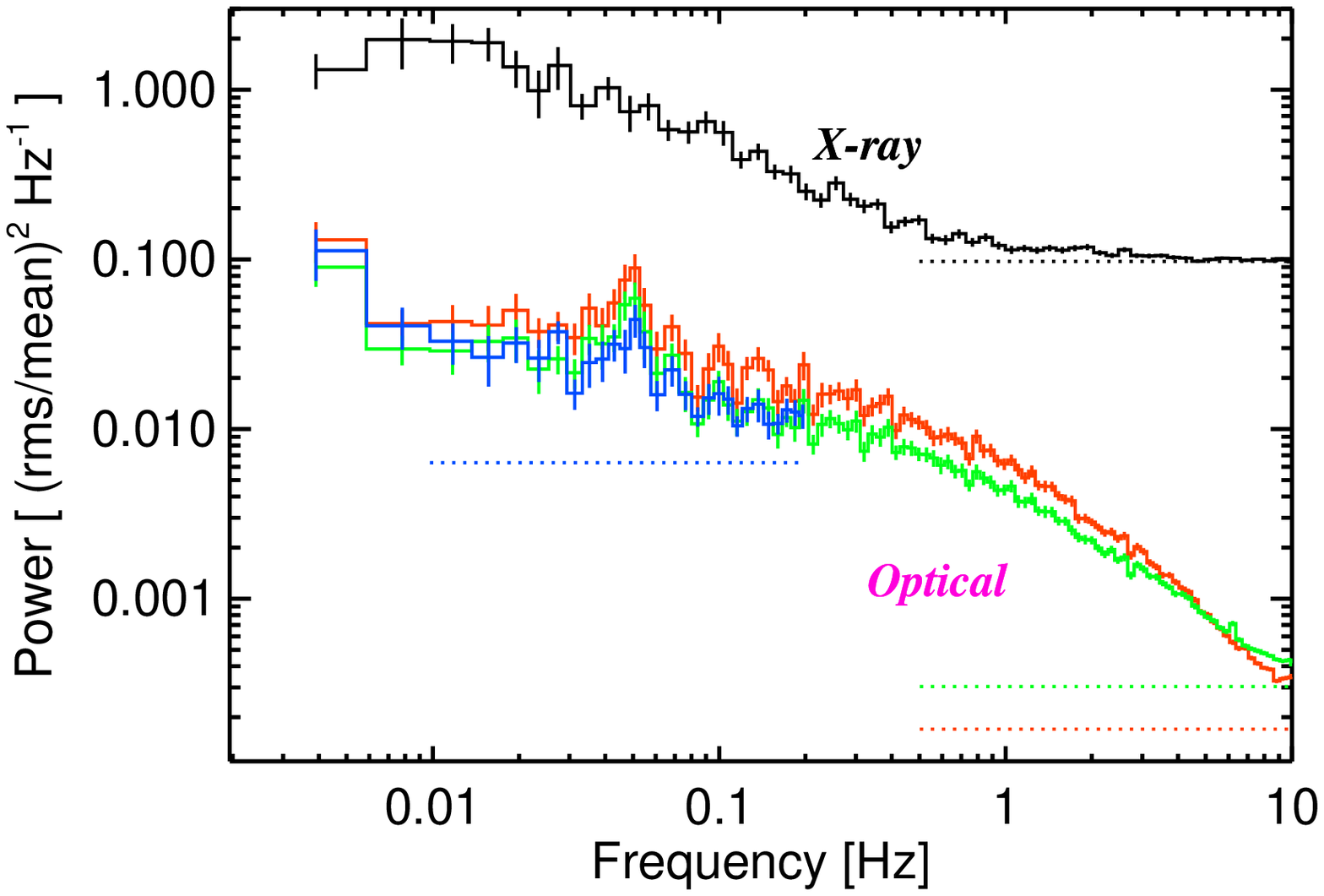}
    \includegraphics[angle=0,width=5.8cm]{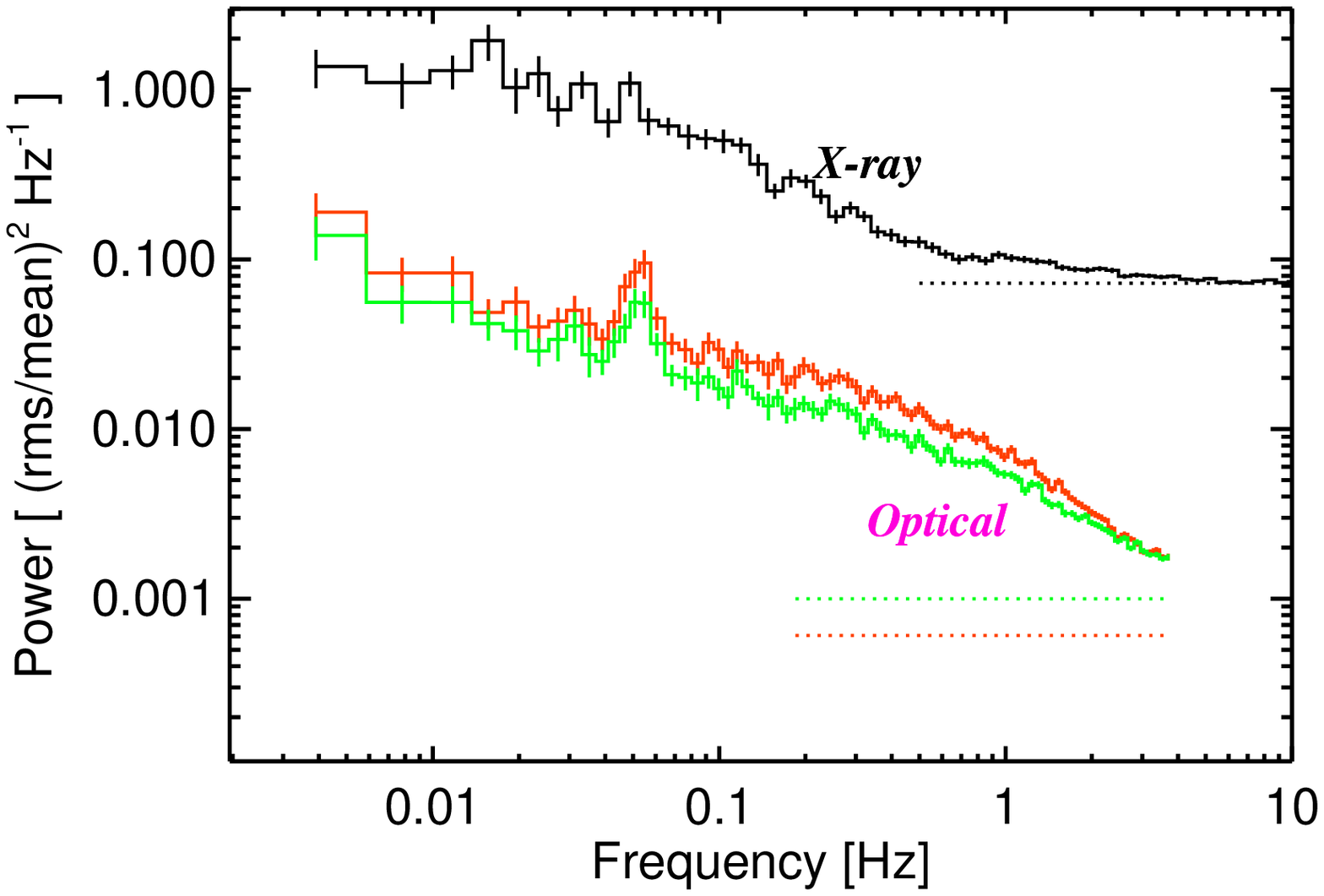}
    \includegraphics[angle=0,width=5.8cm]{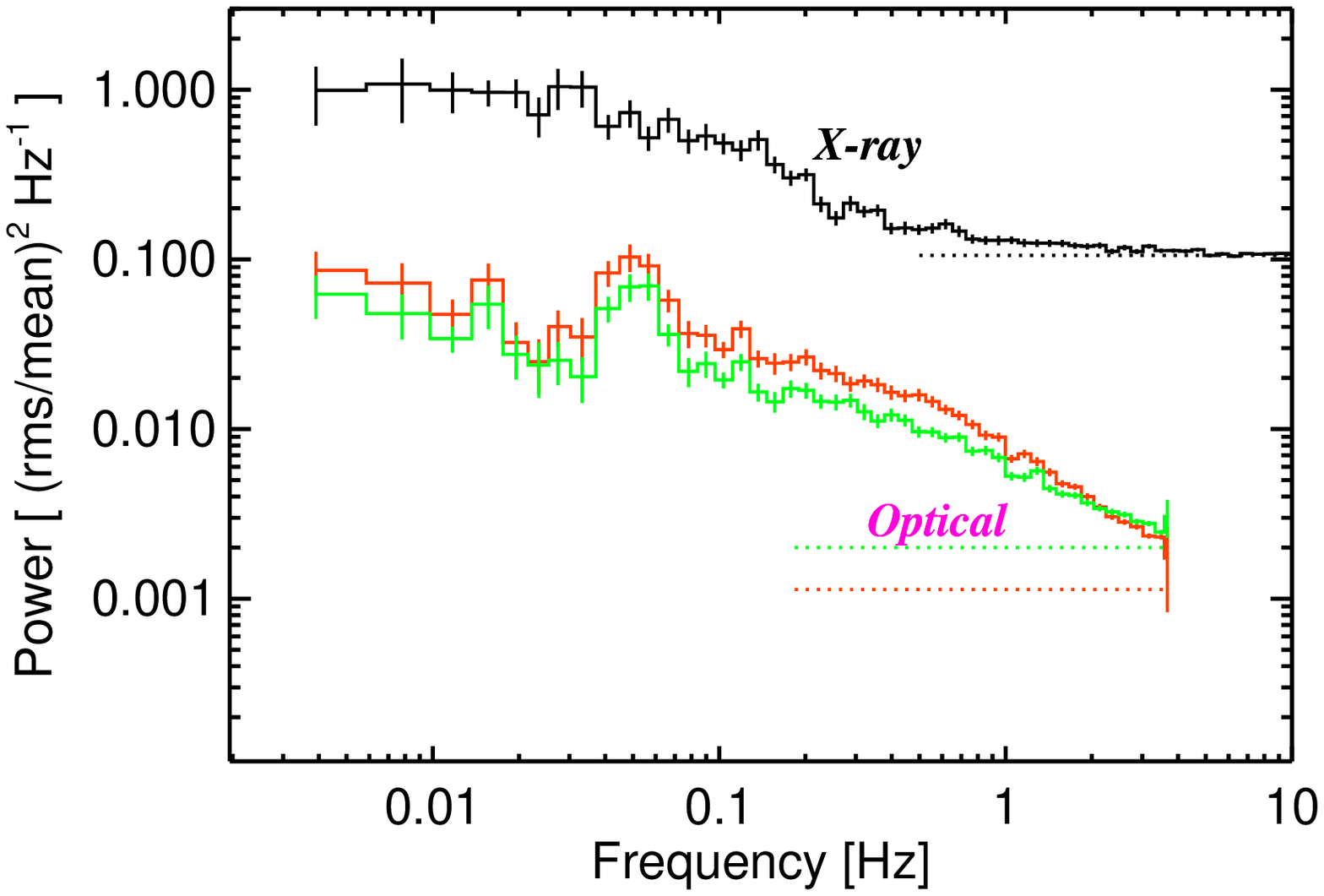}
    \includegraphics[angle=0,width=5.8cm]{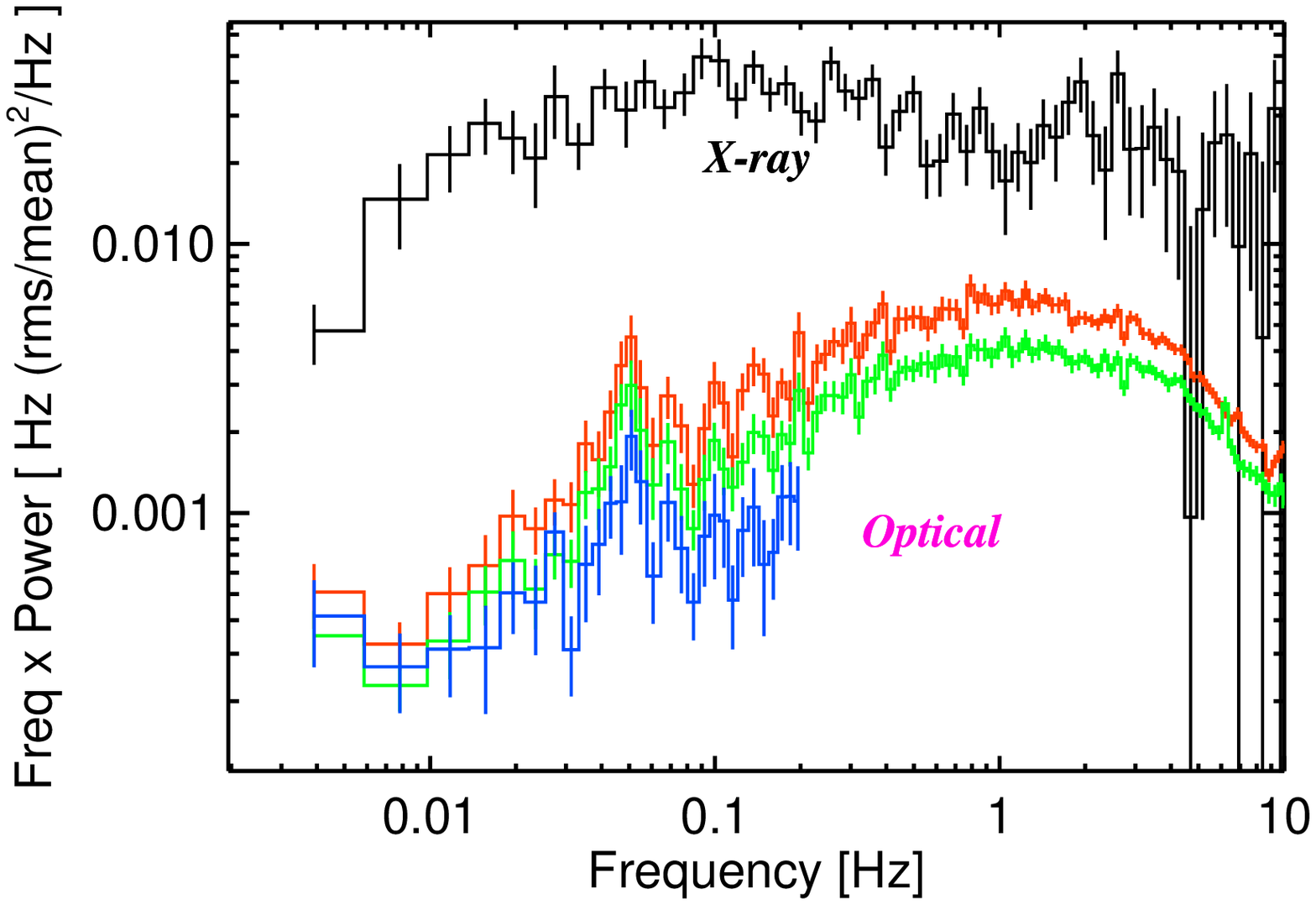}
    \includegraphics[angle=0,width=5.8cm]{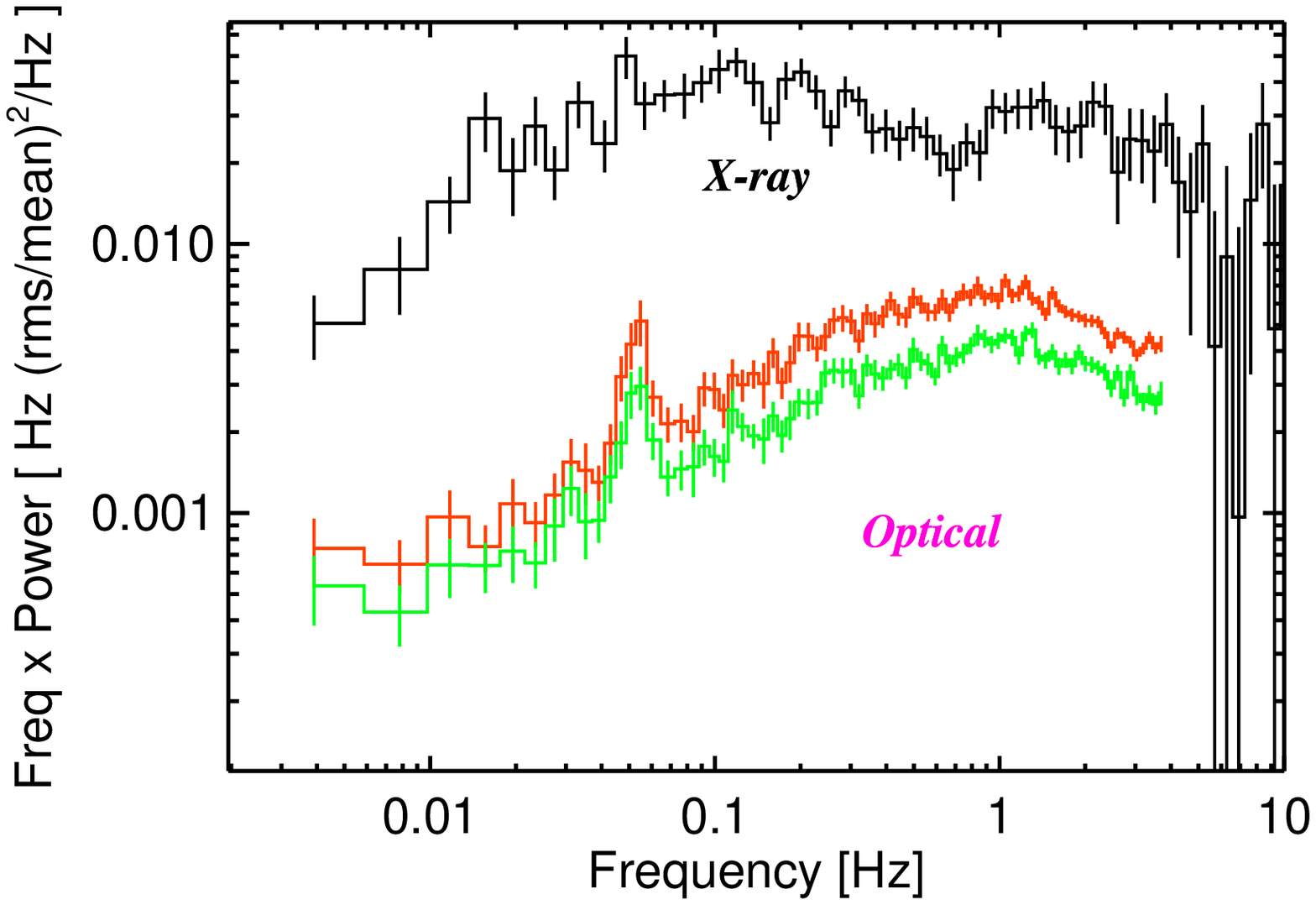}
    \includegraphics[angle=0,width=5.8cm]{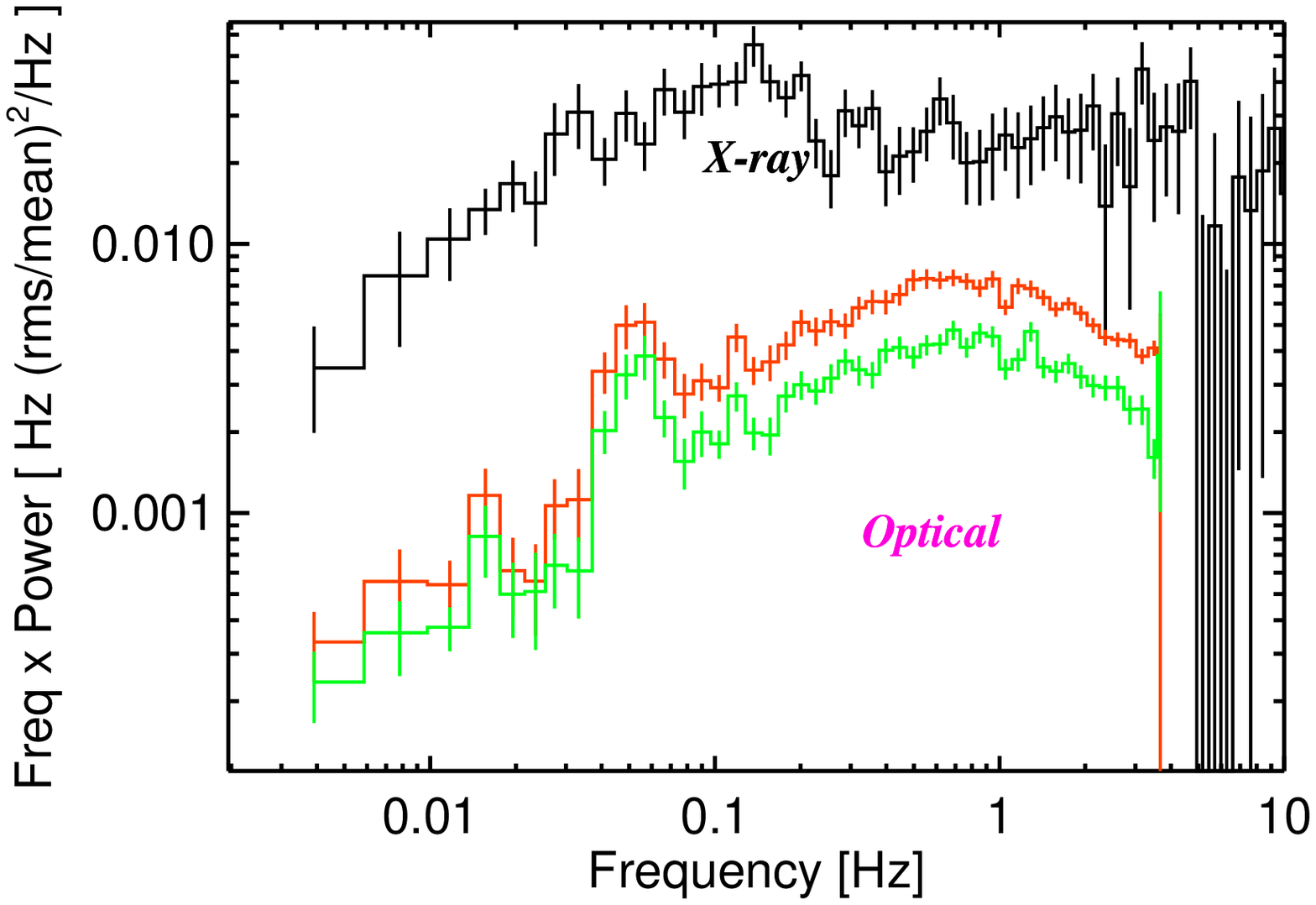}
\caption{Power spectra of \gx339\ for Nights 1--3 (from Left to Right). The X-ray full band PSDs are in black, while the optical ones are in red, green and blue for $r'$, $g'$ and $u'$, respectively. The top row shows the PSDs for the net light curves without subtraction of the white noise contribution, which is indicated by the dotted horizontal lines in the respective colours. Power on the y-axis is per unit Hz, with the standard rms-squared normalisation. In these units, the white noise is a constant, and the integral of the PSD over frequency results in the square of the fractional rms amplitude. The bottom row shows the noise-subtracted PSDs, now in units of frequency $\times$ Power. 
 \label{fig:psds}}
  \end{center}
\end{figure*}

\begin{figure*}
  \begin{center}
     \includegraphics[angle=0,width=8.5cm]{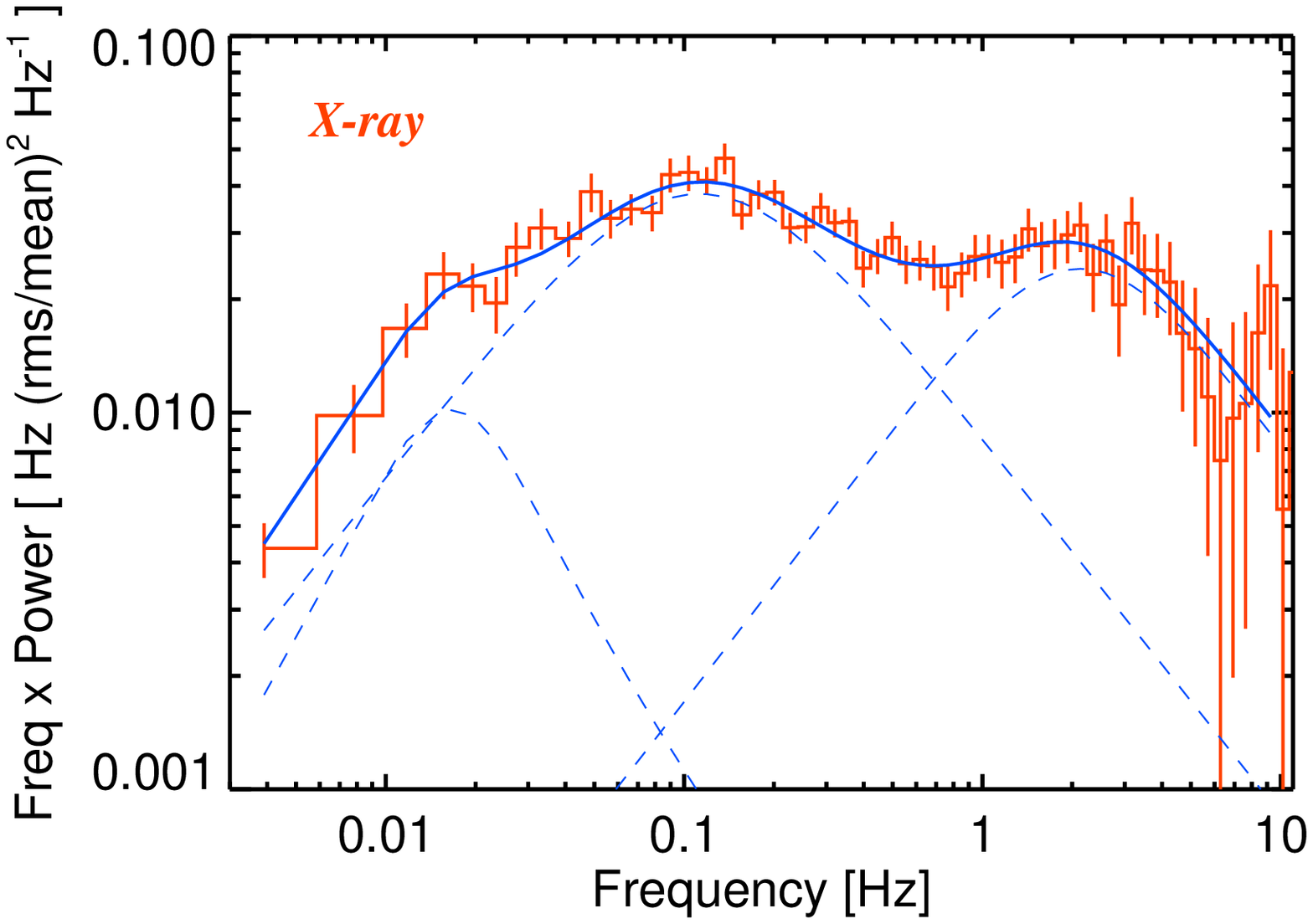}
     \hfill\\
     \includegraphics[angle=0,width=8.5cm]{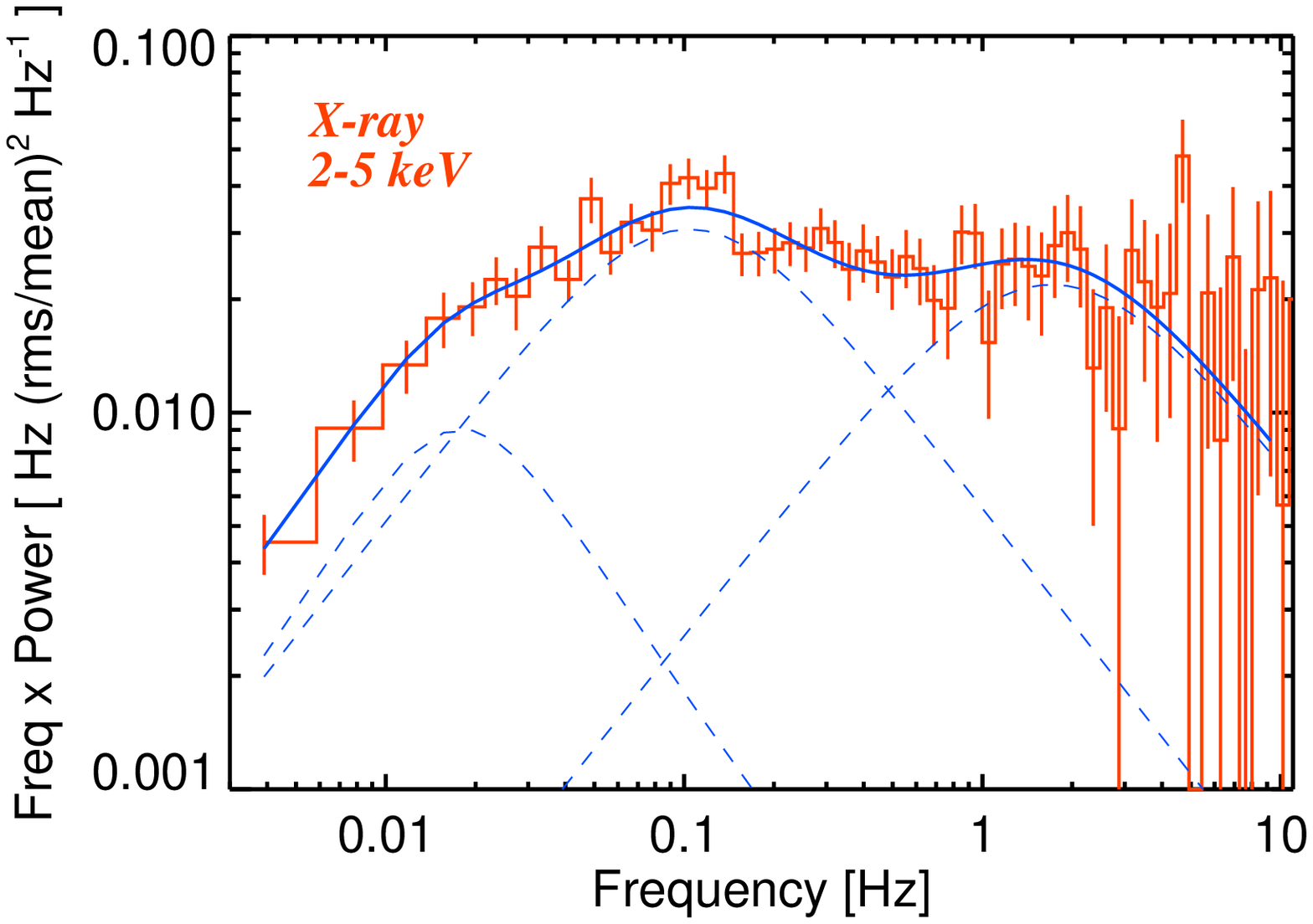}
     \includegraphics[angle=0,width=8.5cm]{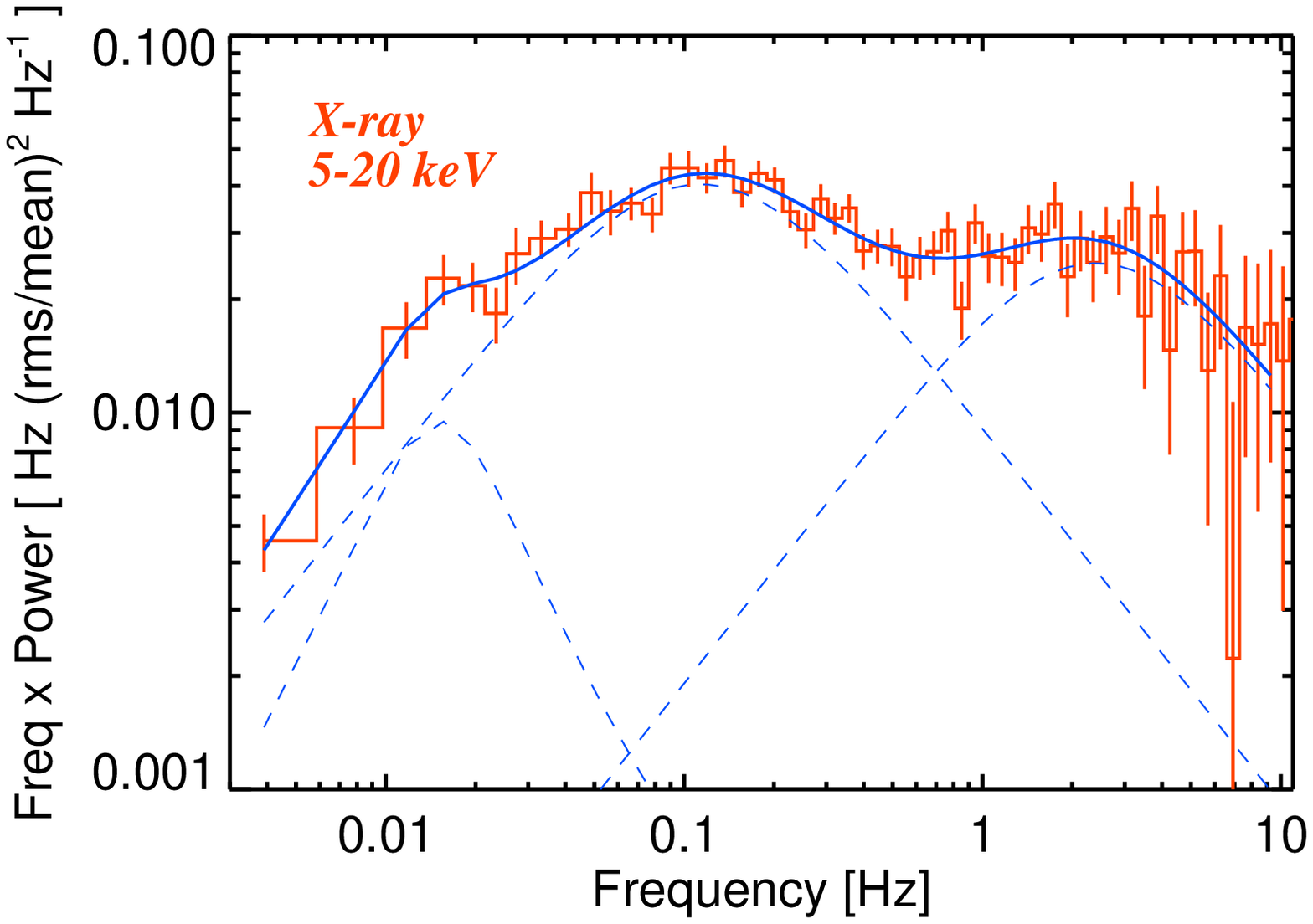}
     \caption{Lorentzian fits to the X-ray PSDs in the full PCA ({\em Top}), the 2--5 keV ({\em Bottom left}) and the 5--20 keV ({\em Bottom right}) bands. The X-ray data are from the average of all three nights. See Table~\ref{tab:xpsdfits} and text for fit parameters and details.
 \label{fig:xpsdfits}}
    \end{center}
  \end{figure*}

\begin{figure*}
  \begin{center}
     \includegraphics[angle=0,width=8.5cm]{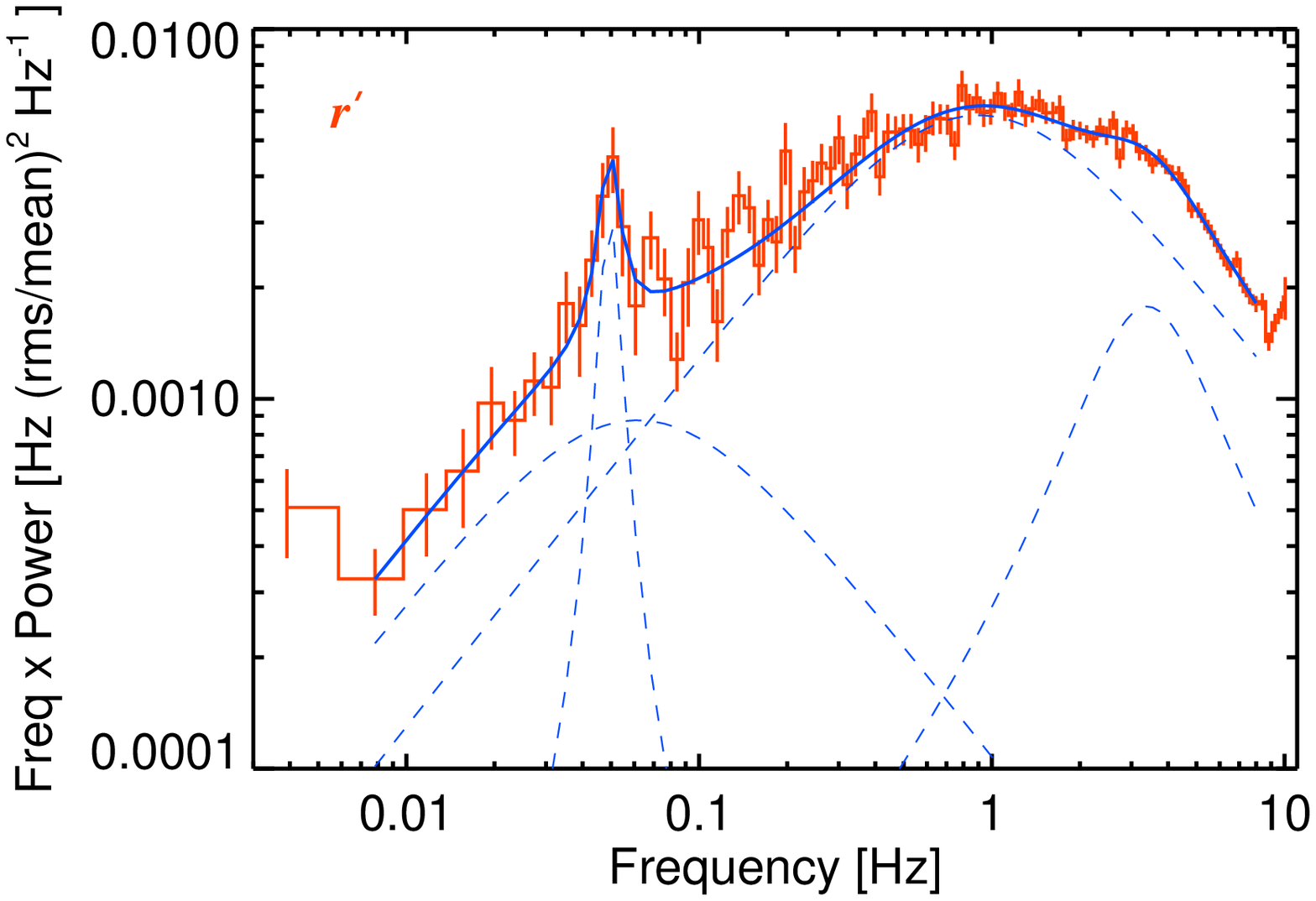}
     \includegraphics[angle=0,width=8.5cm]{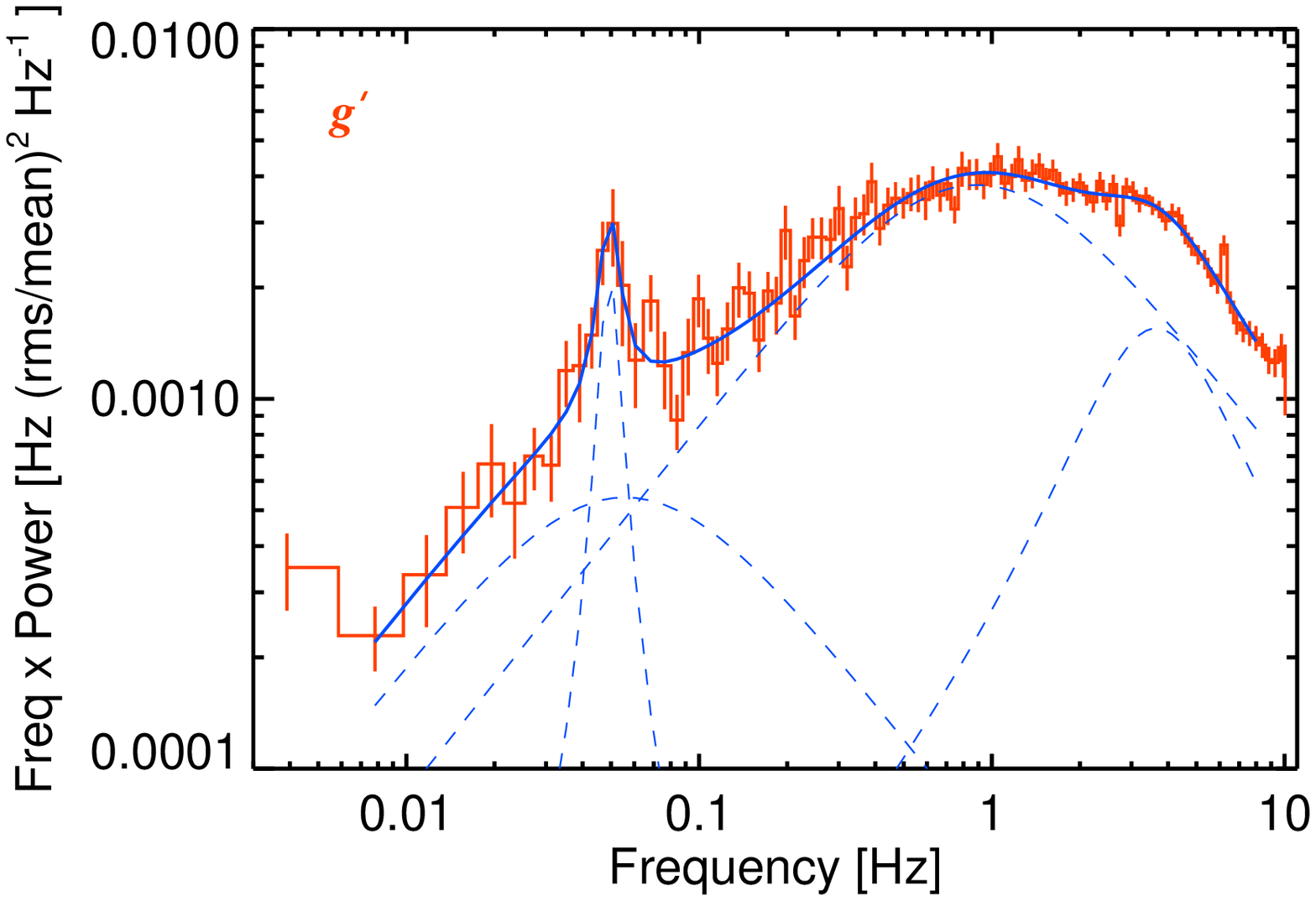}
     \includegraphics[angle=0,width=8.5cm]{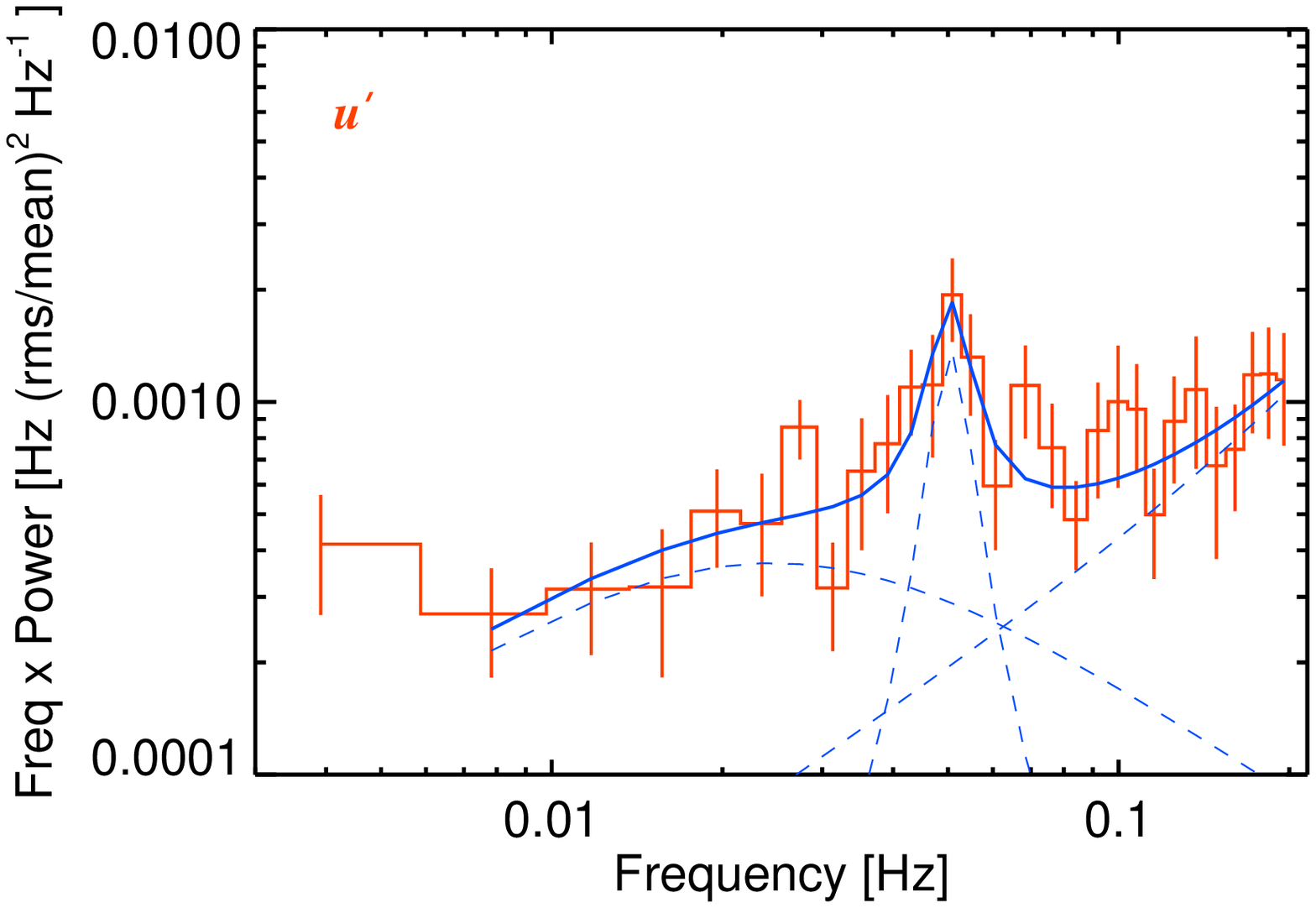}
\caption{Optical PSD fits to Night 1 data. Four component model fits to the $r'$ {\em (Top Left)} and $'g$' data {\em (Top Right)}. The bottom panel is a three Lorentzian fit to the $u'$. See Table~\ref{tab:opsdfits} and text for fit parameters and details. 
 \label{fig:opsdfitsn1}}
  \end{center}
\end{figure*}

\begin{figure*}
  \begin{center}
     \includegraphics[angle=0,width=8.5cm]{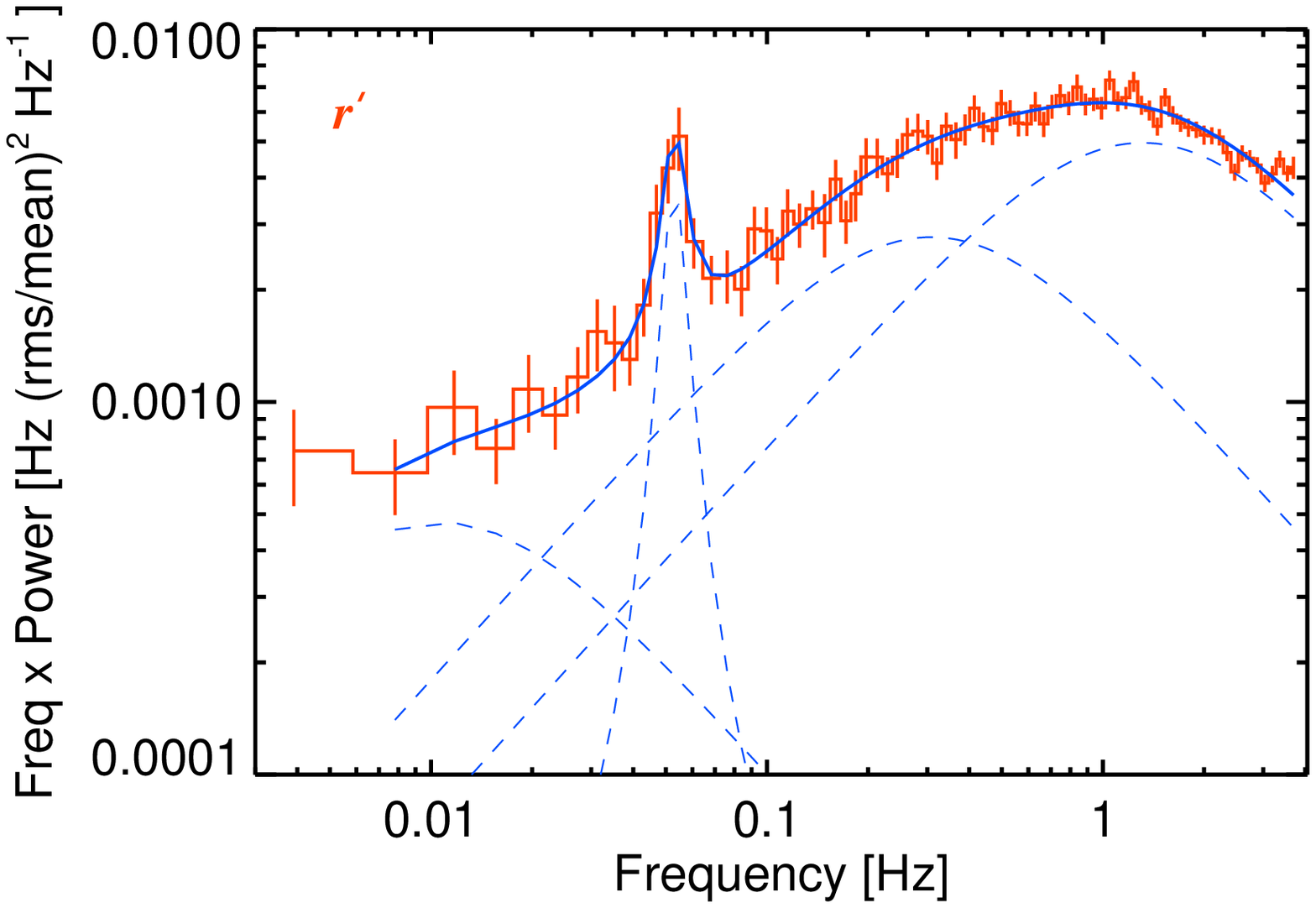}
     \includegraphics[angle=0,width=8.5cm]{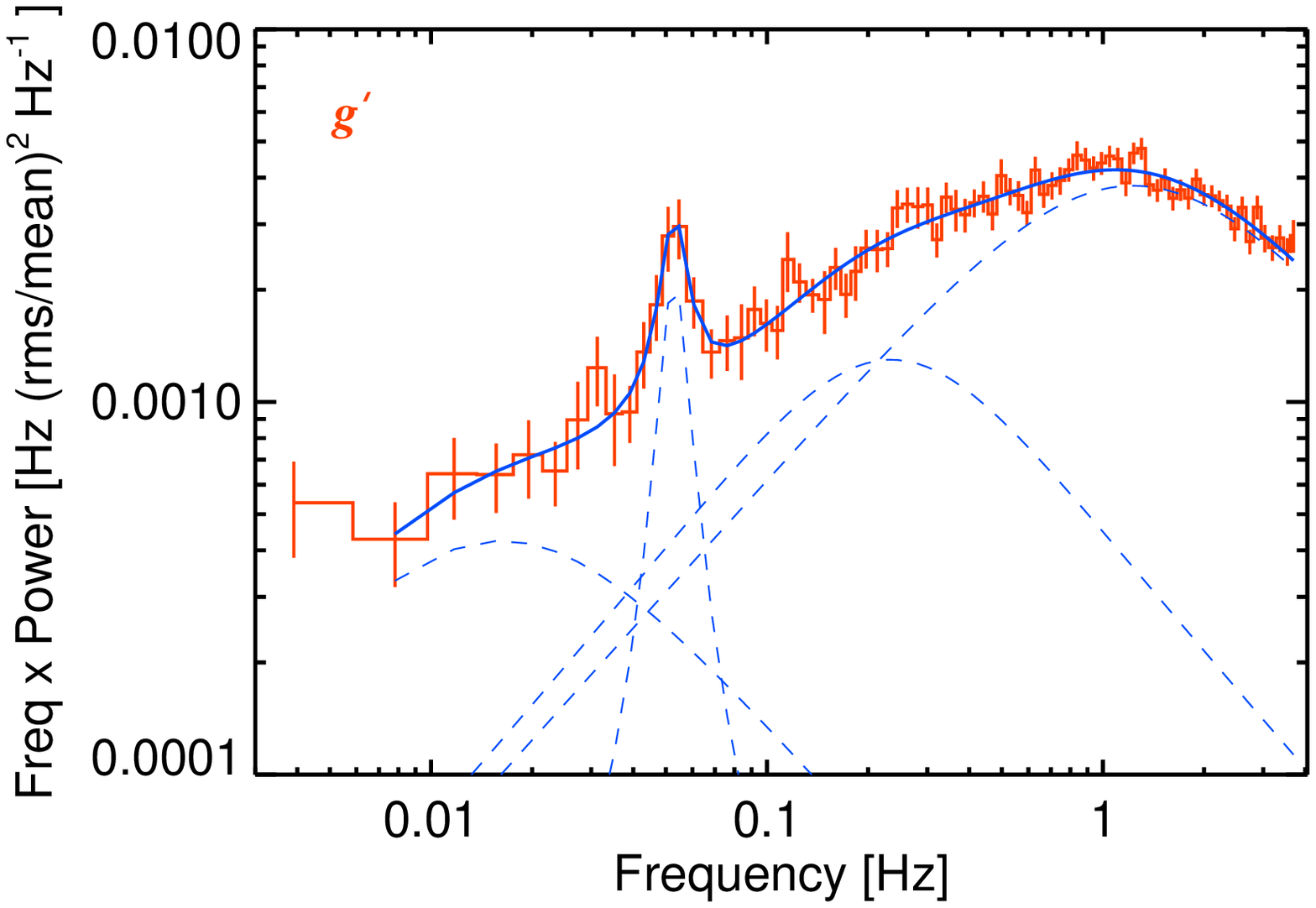}
\caption{Four component model fits to the Optical PSD fits to Night 2 data, for $r'$ {\em (Left)} and $'g$' data {\em (Right)}. See Table~\ref{tab:opsdfits} for fit parameters.
 \label{fig:opsdfitsn2}}
  \end{center}
\end{figure*}

\begin{figure*}
  \begin{center}
     \includegraphics[angle=0,width=8.5cm]{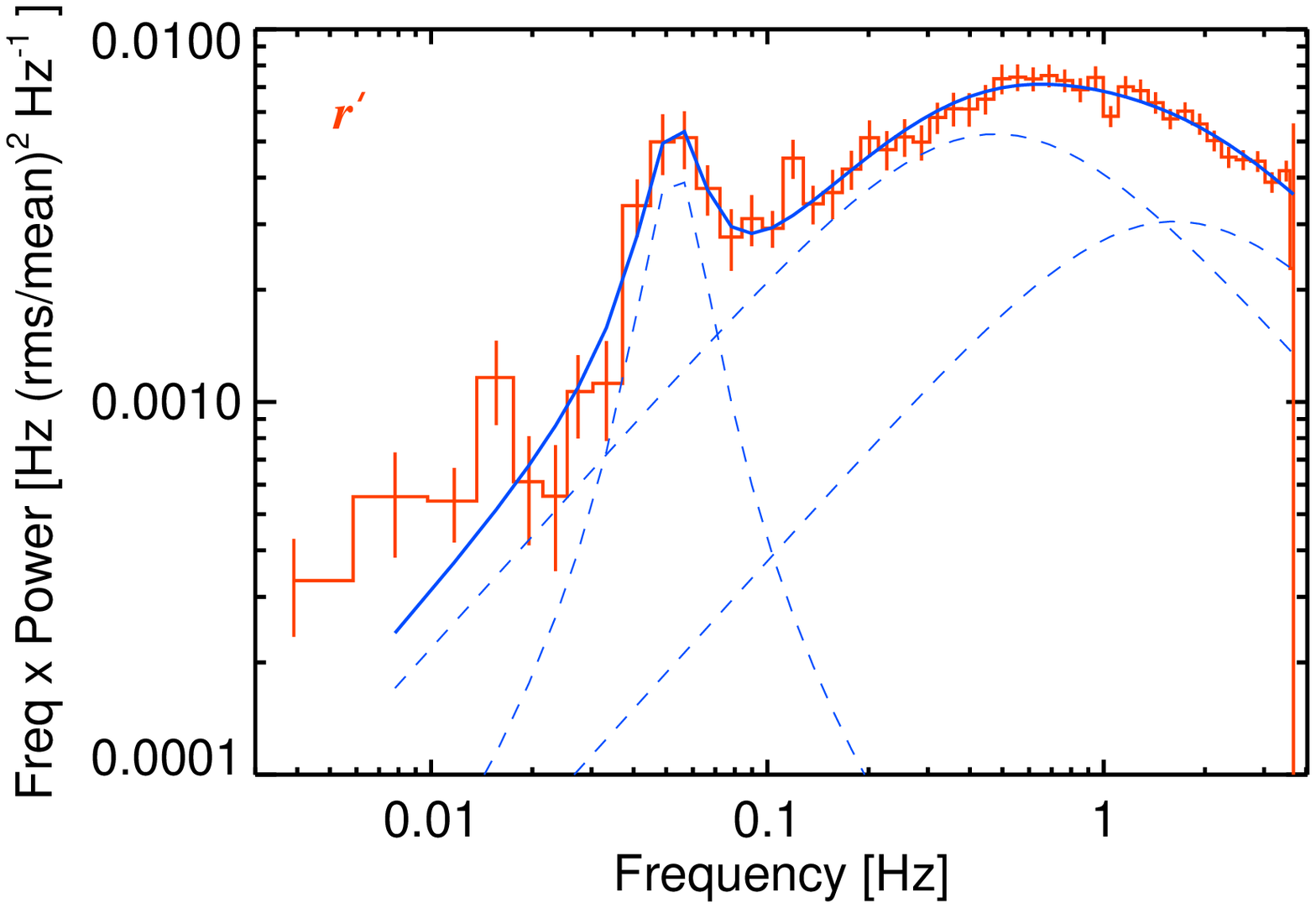}
     \includegraphics[angle=0,width=8.5cm]{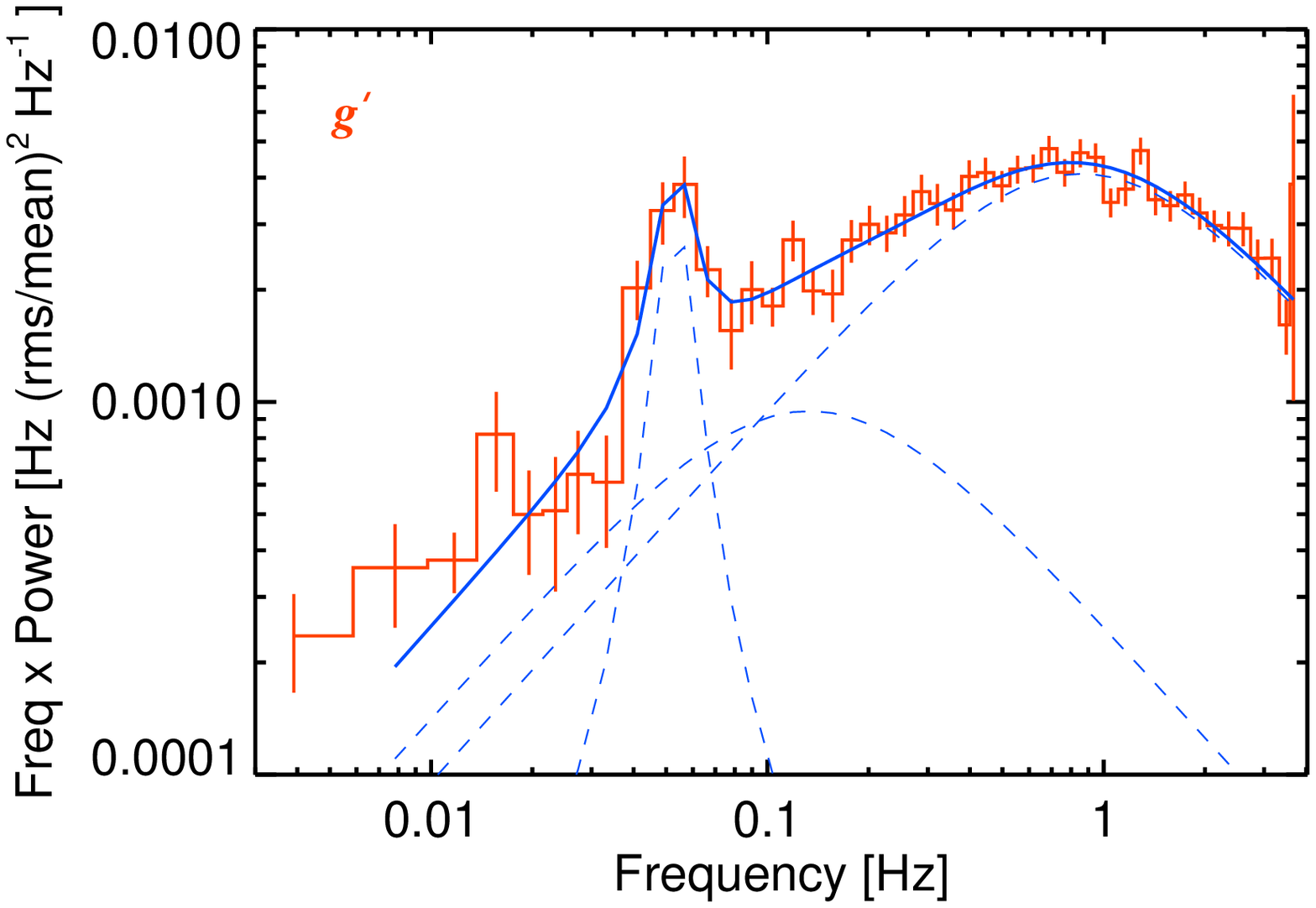}
\caption{Four component model fits to the Optical PSD fits to Night 3 data, for $r'$ {\em (Left)} and $'g$' data {\em (Right)}. See Table~\ref{tab:opsdfits} for fit parameters.
 \label{fig:opsdfitsn3}}
  \end{center}
\end{figure*}

\subsection{Auto-correlations}
\label{sec:acfs}

The auto-correlation function (ACF) represents a convolution of a light curve with itself, and gives a measure of the effective coherence timescales in the data. For a regularly-sampled time series $l$, a simple (but biased) estimator of the ACF at any given lag $\tau_j$ may be computed as 

\begin{equation}
{\rm ACF}(\tau_j)=\frac{\sum_{i=1}^{N-j_i}{[l(t_i)-\bar{l}][l(t_{j_i})-\bar{l}]}}{\sum_{i=1}^N{[l(t_i)-\bar{l}]^2}}
\end{equation}

\noindent
where the lags are discretised so that $t_{j_i}$=$t_i+\tau_j$. If Poisson noise is strong, auto-correlation of the errors causes a prominent, narrow spike to appear at zero lag (normalised to 1), thereby suppressing the relative contribution at all other lags. Random errors are uncorrelated for all non-zero lags. A simple way to correct for this component is by use of the Wiener-Khinchin theorem, which states that the power-spectral density of a random process and its auto-correlation function are Fourier pairs. This allows removal of the Poisson contribution by correcting the PSD for white noise power, followed by an inverse transform.  

\begin{equation}
{\rm ACF(\tau)}=\sum_{f}{[PSD(f)-n]}e^{2\pi i f \tau}
\end{equation}

\noindent
where the symbols have been described in the previous section, and the ACF is defined to be normalised to 1 at zero lag. Removal of the noise ($n$) is crucial for the X-ray data as expected from the low count rate. The resultant noise-corrected full band PCA ACF lies above the uncorrected one by a factor of $\approx$5.8 (at non-zero lags) on Night 1, for example. This low rate limits the detectability of real, high frequency variable components, and we will consider the implications of the presence of any such fast variability in the Discussion section. On the other hand, white noise is only a minor correction to the fast optical PSDs, as discussed in the previous section. This is a result of the high count rate, which means that $n$ lies much below the mean source variability power over the full Fourier frequency range (see Fig.~\ref{fig:psds} and Table~\ref{tab:avgrates}). The maximum correction necessary is for the slower $u'$ data where the count rate is lower, but even in this case, the PSD lies above the noise level at all frequencies so $n$ does not dominate. The resultant noise-corrected ACFs in this case lie above the uncorrected ones by only a factor of 1.4. The corresponding correction factor is much smaller in the $r'$ and $g'$ bands, at 1.1 and 1.2 respectively.

\begin{figure}
  \begin{center}
    \includegraphics[angle=0,width=7.35cm]{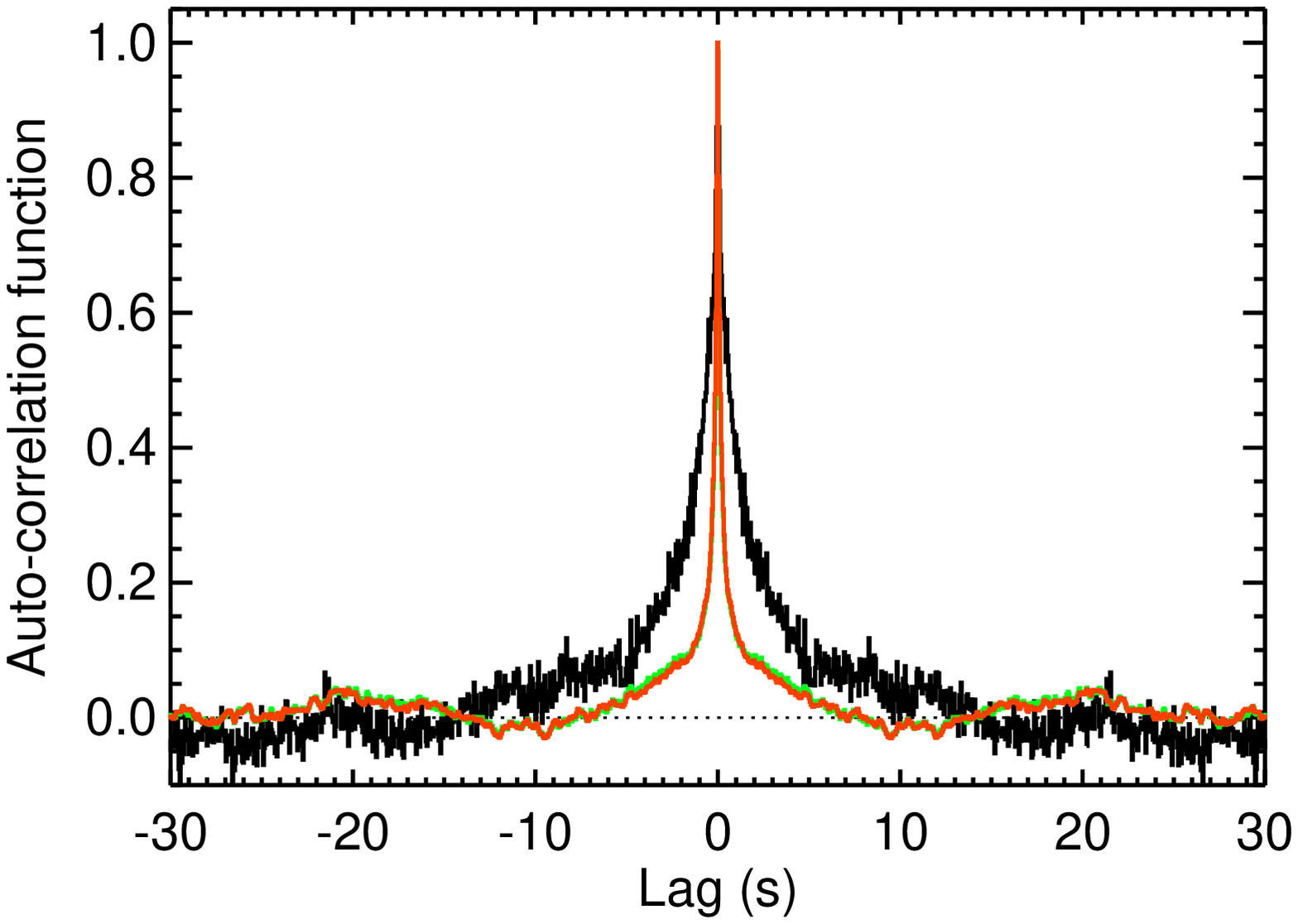}
    \includegraphics[angle=0,width=7.35cm]{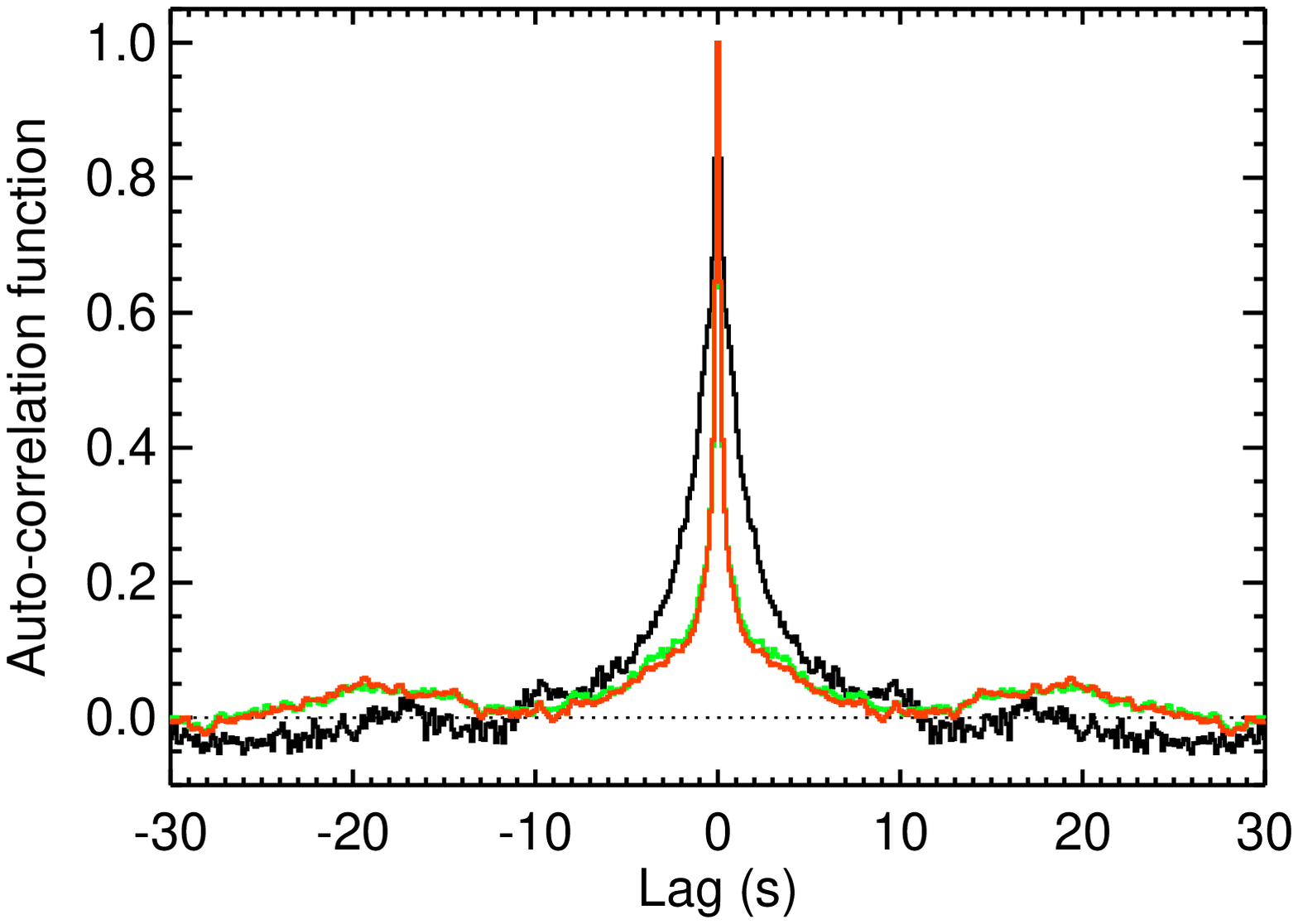}
    \includegraphics[angle=0,width=7.35cm]{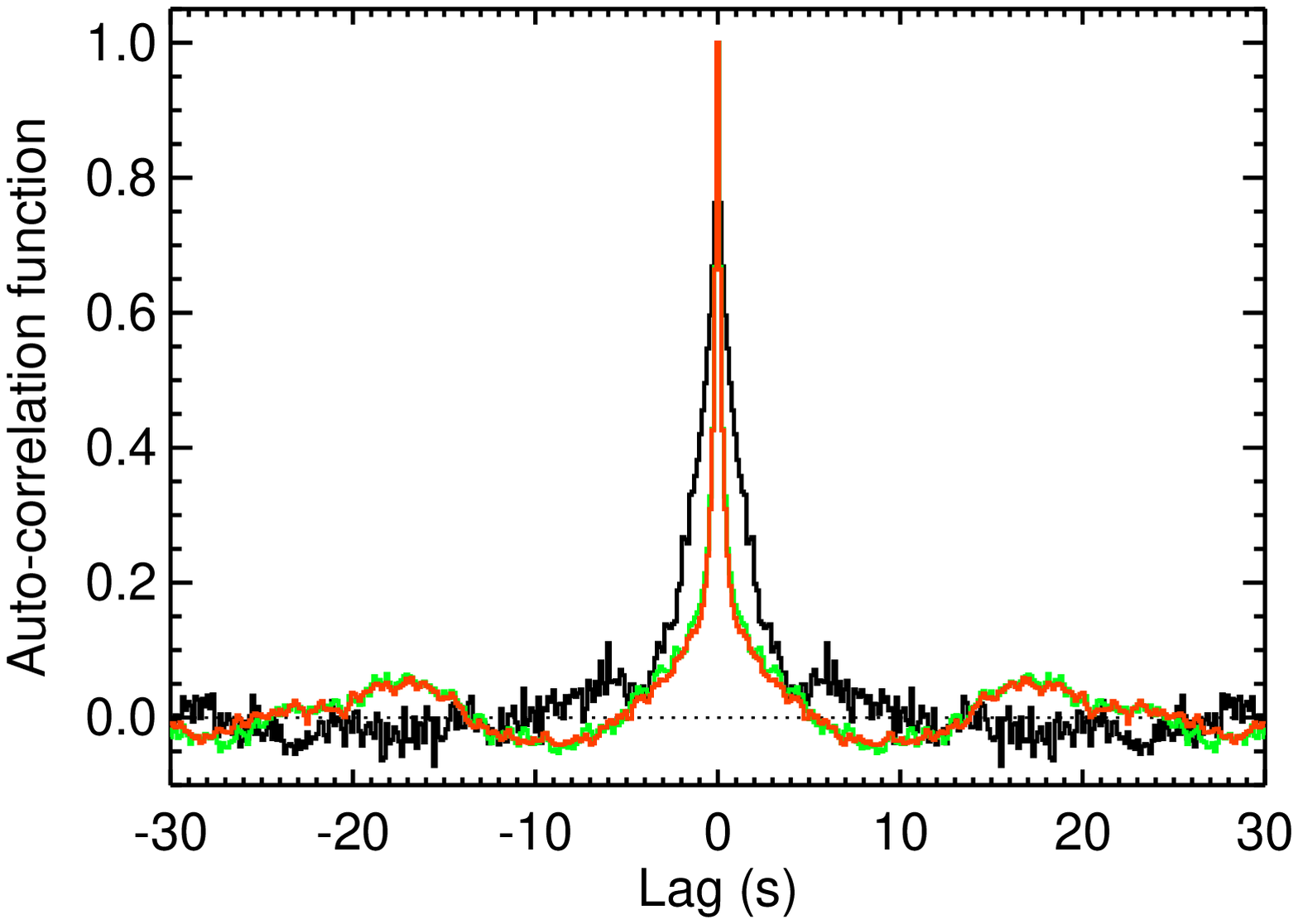}
    \includegraphics[angle=0,width=7.35cm]{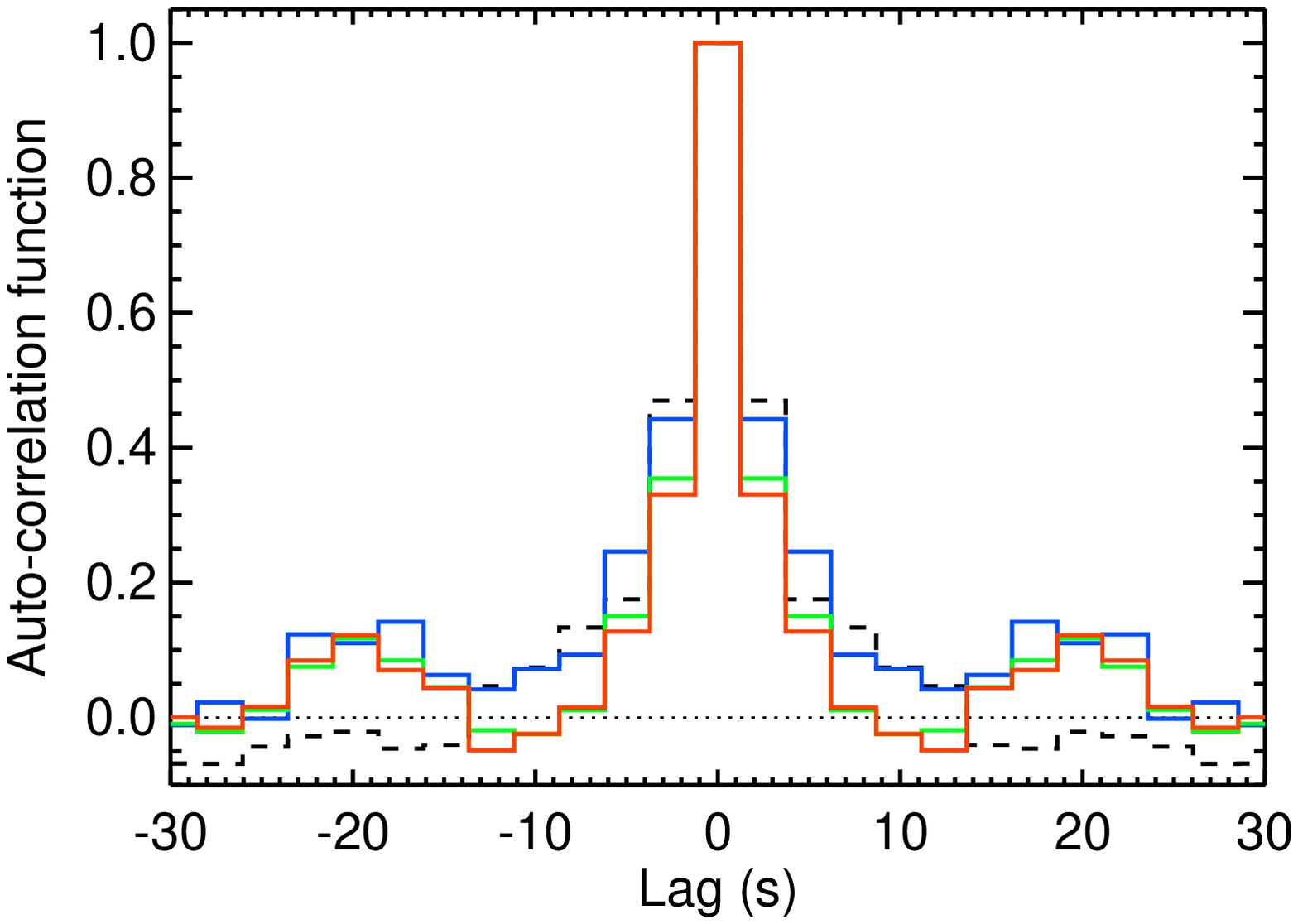}
\caption{The auto-correlation functions for Nights 1--3 from top, going down. In each case, the black histogram is that for the full band X-rays (summed over all the PCUs available on each night), red for $r'$ and green for $g'$. The effect of white noise has been removed in each case (see text). The last panel shows the ACFs for the photometric Night 1, but computed from light curves matching the $u'$ time bins of $\approx$2.5 s. In this case, the X-ray ACF is shown as the black dashed histogram, and the $u'$ one is the blue histogram.
 \label{fig:acfs}}
  \end{center}
\end{figure}

The noise-corrected ACF results for the three nights of data in the different bands are shown in Fig.~\ref{fig:acfs}. For exact comparison, the X-ray ACFs are constructed from light curves binned to the fastest optical time resolution on each night. On all nights, the fast optical data ($r'$ and $g'$) have a narrower central peak relative to the X-ray ACF. The $u'$ light curve is sparser, and the effect of white noise comparatively larger on its PSD. Yet, the $u'$ data show similar characteristics, with a core marginally narrower than the X-ray ACF computed from an identically binned X-ray light curve with 2.5~s time resolution. Such \lq slow\rq\ ACFs for all bands are shown in the bottom panel of Fig.~\ref{fig:acfs}. 

Another feature in all the optical ACFs is the positive hump around lags of 20 s, a result of the optical QPO discussed in the previous section. Some weak humps are also present in the X-ray data between lags of 15--25 s, though their significance is small. These can be quantified simply by first computing the standard deviation of ACF measurements from independent light curve segments, and using this standard deviation as the error estimate at each time lag bin. Then, the positive ACF signal over the 15--25 s is coadded and compared to the propagated error on this coadded value. This results in significance estimates of 3.8$\sigma$, 2.5$\sigma$ and $<$1$\sigma$ for the X-ray humps on Nights 1, 2 and 3 respectively. Some of these may be related to the broad Lorentzian 2 component with \numax$\approx$0.1 Hz found in the X-ray PSD fit (Table~\ref{tab:xpsdfits}); as this is not a QPO, no strong ACF excess is expected. Carrying out the same computation in the optical shows all the $\sim$20 s lag humps to be highly significant ($>>$10$\sigma$).

There is some slight apparent evolution of the ACFs between the various nights. For instance, the optical ACFs on Night 3 are marginally broader, and the X-ray ACF narrower, than in the other nights. Differences in optical weather conditions, the minimum time resolutions, and the overall short observation lengths, must be kept in mind (Table~\ref{tab:obslog}), so detailed inter-night comparisons are more involved, and not considered here. Regardless of these changes, the main ACF structure within lags of \p10 s possess several components on all the nights, with a very narrow central core and a broader component on longer lags. For Night 1, the core has an exponential decay time of close to 0.1 s, a manifestation of the superposed flare analysis shown in \S~\ref{sec:opticaltiming}. In X-rays, this fast decay time is closer to 0.2 s. Lastly, we also note that small systematic changes in the optical PSD white noise level do not affect the overall comparisons presented herein, with the fast optical ACFs still being much narrower than in X-rays if a higher optical PSD noise component (as detailed in the previous section) is subtracted.

\subsection{Optical vs. X-ray cross-spectral behaviour}

\subsubsection{Time-domain cross-correlation functions}
\label{sec:ccfs}

\citet{motch83} uncovered a weak anti-correlation between the optical and X-ray flux around the time of a source transition from the low to the high state. In our simultaneous VLT/ULTRACAM and \rxte\ observations, we confirmed an $r'$ anti-correlation on timescales of a few seconds, both leading and lagging the X-ray peak by a few seconds, and also reported the presence of a sharp positive correlation spike lagging X-rays by only fractions of a second (Paper I). A superposed shot analysis confirmed the broad optical anti-correlation and also revealed fast optical spikes lagging the X-rays by 150 ms.

\begin{figure}
  \begin{center}
    \includegraphics[angle=0,width=7.10cm]{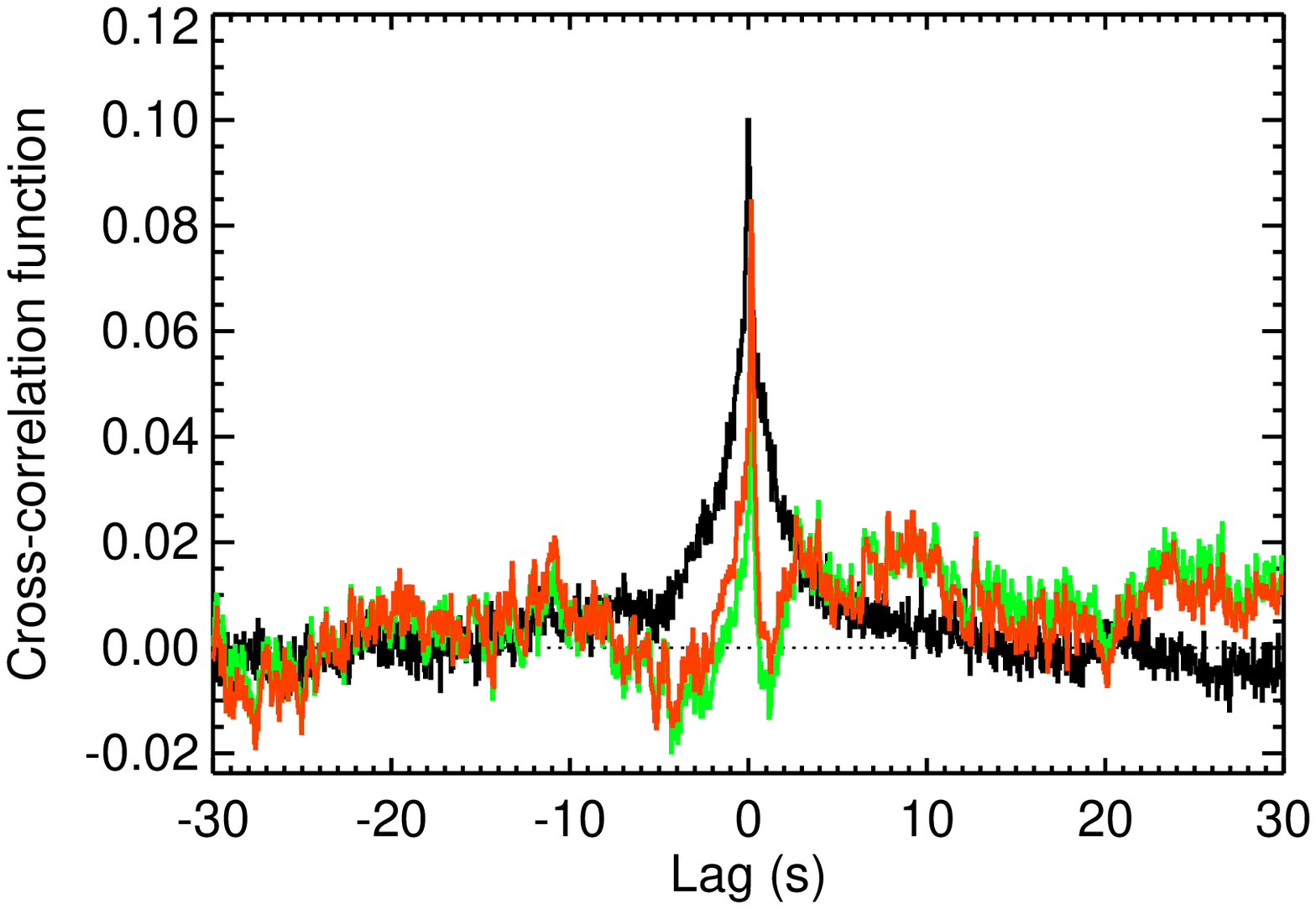}
    \includegraphics[angle=0,width=7.10cm]{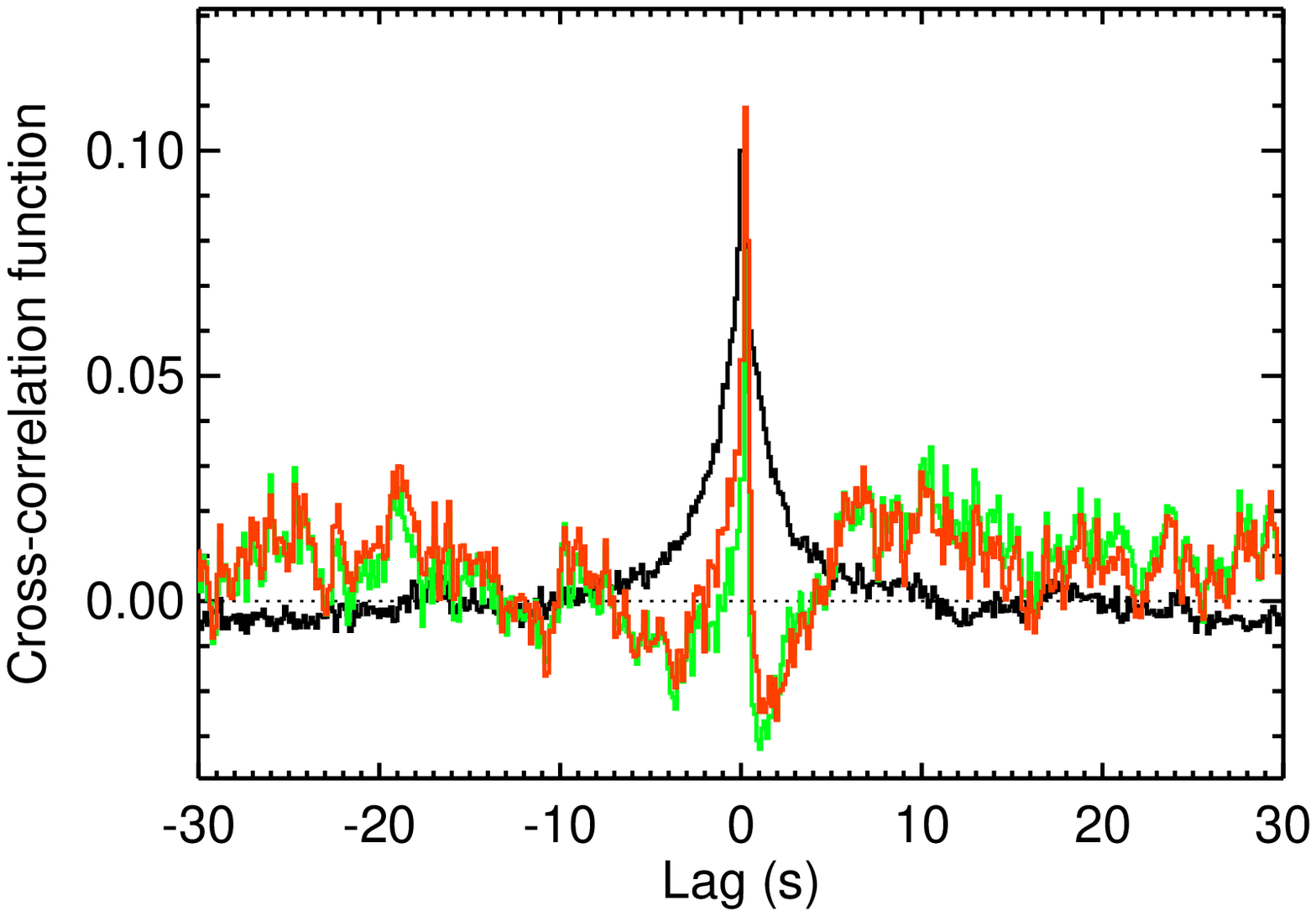}
    \includegraphics[angle=0,width=7.10cm]{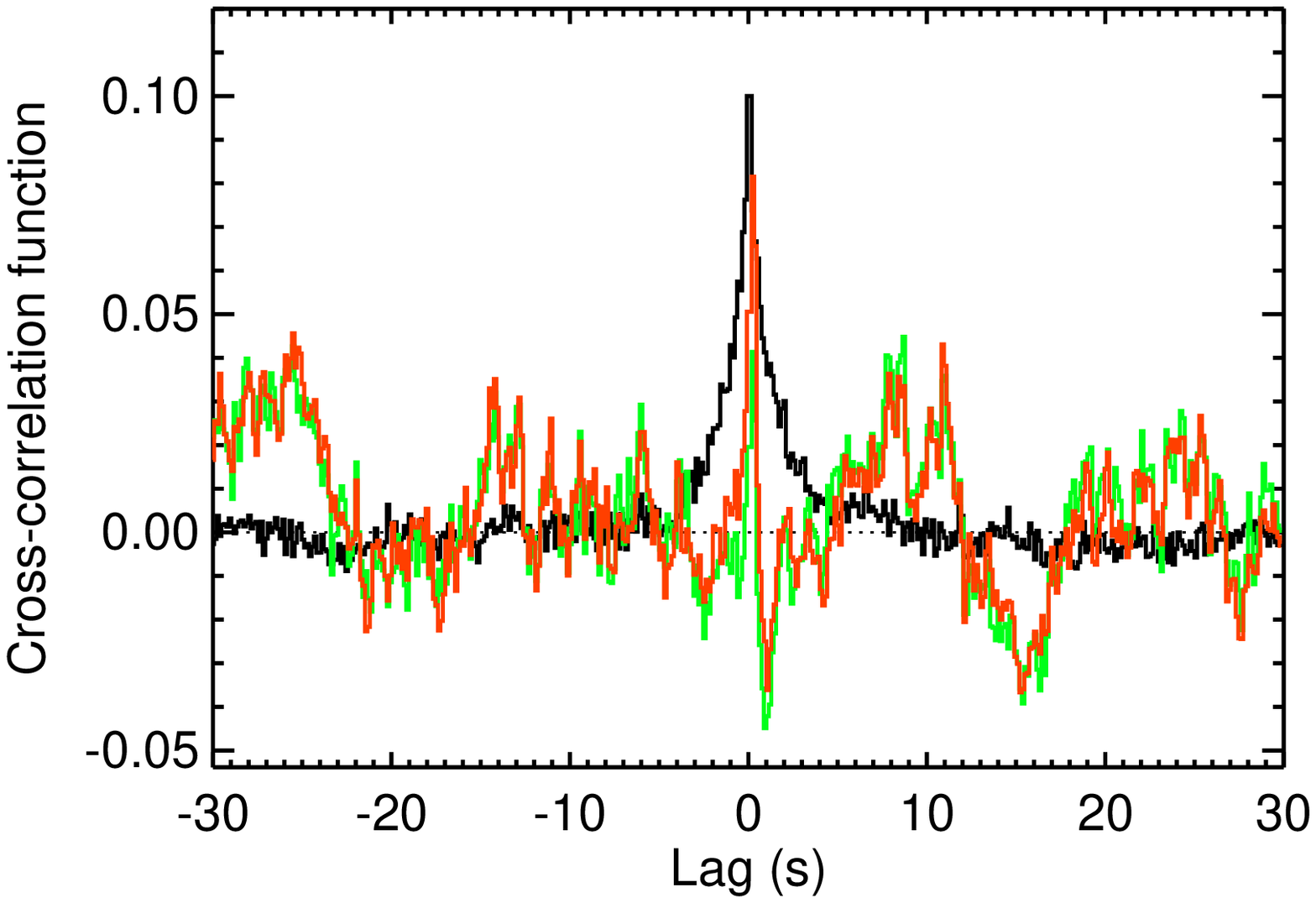}
    \includegraphics[angle=0,width=7.10cm]{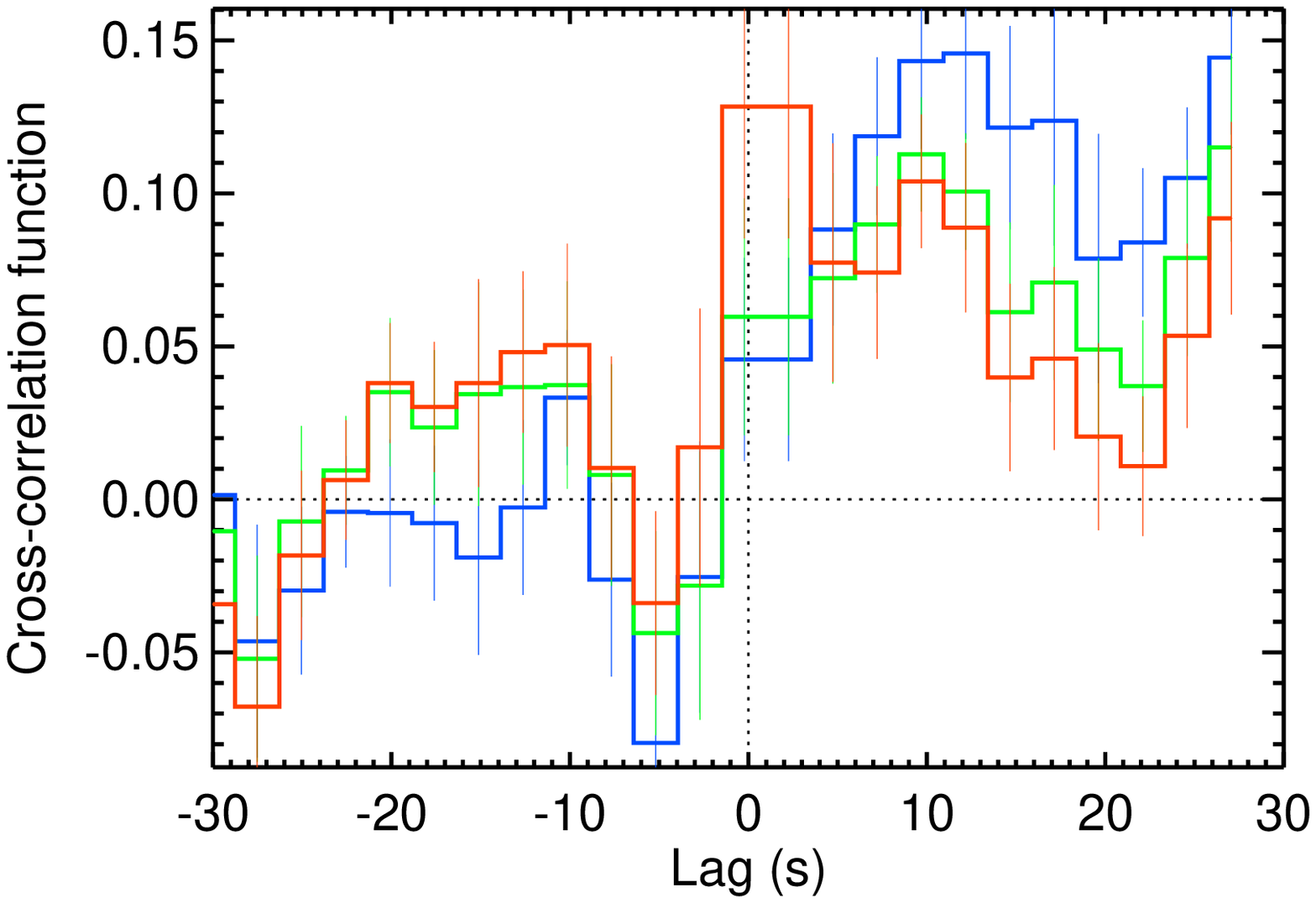}
\caption{The cross-correlation functions (CCFs) between various bands, on Nights 1--3 (from top, down). The red and green curves show the X-ray full band PCA vs. the $r'$ and $g'$ CCFs, respectively. Positive lags mean that the optical is delayed with respect to X-rays. The black curves show the X-ray 2--5 keV vs. 5--20 keV CCFs (at the same time resolution as the optical, and normalised to 0.1 for comparison), with delays referring to the 5--20 keV band. For a zoom-in on the fast peak delay at 150 ms, see Fig.~1 of Paper I. The bottom panel shows slow X-ray vs. optical CCFs, computed from light curves binned at 2.48 s, to match the $u'$ data; in this case, the blue histogram refers to the X-ray vs. $u'$ CCF. The error bars come from error propagation using the standard deviation among independent light curve segments. 
 \label{fig:ccfs}}
  \end{center}
\end{figure}

In Fig.~\ref{fig:ccfs}, we now present the CCFs for all three filters, on the various nights. The CCF represents the convolution of one light curve with respect to the other. For our discrete time series, we used a standard {\sc idl} coding for the time-domain convolution, which is estimated as 

\begin{equation}
{\rm CCF}(\tau_j)=\left\{ \begin{array}{rl}
\frac{\sum_{i=1}^{N-j_i}{[x(t_{j_i})-\bar{x}][o(t_i)-\bar{o}]}}{\sqrt{\sum_{i=1}^N{[x(t_i)-\bar{x}]^2\sum_{i=1}^N[o(t_i)-\bar{o}]^2}}} &  \tau<0\\
\frac{\sum_{i=1}^{N-j_i}{[x(t_i)-\bar{x}][o(t_{j_i})-\bar{o}]}}{\sqrt{\sum_{i=1}^N{[x(t_i)-\bar{x}]^2\sum_{i=1}^N[o(t_i)-\bar{o}]^2}}}     & \tau\ge0
                        \end{array}\right.
\end{equation}

\noindent
where $t_{j_i}$=$t_i+\tau_j$. $x$ and $o$ represent the first and second light curve, in this case the X-ray and optical ones, respectively. The lag $\tau_j$ refers to that of the second light curve, with respect to the first, and is itself discretised. Light curves which are related only by a linear transfer function (e.g. a simple time delay or a multiplicative scaling factor) would give a peak CCF value of 1 at the relevant lag time. 

The above formula may be used if the two light curves have been extracted on an identical time baseline. This was possible because of the event mode of the \rxte\ PCA, with every single event being recorded. X-ray light curves simultaneous to the optical were obtained by binning the fast (2$^{-8}$ s) barycentred X-ray light curves on the optical baseline, separately for each filter and night. We also checked that direct \ftools\ extraction of X-ray light curves with time bins matching the optical ones, followed by minimal interpolation to the common time baseline, produced unchanged results. Such a methodology is standard procedure in cross-correlation AGN light curves \citep[e.g. ][]{gaskellpeterson87}, and has been shown to produce reliable results in the limit of well-sampled data \citep{whitepeterson94}. As an extra check, we also computed the discrete correlation function (DCF) described by \citet[][ their Eq.~2]{edelsonkrolik88}. The DCF does not require any interpolation of the data itself, and may be computed using datasets with differing time resolutions. Rather, a correlation is computed between all pairs of data in the two light curves, and a time lag is attached to each pair. The final DCF is then measured by averaging all the pairs that contribute to a binned time lag series. As expected, the match in the CCF shapes between the two methods was excellent for a variety of time lags and X-ray light curve time resolutions. Note that no additional correction for measurement error was applied (cf. \citealt{whitepeterson94}); in any case, this would only affect the DCF normalisation. 

Each simultaneous light curve was then split into segments 256 s long, which were cross-correlated and then averaged to obtain the final result. The uncertainties on the average at each lag time bin are estimated by error propagation, using the standard deviation of the CCF among these segments. The main CCF structure within absolute lags of several seconds is reproduced on all the nights, with the $r'$ and $g'$ results matching closely in both strength and shape. The most significant other feature is a positive correlation hump around lag=$+10$ s. Using the coadded signal methodology used for estimating hump significance in the ACFs (\S~\ref{sec:acfs}), this time between optical lags of +5 and +15 s, results in a signal:noise of 10--30 for this feature between the various nights and the $r'$ and $g'$ filters. Further structures at lags greater than +20 s and below --10 s are also present, though these change dramatically in strength and width between the nights these are not considered to be sigificant in the full data. The main features have been noted in Paper I with regard to the $r'$ CCF and the overlaid and zoomed-in CCFs from all nights presented in Fig.~1 of that paper make immediately apparent the significance of the main features being discussed (note: in that work, the length of the time segments was not fixed to 256 s, which explains minor differences on long lags with respect to Fig.~\ref{fig:ccfs} here, but all the main features appear in both cases). Again, stringent constraints on the evolution of the CCF between the nights must await better datasets, given the changing weather and time resolution of the present observations. In particular, the prominent variation in the CCF on Night 3 on times of several tens of seconds is an artifact of highly variable transparency. 

The $u'$ data are much slower; at a time resolution of 2.5 s, each 256 s--long $u'$ band segment has only 103 time bins (as opposed to 5120 bins for the other filters). In Fig.~\ref{fig:ccfs}, we compare the X-ray vs. $u'$ CCF from Night 1 with slow X-ray vs. $r'$/$g'$ CCFs, with the latter computed from light curves heavily binned by a factor of 50. Two features stand out from this figure:-- i) The $u'$ shows an anticorrelation at small negative lags ($\sim$--5 s) as in the other filters. The formal significance of this feature is only 2$\sigma$, but this is a result of the large reduction in light curve time bins (and hence independent segments) as compared to the fast data for $r'$ and $g'$. The fact that the location of this trough agrees in all filters means that it is real. ii) The prominence of the $\sim$+10 s lag strength increases in $u'$ (reaching a maximum value close to 0.15 over the plotted lag range) accompanied by a decrease in the fast sub-second lag features apparent in the other filters. The difference between the CCFs computed from fast and from slow light curves may be understood as a result of the very different widths of the respective CCF structures -- the fast and strong peak is confined to very small lags, whereas the $\sim$10 s structure is broad and is therefore enhanced when binned. 

\begin{figure}
  \begin{center}
    \includegraphics[width=8.5cm]{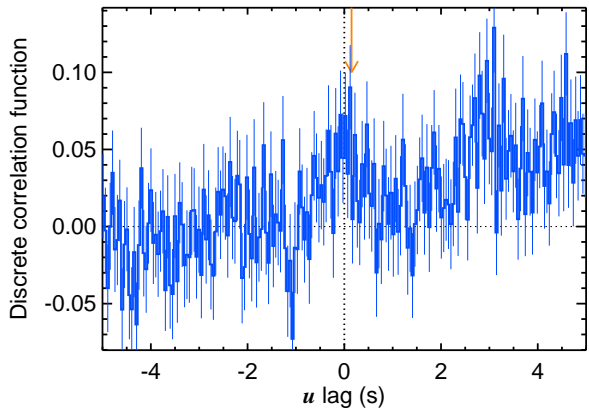}
\caption{Zoom-in of the discrete correlation function (DCF) computed from fast X-ray data (sampled at 50 ms) vs. the slow $u'$ data (2.5 s resolution). The lag refers to the $u'$. A positive correlation around zero lag is seen, in addition to the rising, slower $\sim$10 s component of Fig. ~\ref{fig:ccfs}. The $r'$ and $g'$ fast CCF peak lag time of 150 ms is denoted by the arrow. 
 \label{fig:udcf}}
  \end{center}
\end{figure}

Despite the slow $u'$ time resolution, it is possible to check for the presence of any fast variations correlated with X-rays. We did this by using the slow $u'$ light curves, and the fast X-ray light curves sampled at 50 ms resolution, and (because of the differing time resolutions) computing the DCF between these two. The result is shown in Fig.~\ref{fig:udcf}, where the final DCF is also shown with bins corresponding to the X-ray time sampling. The errors are computed according to the prescription of \citet{edelsonkrolik88}, with the uncertainty on the DCF depending simply on the scatter of the independent lag values that contribute to each bin. Other than the dominant slow component increasing toward 10 s (cf. last panel of Fig.~\ref{fig:ccfs}), there is a weak excess on sub-second lags, just as for the $r'$ and $g'$ filters. This feature appears to be broad (extending over the central $\approx$1 s; note that we cannot claim a significant detection of a 150 ms lagged peak itself), but this is not unexpected given the large $u'$ time bins. The overall significance of the broad feature can be assessed by co-adding bins to obtain a mean DCF and comparing this to the propagated error on the mean (as for the ACF humps), from which we find a significance of 5.8$\sigma$ over the lag range of --1 to 1 s. Alternatively, fitting a simple line profile (e.g. a Gaussian) and comparing this to a constant model fit returns an improvement in the former case at a level of greater than 99.99 per cent, according to an f-test. Similar excesses are also detected in DCF computations over independent light curve segments. Thus, the excess is real. This exercise shows that the overall fast variability patterns appear to be present in all three filters (though with different strengths). 

\begin{figure}
  \begin{center}
    \includegraphics[angle=0,width=8.5cm]{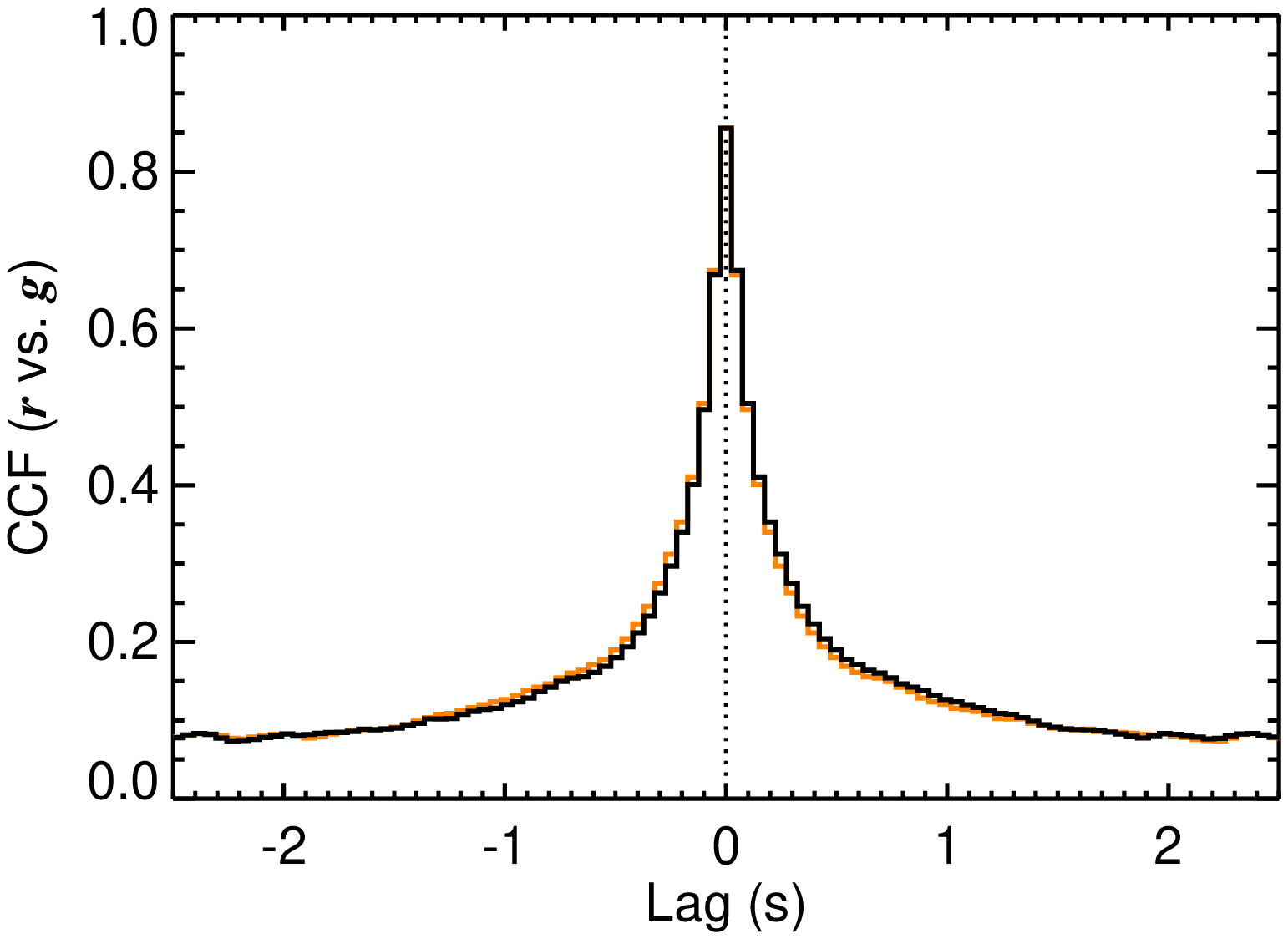}
    \includegraphics[angle=0,width=8.5cm]{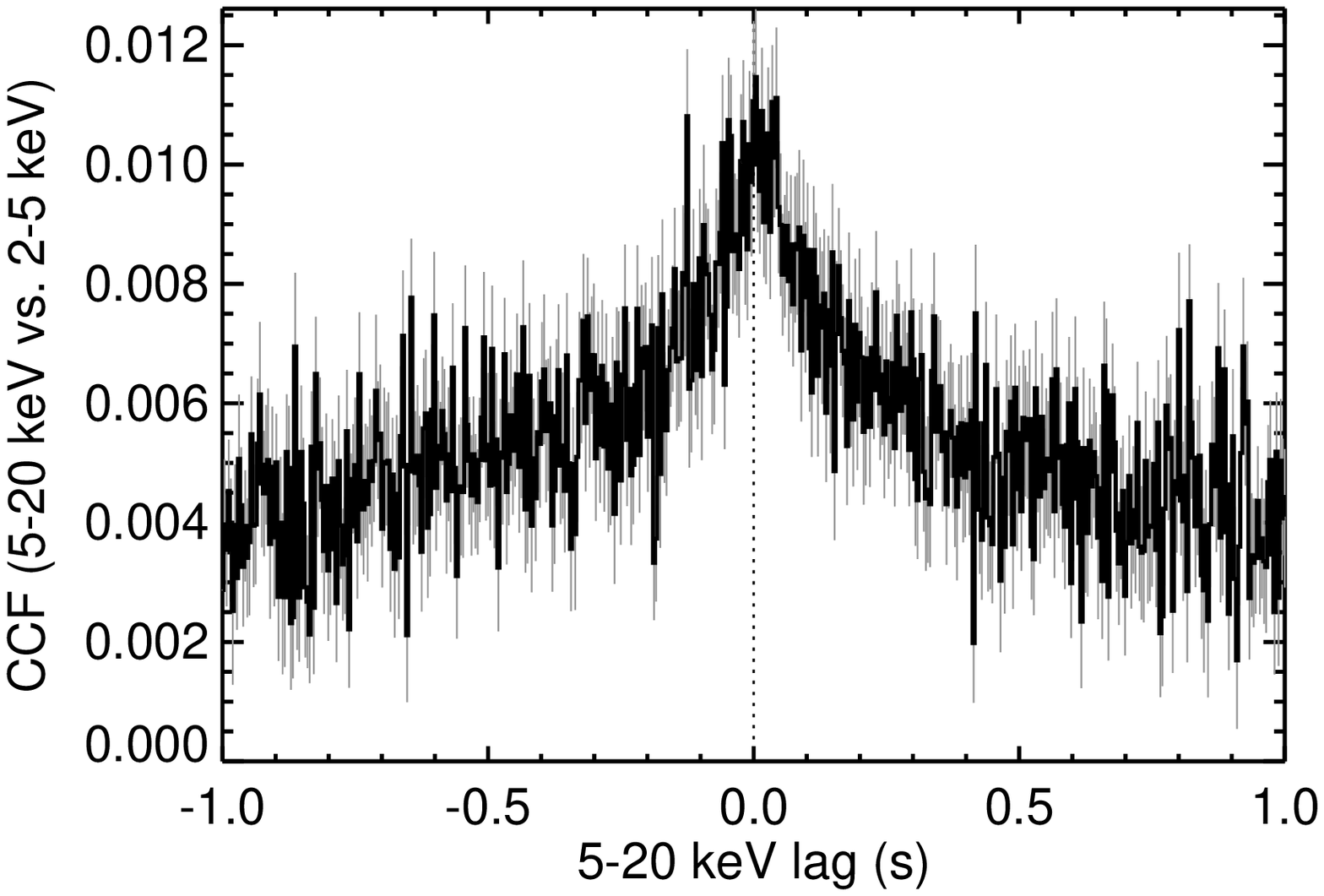}
\caption{Optical-only and X-ray--only CCFs. {\em (Top)} The plot shows the $r'$ vs. $g'$ CCF from Night 1 (light curve time resolution of 50 ms) as the black histogram, with lags referring to the $r'$ band. The orange histogram is the same CCF mirror-imaged about zero lag, i.e. lags referring to $g'$. Some asymmetry is apparent. {\em (Bottom)} This shows the 5--20 keV vs. 2--5 keV CCF from Night 1 -- the same as in the top panel of Fig.~\ref{fig:ccfs} but this time shown zoomed in and computed on a fine 2$^{-8}$ s resolution. The grey error bars denote the standard deviation amongst the CCFs computed from many segments.
 \label{fig:ccfs_oo_xx}}
  \end{center}
\end{figure}

What about the behaviour of one filter vs. another? As is seen from good match between the various optical light curves, the CCFs between the different filters do not reveal any strong correlation characteristics. The $r'$ vs. $g'$ CCF is shown in Fig.~\ref{fig:ccfs_oo_xx} and is seen to be largely symmetric with no lag of the peak. Intriguingly, there is a very small asymmetry discernible in the sense that the CCF is slightly stronger on positive $r'$ lags. This effect is seen on multiple nights, and so is probably real, but minute. We hope to observe the source with finer time resolution in the future, and defer discussion of this for now. 

We also investigated the X-ray--only behaviour. We used the \rxte\ light curves extracted in two separate bands: 2--5 keV and 5--20 keV (see Table~\ref{tab:avgrates}) and cross-correlated them in an identical manner as described above. The results are also shown in Fig.~\ref{fig:ccfs}, with a positive lag referring to the hard energies lagging soft photons. In contrast to the asymmetric X-ray vs. optical CCFs, the soft vs. hard X-ray CCFs are largely symmetric and show no obvious peak lag. This can also be seen in Fig.~\ref{fig:ccfs_oo_xx} where the CCF computed on the fastest X-ray time bin of 2$^{-8}$ s is shown zoomed-in. There are no obviously significant structures to within the sensitivity limits.

\subsubsection{Coherence and time lags}
\label{sec:cohlags}

The time domain cross-correlation functions can be decomposed into their Fourier components -- coherence and phase lags \citep{jenkinswatts69, bendatpiersol86, vaughannowak97, nowak99_cygx1_ii}. In order to compute these, we follow the recipe described by \citet{vaughannowak97}, coded with a combination of custom Fast Fourier Transform routines and a modified version of the publicly available package \idlextract\footnote{http://idlastro.gsfc.nasa.gov/ftp/contrib/rxte/}. The simultaneous lightcurves were split into many segments as with the PSD measurement, and the cross-spectrum between each X-ray and optical segment was computed. A coherence function is then computed from the average of many segments as  

\begin{equation}
\gamma^2(f)=\frac{|\langle C\rangle |^2-n_{xo}^2}{\langle |S_x|^2\rangle \langle|S_o|^2\rangle}
\label{eq:coh}
\end{equation}

\noindent
where $x$ and $o$ represent the X-ray and optical light curves, $X$ and $O$ their Fourier transforms and $C$ the complex-valued cross-spectrum between these two $C=X^*O$ at any Fourier frequency $f$. Poisson noise correction is included via the $S$ and $n_{xo}$ terms.

$|S|^2$ represents the PSD at frequency $f$ after correction for white-noise: i.e. $|S_x|^2=|X|^2-|N_x|^2$ and $|S_o|^2=|O|^2-|N_o|^2$, with $N$ being the Fourier transform of the noise component. The PSDs are typically written in Leahy normalised units so that the ensemble noise level is 2. $n_{xo}$ is a combination of signal and noise power terms: $n_{xo}^2=(|S_x|^2|N_o|^2+|N_x|^2|S_o|^2+|N_x|^2|N_o|^2)/m$ for $m$ independent light curve segments and frequency bins which are averaged over. For all frequency bins satisfying $|S|^2/|N|^2>\sigma/\sqrt{m}$ (for both light curves), $|\langle S_x^*S_o\rangle|^2>\sigma n_{xo}^2/\sqrt{m}$  and $\gamma^2>\sigma n_{xo}^2/(|X|^2|O|^2)$, the final corrected coherence can be analytically estimated from Eq.~\ref{eq:coh} above. The uncertainty is computed as in Eqn.~8 of \citet{vaughannowak97}. This effectively refers to the Gaussian limit and means that it is only valid for significant coherence and PSD power levels, for which we used a threshold value of $\sigma$=3. The computation was performed on geometrically-rebinned (in frequency) cross- and power spectra. 

The phase lag ($\phi$) is simply the phase angle of the complex-valued cross spectrum defined over the range $-\pi$ to $\pi$. If one takes $\gamma_{\rm raw}^2$ to be the raw coherence function before noise correction so that $\gamma_{\rm raw}^2=|\langle C\rangle |^2/(\langle |X|^2\rangle\langle|O|^2\rangle)$, then the phase uncertainty is given by $\delta\phi(f)=\sqrt{(1-\gamma_{\rm raw}^2)/2\gamma_{\rm raw}^2 m}$. We refer the reader to the works cited at the beginning of this section for full details on the theory.

\begin{figure}
  \begin{center}
    \includegraphics[height=20cm]{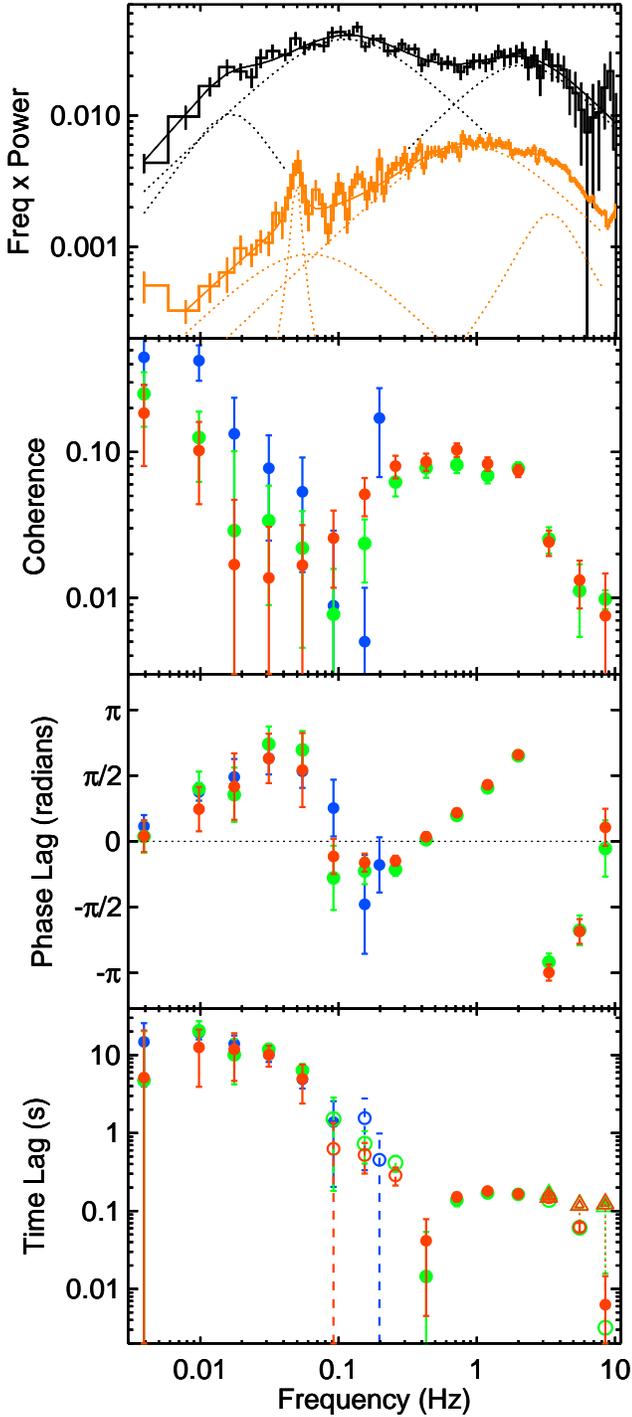}
\caption{X-ray vs. optical multi-band cross-spectrum components. The top plot shows the X-ray and optical ($r'$-band) PSDs and fits from Figs.~\ref{fig:xpsdfits} and ~\ref{fig:opsdfitsn1}. The following panels show the Coherence function, phase ($\phi$), and time lags, respectively, going down. Red, green and blue symbols refer to the $r'$, $g'$ and $u'$ bands respectively. Positive lags imply that optical is delayed. In the time lags plot, circles denote lags computed according to $\phi/2\pi f$. Empty circles with dashed error bars denote negative lags, plotted at their absolute (positive) values on the log--log scale. For the three highest frequency bins, the triangles denote the time lags computed assuming that the phase lag increases smoothly above $\pi$ (see text). Nights 1 and 2 have been averaged over the common frequency range for $r'$ and $g'$. 
 \label{fig:cohlag}}
  \end{center}
\end{figure}

The results are shown in Fig.~\ref{fig:cohlag}. As the coherence function requires averaging over many segments, the data from the best two nights (1 and 2) have been averaged for the high time resolution $r'$ and $g'$ light curves over the common frequency range, and heavily rebinned in frequency. The data from Night 3 (not shown) have a similar pattern, but with a lower mean coherence and larger (weather-related) scatter. There are several small differences in detail in the PSD fits on the two nights (\S~\ref{sec:psdfits} and Figs.~\ref{fig:opsdfitsn1}, \ref{fig:opsdfitsn2}), but given the wide Fourier frequency bins that we use, the small PSD changes between the nights are unimportant. 

At high Fourier frequencies, the coherence function shows a broad profile with a peak around 0.7 Hz, and a maximum value of $\approx$0.1 reflecting the comparatively-weak CCF strength. Its value decreases at both low and high frequencies, but remains significantly above zero over at least the range of 0.1--5 Hz. There is good overall agreement between the two filters. The lower Fourier frequencies exhibit a rise again, and reach a maximum coherence of $\approx$0.4 in the blue $u'$ filter. Around 0.1 Hz, the coherence values are approximately consistent between the filters. 

The argument of the complex-valued cross-spectrum is the phase lag between the bands at any given Fourier frequency. These lags are shown in the third panel of Fig.~\ref{fig:cohlag}. One clear pattern is the smooth increase from $\approx$0.2 Hz up to $\approx$2 Hz. The phase lag is only defined between --$\pi$ and $\pi$, so once it crosses the upper limit, the lag jumps down to --$\pi$. This holds true for all 2$\pi$ multiple intervals of the phase. Thereafter, it should continue to increase again if the trend continues to higher Fourier frequency. The lags rising from $-\pi$ above $\approx$3 Hz indicate that exactly such behaviour is occurring. The corresponding time lags at frequency $f$ can be obtained by dividing the phase lag by $2\pi f$ (Fig.~\ref{fig:cohlag}, Bottom). The rising phase lags correspond to a flat time lag distribution, with value of $\approx$150 ms at the maximum of the coherence function, exactly the peak time delay of the fast CCFs (Fig.~\ref{fig:ccfs}; Paper I). The behaviour of the coherence and lags at low Fourier frequencies is complex, with negative lags over $\sim$0.08--0.3 Hz, a phase lag of $\approx$$\pi$/2 going down to 0.01 Hz, and another reversal with phase lag close to zero in the lowest bin. All these are associated with longer time lags over $\sim$1--20 s (mainly scattered around 10 s). 

The X-ray only behaviour of the coherence and lags (not shown) was found to be typical for XRBs. Computing the coherence and lags between the PCA soft (2--5 keV) and hard (5--20 keV) light curves gave results very similar to the \gx339\ study of \citet{nowak99_gx339}, but now extended to lower source fluxes by about a factor of 2 -- we found a high coherence value of $\approx$0.9--1 over much of the Fourier frequency range, and a logarithmic time lag decline (within the errors) to high frequencies \citep[e.g. ][]{miyamoto92, vaughannowak97, nowak99_cygx1_iii, kotov01}.  

\section{Discussion}

\label{sec:discussion}

The X-ray spectral properties of \gx339\ during our observation are typical of those of the source in its X-ray low/hard state (\citealt{tomsick08}; \citealt{wilms99}), as are the X-ray PSD shape and the hard vs. soft band time lags \citep{nowak99_gx339}. The X-ray rms variability amplitude (Table~\ref{tab:avgrates}) is high, but not unprecedented \citep[e.g. ][]{belloni02}. The optical flux of the source is also within the range observed during this state \citep{buxtonbailyn07}. The multi-filter optical PSD analysis and fits across three decades in frequency, and extending to $\sim$10 Hz, is the most detailed so far for this source. The detection of an optical low-frequency QPO is not new (cf. \citealt{motch83, imamura90, steiman-cameron90}), but is now confirmed in several filters. 

The origin of the source emission at multiple wavelengths remains uncertain, and the disc, relativistic jet plasma, or non-relativistic cyclosynchrotron from a corona are all proposed as giving rise to the optical flux \citep[e.g. ][]{dimatteo99_gx339, corbel02, homan05}, while the X-rays are thought to originate as a result of Compton upscattering of seed photons either from the disc or from the jet \citep{reis08, markoff05}, or as a direct result of optically-thin jet synchrotron \citep[e.g. ][]{maitra09}. What new information do we have on the state of the source during our 2007 post-outburst low/hard state observations from the broad-band timing and spectral analysis above? 

\subsection{Spectral energy distribution model comparisons}
\label{sec:sedmodels}

\begin{figure}
  \begin{center}
    \includegraphics[width=8.5cm]{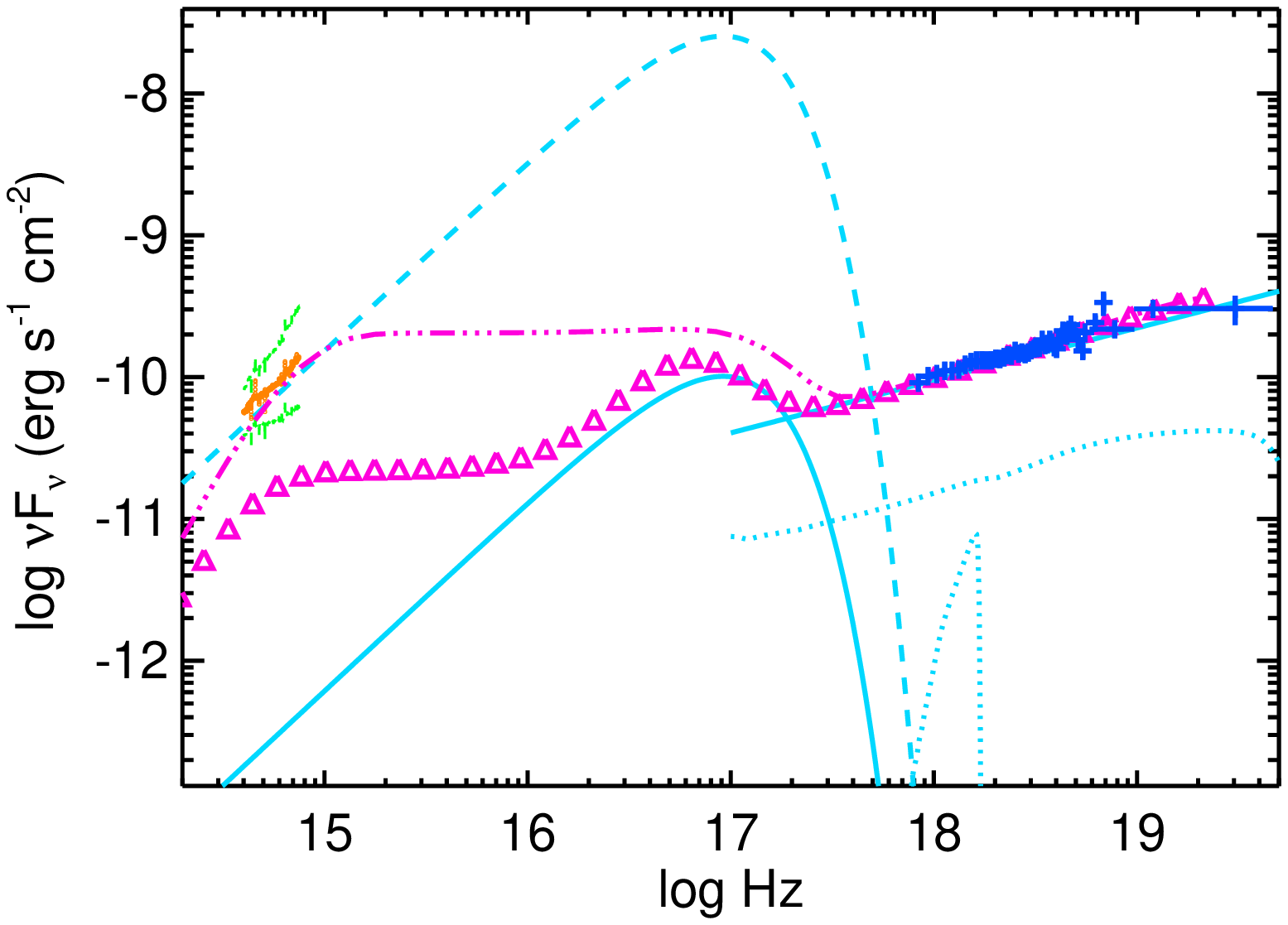}
    \includegraphics[width=8.5cm]{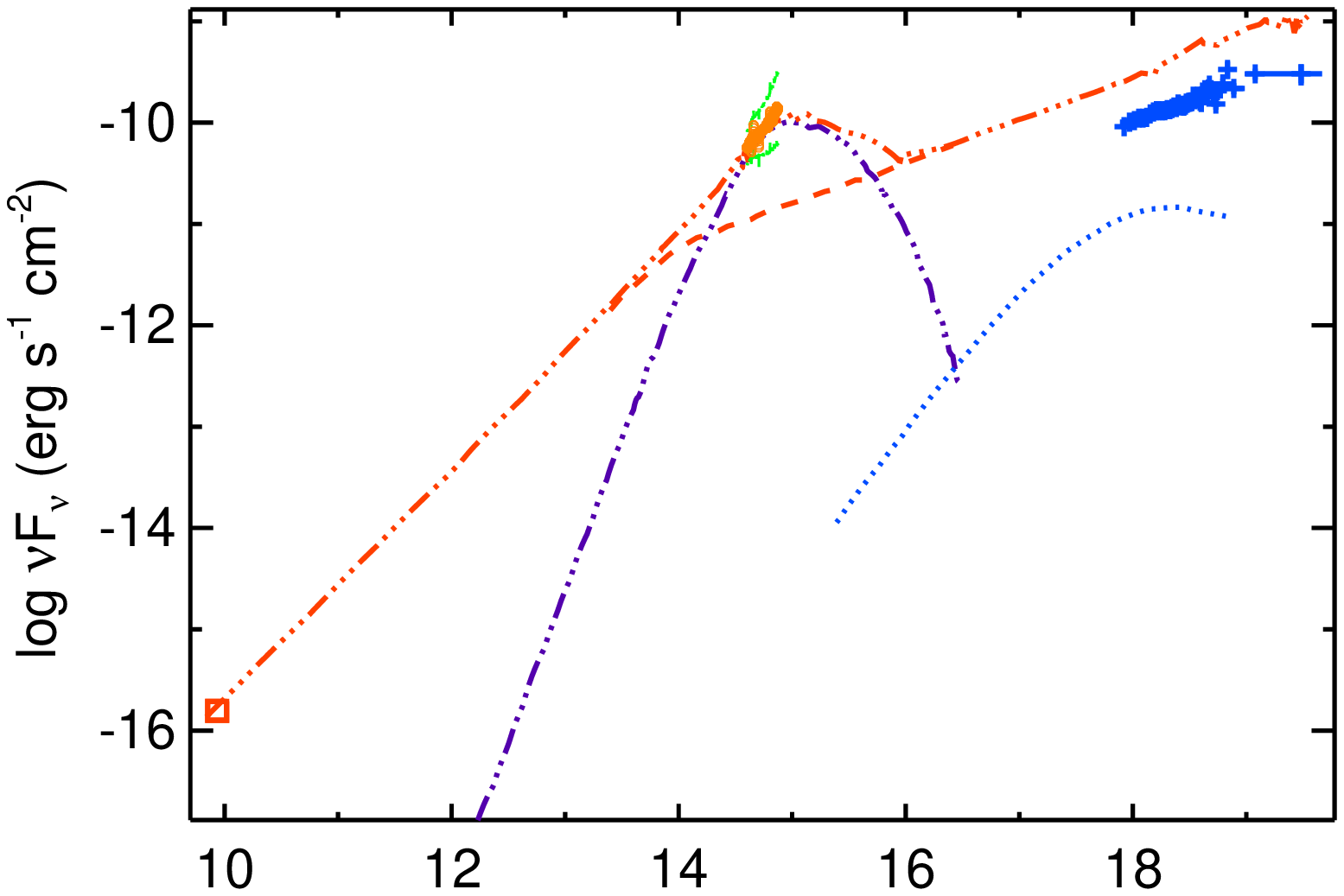}
\caption{Absorption-corrected SED (cf. Fig.~\ref{fig:sed}) compared with various models. ({\em Top}) The solid blue curves show the disc and power-law of \citet{tomsick08}, and the dotted curves are their blurred {\tt pexriv} and gaussian line reflection components. The dashed blue curve is the disc scaled up by a factor of 250. The pink triple-dot-dashed curve is an extreme \diskir\ model (\lq diskir1\rq), whereas the triangles denote a more typical \diskir\ model (which we call \lq diskir2\rq), taken from \citet{diskir} for the case of XTE~J1817--330. ({\em Bottom}) Jet model (\lq 2\rq) from \citet{maitra09} from radio to X-rays: the red dashed and the purple triple-dot-dashed curves represent post-shock (outer jet) and pre-shock (jet base) synchrotron respectively, and the dotted dark blue curve is the self-Comptonised pre-shock synchrotron. The red triple-dot-dashed curve is the sum of these. The plotted radio flux is a prediction based on the X-rays, assuming the low/hard state relation between the 3--9 keV integrated and the 8.6 GHz monochromatic fluxes, found by \citet{corbel03}. 
 \label{fig:sedmodels}}
  \end{center}
\end{figure}

A blue optical spectrum and several emission lines are seen in optical spectroscopy (Fig.~\ref{fig:optspec}, \S~\ref{sec:optspec}). The spectral slope is bluer than that expected from a canonical Rayleigh-Jean's tail, but is within the range observed in many other XRB systems \citep{shahbaz96}, where it is attributed to the outer disc. But the high source optical:X-ray flux ratio during the low/hard state that we probed does not support a disc-only scenario energetically. This is illustrated in Fig.~\ref{fig:sedmodels}, where the top panel shows the absorption-corrected \swift\ spectrum of the source reported by \citet[][ their spectrum 2 was quasi-simultaneous to our observations]{tomsick08}. It is clear that the multi-colour disc (using the \diskbb\ implementation in \xspec) vastly underestimates the flux of the optical counterpart. Constraints on this disc component are weak, as the sensitivity of the \swift\ X-Ray Telescope (XRT) drops fast below 1 keV, in addition to the unknown absorption. But these uncertainties cannot account for the large excess suggested for \diskbb\ (dashed line in the plot). The implied energy budget of a disc scaled up to the observed optical flux is about 50 per cent of the Eddington luminosity for a 6 \Msun\ BH, completely inconsistent with the low/hard state. Instead, one may consider the \diskir\ model \citep{diskir}, in which the disc is irradiated by central X-rays, and a fraction of the incident energy is thermalised to lower energies spanning the ultraviolet to the infrared. Fig.~\ref{fig:sedmodels} shows such a model (hereafter referred to as \lq diskir1\rq) with an inner disc temperature and high energy Comptonisation tail parameters matching those found by \citet{tomsick08}. The main parameters are the fraction of reprocessing occurring in the outer disc ($f_{\rm out}$=0.1) and the Comptonised-to-disc luminosity fraction ($L_c/L_d$=10). Other parameters include the outer disc radius ($r_{\rm out}$=10$^{4.5}$$r_{\rm in}$) and a low fraction of reprocessing in the inner disc. But the problem with such a disc-only scenario is still the high optical:X-ray flux ratio, which requires a large fraction of reprocessing to occur in the outer disc. The above value of $f_{\rm out}$ implies a very high efficiency of reprocessing in the outer disc, orders of magnitude more than in other systems \citep[e.g. ][]{diskir, chiang10}. Additionally, the spectral curvature of the outer disc drop-off curves in an opposite fashion to that of the broken power-law optical continuum (cf. Fig.~\ref{fig:optspec}). Finally, the very fast time variability, and anti-correlations (discussed further below) also point to the presence of other physics.

Alternatively, pure jet models have recently been discussed in the literature. In Fig.~\ref{fig:sedmodels}, we show the results of one of the latest of these, by \citet{maitra09}, in which a bright hard state outburst SED from 2002 reported by \citet{homan05} is modelled (their \lq Model 2\rq). \citeauthor{maitra09} found that the source jet power, which could fully account for their observed optical flux, was higher than on any prior occasion when simultaneous broad-band observations were carried out. But the model comparison with our data in Fig.~\ref{fig:sedmodels} is seen to over-predict the observed X-ray flux by a factor of $\approx$3 at 10 keV, when normalised to the optical data (only a slight overall renormalisation by a factor of 0.4 has been applied to match to our 2007 optical flux). In other words, the source optical:X-ray flux ratio in our observations is {\em higher} than that seen when the source was previously found to be fully jet-dominated. We note that no simultaneous radio measurements of \gx339\ around our observing dates have been published (though a compact jet was detected; cf. \citealt{tomsick08}), so we have used the radio/X-ray correlation of \citet{corbel03} to extrapolate the 3--9 keV X-ray flux to 8.6 GHz, and this monochromatic flux is also plotted in the figure and is found to agree with the normalised jet spectrum. Complete jet modelling is beyond the scope of the present paper, but the plot clearly shows that to be successful, it will have to incorporate a decreased contribution of the power-law tail electrons at higher energies. Additionally, a separate ionized gas component cool enough to produce the observed optical recombination and fluorescence emission lines would be required. 

From the above comparisons, we find that 1) pure disc (re)emission is unlikely to explain the optical flux energetically, and 2) the optical:X-ray flux ratio is higher than found by previous jet modelling when the source was already optically very bright compared to its X-ray flux.

\subsection{The timing behaviour and its implications}
\label{sec:timingimplications}

The detailed analysis of the optical vs. X-ray timing behaviour presented herein is completely new for GX~339--4. In particular, the Fourier lags indicate the presence of at least two components to the optical variability. 

\subsubsection{High Fourier frequencies}

Above $\sim$0.2 Hz, phase lags ($\phi$) increase almost linearly as a function of Fourier frequency up to $\approx$2 Hz, then switch to a lower limit of $\approx$$-\pi$ and continue increasing again. The corresponding time lags have a complex shape, with sharp rise from 0.4 to 0.7 Hz, followed by a gentle roll-over to about 3 Hz, and a quicker decrease thereafter. But the phase lag switch in this case is simply the result of the fact that it is defined within the range of --$\pi$ to $\pi$ (\S~\ref{sec:cohlags}). For a smooth increase above the upper limit of $\pi$, we can transform these negative $\phi$ values at high Fourier frequency to an absolute phase value of 2$\pi$+$\phi$. The recomputed time lags [i.e. (2$\pi$+$\phi$)/2$\pi f$] then have similar values to the other high frequency time lags, and are plotted as the triangles in Fig.~\ref{fig:cohlag} (bottom). The distribution suggests an approximately constant value of the time lags in this regime, close to the peak of the CCF time delay at 150~ms (Fig.~\ref{fig:cohlag}). Fitting a constant to the time lags above 0.4 Hz results in a compatible lag of 0.145\p0.004 s and 0.148\p0.004 s for $r'$ and $g'$ respectively. The time lag uncertainties in this regime are very small, with the result that a simple constant is not statistically a good fit. But without more insight as to good physically-motivated models to try, we do not fit other detailed forms here. The important points are that: i) the range of variation of high Fourier frequency time lag values is only a factor of $\approx$3 or less, which is small compared to the variation seen at lower Fourier frequencies; and ii) the time lag at the peak of the coherence function itself (at a Fourier frequency of 0.7 Hz) is measured to be 0.15\p0.02 s and 0.14\p0.02 s for $r'$ and $g'$ respectively, matching the peak of the CCF time delay.

A constant time lag can be produced if a simple propagation delay operates between two signals. If one assumes that this corresponds to a simple light-travel time, the distance between the X-ray and optical emitting regions is $\approx$$5000$ \gravrad\ [$\equiv GM/c^2$] for $M$=6~\Msun\ (or a range of 2000--6000 \gravrad\ for the time lag variation range noted above). The narrow optical ACFs mean that the optical emitting region must be compact, ruling out the outer portions of the accretion disc as a possibility. Assuming that the optical fluctuations arise at the base of the jet (see Fig.~\ref{fig:sedmodels}), the above distance is too large for typical jet acceleration zones, which are thought to be $\ltsim$100 \gravrad\ \citep{markoff05}. This association is much more reasonable if the fluctuations propagate instead at the sound speed ($c_s$) of the corona. In the case of a magnetized corona, this may be the Alfv\`{e}nic velocity, which is expected to be only a fraction of the speed of light ($c$). For typical values of low/hard state plasma optical depths ($\tau$$\sim$1) and magnetic compactness ($l_B$$\sim$0.5), $c_s$$\sim$0.01$c$ \citep[cf. ][ for constraints on Cyg X-1]{malzacbelmont09}. In this case, the jet base lies at a distance of $\approx$50 \gravrad\ (with the time lag variation implying a range of 20--60 \gravrad). On the other hand, the red variability rms (Table~\ref{tab:avgrates} and Fig.~\ref{fig:rg}) may instead be arguing for the fluctuations originating in the optically-thin jet. In this case, the delay is a result of the distance between the optical and X-ray emission regions traversed by fast relativistic jet plasma. The resultant lower limit on the jet elongation is then several thousand gravitational radii, as above. 

\subsubsection{Low Fourier frequencies and QPO}
\label{sec:qpodiscussion}

Below $\sim$0.2 Hz, the lags again rise towards low Fourier frequencies, with the additional appearance of a 0.05 Hz QPO in the optical PSD associated with low coherence values. The long timescales at the lowest Fourier frequencies below 0.01 Hz are probably associated with weak reprocessing in the outer disc, as we discuss further in \S~\ref{sec:reprocessing}. Other contributions are also likely to exist over the intermediate frequency interval. For instance, the phase lags of $\pi$/2 below 0.1 Hz may contribute an anticorrelated or \lq differential\rq\ response component \citep[as suggested for \xtej1118; cf. ][]{malzac03, malzac04}. In addition, time lags of a few to $\sim$20 s are too short for viscous effects to dominate, but may fit in with thermal instabilities at radii of several hundred \gravrad, although there is no obvious characteristic timescale to associate with the QPO. 

QPOs in the optical are relatively rare, so no general classification exists. But how does this feature compare with the extensive QPO detections in X-rays? The characteristic frequency of 0.05 Hz, $Q$ values of a few, and integrated rms values of $\approx$2--3 per cent (at least on the better Nights 1 and 2), suggest that this feature has properties similar to \lq Type C\rq\ low frequency QPOs classified by \citet{casella05}. The QPO frequency lies below the range studied by \citet{wijnandsvanderklis99}, where they found a close relation between the low frequency broad-band break ($\nu_{\rm b}$) and QPO characteristic frequency. If our QPO agrees with this relation, then a break at $\nu_{\rm b}\approx$0.005 Hz must be present. We note the presence of a weak upturn at the lowest frequencies in the optical PSDs in all filters (Fig.~\ref{fig:opsdfitsn1}) consistent with this, though it is clearly important to sample lower Fourier frequencies with longer observations.

A dynamical time $\approx$20 s corresponding to the QPO characteristic frequency would mean a perturbation at a radius of $\sim$2000 \gravrad, which is too large for the expected inner radius of the disc, even if it is recessed \citep[cf. ][]{tomsick08}. Epicyclic frequencies corresponding to Lense-Thirring precession \citep{ingram09} could work. But one would then expect the QPO to appear in X-rays as well, unless it is somehow isolated from the coronal flow and strictly associated with the disc, to which \rxte\ is insensitive (although note that \rxte\ did detect a low frequency QPO in \xtej1118; \citealt{hynes03_xtej1118}). We do not detect an X-ray QPO, but have already noted in \S~\ref{sec:results_qpo} that we cannot rule out the presence of an X-ray QPO identical to the optical one. On the other hand, the optical QPO carries $\approx$20-25 per cent of the total source rms variability detected (cf. $r_{\rm Optical}^{\rm total}$=0.15--0.12; Table~\ref{tab:avgrates}), whereas our derived limit of $r_{\rm X-ray}^{\rm QPO}$$\approx$0.075 on any X-ray QPO implies an upper-limit of only 16 per cent of the corresponding X-ray source rms. Thus, the process generating any QPO common to both energies must contribute a larger fraction to the optical variable flux, as compared to X-rays. 

With regard to other simultaneous optical and X-ray observations of the source, no X-ray QPO detection has been reported at X-ray flux levels similar to ours. The timing study of \citet{nowak99_gx339} carried out at a flux level only two times higher than ours (\S~\ref{sec:previouspsds}) should have been much more sensitive because of the fact that all five PCU detectors on the PCA were operational. Although no targeted search at characteristic frequencies around $\approx$0.05 Hz was performed, there is no obvious QPO in their presented figures. On the other hand, a prominent low frequency X-ray QPO was found by \citet{motch83} using the {\sl Ariel 6} satellite in conjunction with their optical QPO, at a low/hard state X-ray flux level $\approx$5 times higher than ours. The X-ray QPO amplitude was reported to compare well with that seen in the optical. So clearly, the X-ray QPO properties are changeable even within the low/hard state. The LAXPC instrument on board the upcoming \astrosat\ satellite has an effective area of about 3 times that of the \rxte/PCA\ at hard X-ray energies \citep{astrosat} and will be an excellent tool for identifying any weak low-frequency QPOs during X-ray faint states.

\subsubsection{Overall coherence and PSD change patterns}

The high Fourier frequencies show a significant optical vs. X-ray coherence, mainly associated with the $\nu_{\rm max}$$\approx$0.9 Hz Lorentzian component (top panel of Fig.~\ref{fig:cohlag} and Table~\ref{tab:opsdfits}). A slight bump between 2 and 3 Hz also seems to coincide with the highest frequency Lorentzian above 3 Hz. At the very highest frequencies, the lack of significant X-ray power (above PSD white noise) is responsible in part for the loss of coherence. At lower Fourier frequencies, the cross-spectrum behaviour is complicated. The coherence decreases sharply from $\approx$0.3 to 0.1 Hz and the phase lags in this regime also undergo a sign reversal, in the sense that the optical {\em leads} the X-rays. The uncertainties are smaller than the absolute phase values. These effects are seen to coincide with a change in the dominant PSD variability components, with the coherence decline associated with the decrease of the 0.9 Hz optical Lorentzian. The negative lags occur in conjunction with the rise of the strongest X-ray PSD component around 0.1 Hz, and also the emergence of the lowest frequency zero-centred optical component ($\nu_{\rm max}$$\approx$0.06 Hz). Such a changeover in fluctuations can naturally cause a loss of coherence \citep[e.g. ][]{vaughannowak97}, especially if the origin of the low and high frequency Lorentzians is independent, which is likely to be the case (e.g. \citealt{wilkinson09}). 

\subsection{RMS spectrum and broad-band constraints}
\label{sec:rmsspectrum}

\begin{figure}
  \begin{center}
    \includegraphics[width=9.5cm]{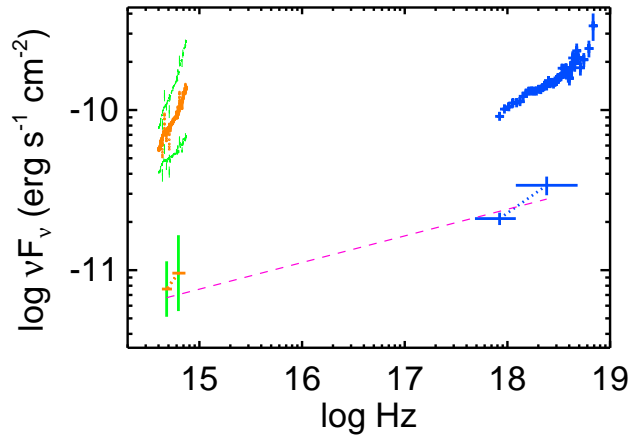}
\caption{Optical and X-ray absolute 0.4--4 Hz rms spectral fluxes (lower four points) compared to the total observed spectra. The fractional rms values derived from the PSDs in the $r'$ (centred at the effective wavelength of 6250 \AA), $g'$ (4800 \AA), 2--5 keV (3.5 keV) and 5--20 keV (10 keV) bands, from left to right, are 0.11, 0.09, 0.23\p0.02 and 0.24\p0.03, respectively. Systematic dereddening uncertainties to the optical are shown in green. The pink dashed line is a fit to these four points and has a slope of 0.17\p0.04 in the plotted log-log units. The orange and blue dotted lines connect the two optical and two X-ray rms points respectively; the slopes of these two are flatter than that of the pink line.
 \label{fig:rmsspectrum}}
  \end{center}
\end{figure}

What does the energy spectrum of variations look like? This can be examined by combining the results from the spectral and timing analysis as follows. One first computes the fractional variance by integrating the noise-subtracted X-ray and optical PSDs over any desired Fourier frequency range (from $f_1$ to $f_2$). The observed fluxes ($F_{\rm band}$) in all bands are then multiplied by the respective fractional rms values (i.e. the square-roots of the variances) to give the broad band rms spectral energy distribution (see, e.g., \citealt{hynes03_xtej1118}): 

\begin{equation}
F_{\rm rms}=F_{\rm band}\sqrt{\int_{f_1}^{f_2}{P(f)df}}
\label{eq:rmsspectrum}
\end{equation}

\noindent
where $P$ represents the noise-subtracted PSD model fits for each band in rms-squared normalised units. We choose the Fourier frequency range of 0.4--4 Hz for this exercise because this is associated with a relatively uniform optical/X-ray coherence (Fig.~\ref{fig:cohlag}). This range is dominated by the strongest optical PSD Lorentzian component, and in X-rays, includes the peak of the high frequency Lorentzian. Fig.~\ref{fig:rmsspectrum} shows the absolute rms fluxes for both the fast $r'$ and $g'$ light curves, and for the 2--5 keV and 5--20 keV ranges in X-rays. The respective mean fractional rms values were measured to be 0.11, 0.09, 0.23\p0.02 and 0.24\p0.03, where the quoted errors are statistical 1$\sigma$ uncertainties determined from Monte Carlo sampling of the PSD fits to obtain an ensemble of rms measurements. These errors are very small in the optical, where systematic dereddening uncertainties instead completely dominate (overplotted in the figure). The dotted line is a power-law fit to the four rms points, and yields a power-law index of 0.17\p0.04 in the plot units of the figure ($\chi^2$/dof=3/2), or an index of --0.83 in the commonly-used spectral energy density ($\equiv$ $F_\nu$) units. Instead of the 0.4--4 Hz Fourier frequency range, if we use the full PSD range from Figs.~\ref{fig:xpsdfits} and \ref{fig:opsdfitsn1} for this exercise, we obtain a broad-band slope of --0.78\p0.04. These values are significantly steeper than the slope inferred from spectral fitting to the full X-ray spectrum (\S~\ref{sec:xspec}), where we found an energy index of 1--$\Gamma$=--0.63(\p0.01) or --0.66 (\p0.04), depending on the model used. They are also much steeper than the slopes found in the optical (energy indices of 0 to +1; \S~\ref{sec:optspec}). 

A power-law was also found by \citet[][see also \citealt{hynes06}]{hynes03_xtej1118} for the rms spectrum of \xtej1118, and ascribed to a single optically-thin jet synchrotron component extending from the near-IR to X-rays. Our findings suggest that if there is a such a single broad-band varying optically-thin component in \gx339, it must have a spectrum different to that describing the mean emission in both optical and in X-rays. Additional components with different spectral and variability characteristics, such a jet base, magnetic coronae or inner hot accretion flows in the optical \citep[e.g. ][]{merloni00, yuan05, maitra09}, and Compton upscattering of jet and disc seed photons in X-rays \citep[e.g. ][]{markoff05, makishima08}, may then account for the remaining spectrum. 

But the problem with such an interpretation is the low absolute value of the optical:X-ray coherence of $\approx$0.1, which means that a single component is not necessarily a good representation of the broad-band variability. Were such a component to dominate the variability, an optical:X-ray coherence of close to unity would be expected. Instead, it is to be noted that the rms values in both X-ray bands are almost identical, meaning that the X-ray--only rms spectral slope is close to the mean flux X-ray spectrum, and is flatter than the single broad-band rms spectral slope inferred above. In the optical, the $g'$ rms is lower than that in $r'$, indicative of the red rms slope discussed in \S~\ref{sec:opticaltiming} and Fig.~\ref{fig:rg}; still the $r'$--$g'$ slope is again significantly flatter than the broad-band one. The dotted lines in Fig.~\ref{fig:rmsspectrum} denote the X-ray--only and optical--only rms energy slopes. These flat slopes support the scenario whereby separate physical components dominate the variability in the optical and in the X-ray bands, with the positive CCF implying some underlying connection between the two. Note that this inference holds true in the optical despite the large dereddening uncertainties, i.e. for all \av\ values ranging over 3--4 mags at least.

In X-rays, the variable component describes the mean flux spectrum well. This may be powered either by an optically-thin jet, or Compton upscattering. We cannot distinguish between these, but we note that a disc extending down to an inner radius to $\sim$10 \gravrad\ is seen to be present in soft X-rays below 1 keV by \swift\ \citep{tomsick08}. This, as well as the prominent jet base or inner coronal flow which likely dominates the optical (see below), can all easily serve as seed photon sources for Comptonisation \citep[cf. ][ for a similar discussion with regard to Cyg X--1]{makishima08}.

In the optical, a straightforward cause of the lower $g'$ rms variability as compared to $r'$ may be a steady component such as an irradiated disc, which is expected to emerge mainly at blue wavelengths. An estimate of the disc's maximal contribution at $g'$ can then be simply computed by using the excess $r'$ rms (over $g'$), and turns out to be $\approx$20 per cent. This is likely to be an upper limit, because other possible components such as a jet base, magnetic coronae and a inner hot accretion flow \citep[e.g. ][]{merloni00, esin01, yuan05} which may be present also peak in the blue. \diskir\ parameters more typical than those required by the extreme diskir1 model discussed in \S~\ref{sec:sedmodels} can easily produce a component consistent with this. In Fig.~\ref{fig:sedmodels}, we overplot a model from the high end of the range found by \citet{diskir} for the source XTE~J1817--330, with the fraction of reprocessing on the outer disc $f_{\rm out}$=0.01 and the irradiation fraction at the inner disc of $f_{\rm in}$=0.1 (model hereafter referred to as \lq diskir2\rq). This results in an optical disc:total flux ratio $\approx$10 per cent which is also consistent with a canonical disk heating scenario discussed in next section, and satisfies the \swift\ soft X-ray disc constraints of \citet{tomsick08}. Additionally, in this scenario, the broken optical spectral continuum detected by FORS may be explained by the appearance of the disc at high frequencies, for which the power law fits of Fig.~\ref{fig:optspec} suggest a contribution of $\approx$16 per cent at the bluest wavelengths of 4000~\AA. 

Further testing for the existence of any additional underlying single broad-band component will require extending such an rms analysis to low energies in the near-IR (cf. \citealt{hynes03_xtej1118, casella10}), because this regime is not expected to have strong contributions from the disc or inner hot flow and jet base discussed above.

~\par
\noindent
That the optical and X-ray CCFs show a positive peak means that the variability in the bands is connected. At the same time, the rms analysis in this section suggests that distinct physical components dominate the variability in each band. A self-consistent scenario that satisfies the main constraints of our work may then be as follows. 

X-ray and optical studies have now shown that perturbations in the outer parts of the accretion flow propagate towards the centre, seeding and coupling fluctuations in all inner regions, and resulting in a linear relation between rms and flux \citep[e.g. ][and references therein]{uttley05, g09_rmsflux}. If the disc (required by the \swift\ data) is magnetically-threaded and anchors a jet (which could explain the bulk of the optical flux), then perturbations will travel along the field lines leading to line reconnection accompanied by X-ray flares in the innermost regions. These will be Comptonised to high X-ray energies with negligible time delay to produce the hard X-ray power-law and high X-ray--only coherence values that we observe. Upon reaching the jet, the perturbations are expected to trigger (cyclo-)synchrotron emission, leading to the optical emission and hence the correlated optical:X-ray CCFs that we detect. At the same time, the slower and broader CCF anti-correlation humps can result from decreased coronal cyclo-synchrotron caused by magnetic field dissipation on the X-ray flaring timescales (see further discussion in Paper I). Thus, such a propagation sequence can generate the CCF features that we observe. 

In this picture, the low peak values of the coherence function and CCFs are a result of interaction between at least two distinct components: the jet (generating the optical power) and the corona (emitting X-rays). It may be that propagation and energy transfer events between the two components are episodic, or that the transfer occurs via some \lq filter\rq\ (e.g. magnetic reconnection, or change in field line dynamics) which ends up diminishing coherence (see also Paper I; Durant et al. 2010 in prep.). Propagation of perturbations will also naturally lead to linear rms--flux relations in both X-rays and in optical. The perturbations may arise directly within the hot flow itself \citep[e.g. ][]{done07} as long as the flow is large enough to encompass the broad range of timescales required by the rms--flux relation. 

If some form of cyclosynchrotron is indeed important for the optical variability (whether in a pre-shock jet or a corona), one can estimate a minimal magnetic field ($B$) energy density from one of the observed prominent flaring episodes, assuming that the $B$ field is in equipartition with the radiation field. For this estimation, we use Eq.~1 of \citet{dimatteo97} with the following assumptions: 1) dissipation of energy within the disc is negligible compared to that in the jet base or corona, as should be true for the low/hard state; 2) a single emission region is responsible for powering the flare, and can be described as a uniform electron cloud with diameter $\sim$100 milli-light-seconds, equal to the fast decay times that we observed (Fig.~\ref{fig:optflares}); and 3) an integrated radiative peak flare output of $\sim$10$^{36}$ erg s$^{-1}$ $\approx$0.001$\times$\ledd. This yields a $B$ field density of 5$\times$10$^3$ G, which is likely to be a very conservative lower limit, given that much faster optical flaring timescales have been detected previously \citep{motch82}, and that the corona may be patchy. Both these effects will considerably decrease the emitting region size, and in turn increase the $B$ field density. Hence, the observed optical flaring radiative losses are easily consistent with strong field energy dissipation events \citep[e.g. ][]{wardzinski00, dimatteo97, fabian82}.

\subsection{On the reprocessed component}
\label{sec:reprocessing}

\subsubsection{The fast variability}
\label{sec:reprocessedcomponent_fast}

The PSD plots of Fig.~\ref{fig:psds} show that there is consistently higher power in the X-ray variability as compared to the optical. So could the optical variability be a result of reprocessing of faster X-ray variations at frequencies above the optical PSD peak (or even higher than those probed by us due to white noise limitations)? In Paper I, we argued against reprocessing dominating the rapid optical variability on times of $\ltsim$ 1 s, at least as described by simple linear transfer functions. The reasons are several. 

Most obviously, the anti-correlation between optical and X-rays is the opposite of the expectation under a reprocessing scenario, unless one is willing to invoke negative transfer functions \citep[][]{hynes03_xtej1118}. 

Then there is the very short time lag and narrowness of the peak of the positive CCF component for the fast light curves. For an orbital period of 1.7 days, the corresponding binary separation is $\approx$25 light-seconds. Our observations on the three alternate nights were carried out with an inter-observation spacing of about 47 and 49 hrs, respectively (see Fig.~\ref{fig:finaltimes}), or relative phases of 0.15 and 0.38. For X-rays irradiating the companion star followed by an optical light echo, the expected time lags are then much longer than the 150 ms peak delay observed on all the nights. In any case, \gx339\ is thought to be a low inclination binary which increases the minimum expected time delay to $a$(1-sin$i$)=19 s for $i$=15 degrees \citep{wu01}. Reprocessing in the outer disc, instead of the companion star, ought to peak on much longer times and also be smeared out, and so is clearly unrelated to this fast variability. 

Another clue is the strength of the X-ray vs. optical CCF on the shortest lags, which is strongest (weakest) in the $r'$ ($u'$) filter, the opposite of the expectation based on disc reprocessing.

Finally, the optical ACFs are wider than the X-ray ones, in all optical filters (see discussion in \S~\ref{sec:acfs}). This is another way of saying that the peak of the optical variability power is skewed towards higher frequencies as compared to the X-ray peak. The PSD decomposition in \S~\ref{sec:timing} showed a dominant optical component with \numax$\approx$1 Hz, whereas the X-ray PSD peaks around 0.1 Hz. 

None of these arguments, by themselves, rule out a more complex reprocessing origin for the fast optical peak delay, especially given that the low X-ray source count rates limit our ability to detect high frequency variability power. But such an origin also requires that the reprocessor be 1) compact; 2) stable over several days at least; and, 3) able to isolate the fast X-ray variable component from the slow one. This can be discerned directly by eye from Figs.~\ref{fig:lcsection} and \ref{fig:lcsection1s}, where the X-ray light curves have very significant broad flaring components during which the average source flux is increased by factors of a few, over times of several seconds. The simultaneous optical section shows no obvious positive response to this. If anything, the optical shows local minima over a period of several seconds around the X-ray flares in both figures. So if part of the X-rays are being reprocessed on sub-second timescales on some reprocessing region, the slower and broader flares are not simultaneously seeing the same reprocessing region. This scenario is inconsistent with canonical disc reprocessing models, at least.

\subsubsection{The slow variability}
\label{sec:reprocessedcomponent_slow}

Disc reprocessing is expected to be stronger at bluer wavelengths, because the hot inner regions (of the outer disc) can intercept a larger fraction of incident photons from a central X-ray source. So bluer filters ought to show stronger signal on timescales commensurate with the outer disc. This is consistent with the behaviour of our slow CCFs in Fig.~\ref{fig:ccfs}, and also the high $u'$ coherence values and long time lags of Fig.~\ref{fig:cohlag}. So it is natural to associate this slow component with disc reprocessing. Such a component increasing towards the ultraviolet has also been identified in the case \xtej1118\ by \citet[][see also \citealt{malzac04}]{hynes03_xtej1118}. The Bowen blend that we observe in the optical spectrum is also thought to be a result of reprocessing (fluorescence), though its location, whether in a disc, or wind, or a low density gaseous component, is unclear \citep[cf. ][]{schachter89, wu01}. Echo tomography of the line during subsequent active low/hard states should help to settle the issue \citep[e.g. ][]{obrien02, munoz-darias07}. It is worth mentioning that if the Bowen blend arises in the disc, our limits on the optical disc contribution (\S~\ref{sec:rmsspectrum}) imply that the equivalent width of the blend against the disc continuum must be at least 5 times larger than the observed equivalent width. Finally, we note that exactly which Fourier frequencies correspond to reprocessing is also unclear because of the changing phase lags below 0.1 Hz (see Fig.~\ref{fig:cohlag}); only the very low Fourier frequency bin shows an optical phase lag of close to zero with respect to X-rays.  

Energetically, the reprocessed (disc) component must be weak unless the reprocessor is highly atypical, as already discussed in \S~\ref{sec:sedmodels} with regard to the SED models of Fig.~\ref{fig:sedmodels}. Under canonical disc heating scenarios, what is the average optical flux expected? \citet{vanparadijsmcclintock94} have shown that irradiation by a central X-ray source on the outer portions of an optically thick disc, where visual surface brightness varies as $T^2$ ($T$ being the local disc temperature), leads to the optical luminosity varying in proportion to $\Sigma$=(\lx/\ledd)$^{1/2}$$P_{\rm h}^{2/3}$, where $P$ is the orbital period in hours. For our observations of \gx339\ carried out at a low Eddington rate, $\Sigma$$\approx$1. For this value of $\Sigma$, an absolute visual magnitude of $M_V$=1.6 is predicted as a result of reprocessing. Our inferred $M_V$=--1.02 (\S~\ref{sec:optspec}), an excess flux of a factor of 11. The implied reprocessed flux is close to that of the irradiated disc diskir2 model shown in Fig.~\ref{fig:sedmodels}. Dereddening uncertainties only change this by a factor of two, while distances of 5--15 kpc imply a range for the excess flux factors (above the reprocessing prediction) of 7--21. 

\subsubsection{Lags longer than $\sim$ 10 s?}

\begin{figure}
  \begin{center}
    \includegraphics[height=22cm]{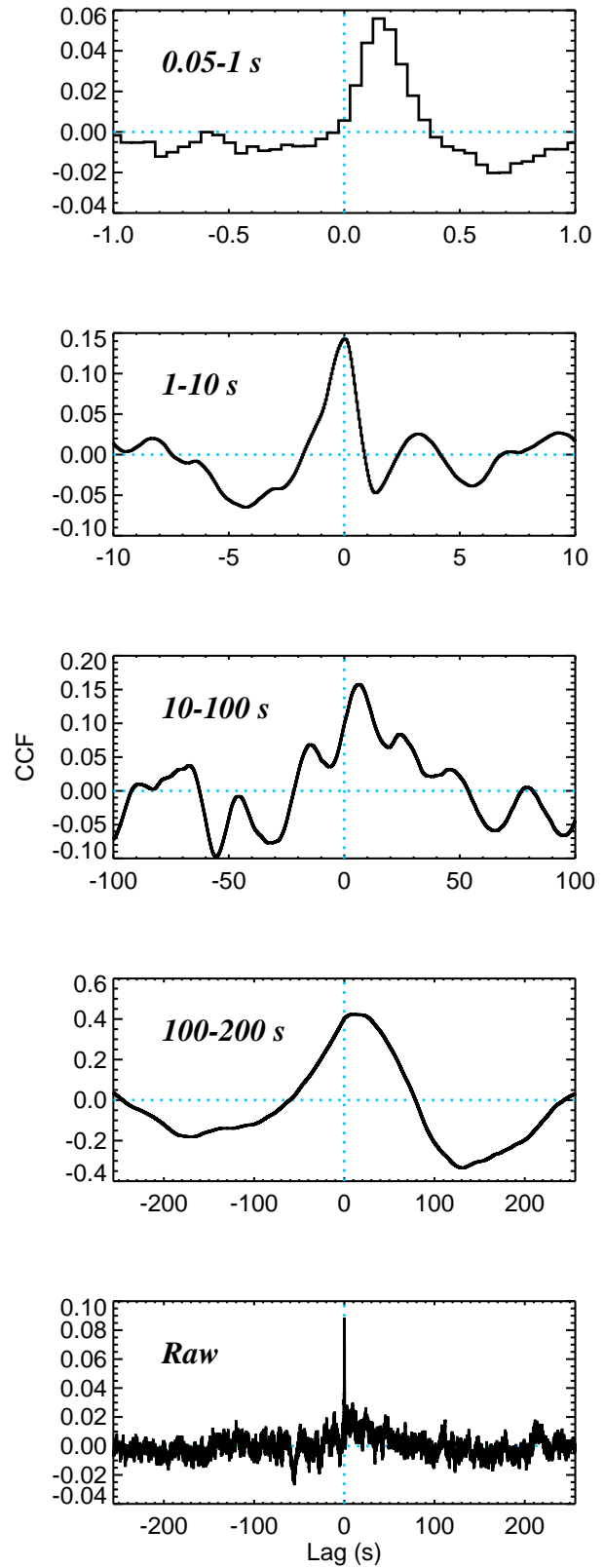}
\caption{X-ray vs. $r'$ CCFs filtered in order to highlight various timescales. Only times within the range labelled at the top left were retained in each case. Times shorter than the smallest labelled ones were smoothed out using box car filtering. Those longer than the longest labelled one were filtered by piecewise linear normalisation. Note that smoothing can artificially increase the absolute correlation strength on the y-axis. 
 \label{fig:filteredccfs}}
  \end{center}
\end{figure}

It is clear from the discussion so far that at least two timescales, associated with the fast ($\sim$150 ms) and the slow ($\sim$10 s) variability lags respectively, and one more related to the QPO characteristic time ($\sim$20 s), are important in GX~339--4. What about other, even longer timescales? One way to highlight these may be to filter the light curves on timescales better matched to the time delay being sought. We did this by following the procedure of \citet{malzac03}. In short, the CCFs were computed after keeping only the time-scales in a specific range of frequencies. High frequency noise was removed by smoothing with a box car filter. Low frequency noise was removed by dividing the light curve by a piecewise linear trend, i.e. a linear interpolation of the light curve on the longest time-scale retained. The results are displayed in Fig.~\ref{fig:filteredccfs}. 

This exercise highlights various features seen in the raw (unfiltered) CCF shown at the bottom. In particular, the main positive peak appears at longer times, matching the expected delay on several seconds expected from reprocessing. Some wiggles due to the 20 s QPO are also apparent. On the other hand, the negative correlation components are also seen to shift smoothly to longer (absolute) time delays, which is not expected. Such \lq self-similar\rq\ behaviour was also noted by \citet{malzac03} for \xtej1118, and used a basis for the stochastic fluctuation model of magnetic reservoir energy release which successfully explains the spectro-temporal characteristics of that source \citep{malzac04}. But, unlike \xtej1118 where all time lags from 0.01--10 s contributed smoothly over the full Fourier frequency range (e.g. Fig.~8 of \citealt{malzac03}), the main contributions in GX~339--4 are the 150 ms, and the $\sim$10 s component (bottom panel of Fig.~\ref{fig:cohlag}). No other important features are apparent on timescales of $\sim$ 100 s, at least. 

\subsection{Comparison with other sources}

Several distinct components (including both reprocessed and direct variability) contribute to the optical power of XRBs in general (e.g. \citealt{russell06}). There are indications of non-reprocessed optical and IR (OIR) variability in several sources now \citep{eikenberry98, uemura02_v404cyg, kanbach01, durant08, g08}. \gx339\ has one of the highest fractional variability amplitudes and strongest non-linear flaring amongst these \citep{g09_rmsflux}. Fast variability was typically found in the low/hard state in all sources, though OIR observational coverage in other states still remains very sparse on rapid timescales. In all cases, the sources exhibited bright counterparts, and rather high OIR:X-ray flux ratios. Which low/hard state property drives this non-reprocessed component is currently unclear. The jet seems to play a key role in both \xtej1118\ and \gx339, but \swiftj1753\ possesses only a weak jet, with the corona and inner disc playing more important roles \citep{durant09}. On the other hand, the optical ACFs of both the former sources were narrower than the corresponding X-ray ones, which is not true for \swiftj1753. This does suggest that the very fastest flaring may be associated with the jet. Finally, the possibility of jets with strong magnetic fields and corresponding OIR synchrotron emission, but with a quenched radio emission (hence making their detection less secure observationally), cannot be discounted \citep{casellapeer09}. It may also be the case that a direct OIR variable component is generally present in other XRBs as well, but is smothered by the typically brighter reprocessed component. Observations in the high/soft state sensitive enough to be probe beneath the expected reprocessing level are needed to confirm this. 

\xtej1118\ is the only other source with detailed rapid optical vs. X-ray coherence and lag analyses published so far. As mentioned in the previous section, one key distinction between \gx339\ and \xtej1118\ is that in the latter case, a continuous range of time lags contribute to the coherence function, whereas we find only a fast ($\sim$150 ms), and a slow ($\sim$10 s) timescale to be important for \gx339. In terms of their physical characteristics, the most important difference between the sources is their orbital periods: \gx339\ is a long period binary with $P$=1.75 d whereas \xtej1118\ has a short period of only 0.17 d \citep{mcclintock01}, which implies a correspondingly larger disc extent for the former, assuming that the Roche lobe is filled. Whether or not a larger disc can somehow suppress the contribution from a continuous range of timescales in the case of \gx339\ remains to be seen. One way to test this would be to compute the coherence and lags for other systems: e.g. \swiftj1753\ has the shortest period currently-known for any XRB \citep{zurita08}.

Finally, the detection of a low frequency QPO only in the optical and not in X-rays (to within the limits discussed in \S~\ref{sec:qpodiscussion}) may be unique to \gx339. \xtej1118\ showed a QPO in both optical and X-rays during its 2000 outburst \citep[][ Spruit et al. 2010 in prep.]{hynes03_xtej1118}. That the QPO in \gx339\ may be more stable in the optical, as opposed to X-rays, has also been discussed by \citet{motch85}. This argues against an association with some instability at a radius characterising the edge of an inner hot flow, as these would produce X-ray QPOs as well. Instead, the QPO origin may be related exclusively to the optical emission region. If this is the jet, then the characteristic time (and hence size) scale of the QPO may be telling us about the extent of the region where perturbations are being pumped into the jet. In this case, searches for similar oscillations at radio frequencies with future sensitive telescopes (e.g. the EVLA) could be a good discriminator. We note that \citet{pooleyfender97} found radio QPOs in the strong jetted source GRS~1915+105 on timescales associated with IR flares in this system. On the other hand, GRS~1915+105 also exhibited soft X-ray oscillations on similar times, so the mechanism is not unique to the radio in this case.

\section{Summary}

Our main results may be summarised as follows:

\begin{enumerate}

  \item We have carried out rapid timing observations of \gx339\ simultaneously in optical and X-rays, on a best optical time resolution of 50 ms. These are the fastest such observations of the source in the optically-dim ($V$$\sim$17) low/hard X-ray state. 

  \item Simultaneous optical spectroscopy shows the presence of a blue, broken power-law continuum, and a host of recombination lines as well as the Bowen fluorescence blend (\S~\ref{sec:optspec}).

  \item The light curves reveal the presence of fluctuations on a range of timescales in both optical and X-rays (Fig.~\ref{fig:lcsection}). The fastest optical flares that we detect have decay times as short as $\sim$100 ms, while the strongest flares can increase the observed flux to more than twice the mean level on these short times (Fig.~\ref{fig:optflares}). The multi-filter rms values suggest that flaring is redder than the absolute flux spectrum (Fig.~\ref{fig:rg}).

  \item Simultaneous multi-filter power spectrum fits extending over three decades in Fourier frequency, and to well beyond 1 Hz in $g'$ and $r'$ (and slower in $u'$) are presented for the first time, and show the presence of a low-frequency QPO in all filters (\S~\ref{sec:psds}; Figs.~\ref{fig:xpsdfits}--\ref{fig:opsdfitsn3}). No corresponding sharp feature carrying a similar fraction of the total X-ray variability is seen in the simultaneous X-ray data (\S~\ref{sec:qpodiscussion}). Most of the optical variability power emerges around $\sim$1 Hz. The X-ray rms is larger than in the optical, but peaks at lower frequencies around $\sim$0.1 Hz (Fig.~\ref{fig:psds}). 

  \item Auto and cross-correlation analyses shows consistent results over several nights and in all filters, with the optical ACFs being narrower than the white-noise--corrected X-ray ones (Fig.~\ref{fig:acfs}). The CCFs in $r'$ and in $g'$ show a sharp and fast delayed peak with respect to X-rays (Fig.~\ref{fig:ccfs}). The $u'$ data is slower, but also shows evidence of the presence of fast fluctuations (Fig.~\ref{fig:udcf}). 

  \item Other than \xtej1118, our work presents the first detailed analysis of the fast optical vs. X-ray time lags and coherence function in an XRB (\S~\ref{sec:cohlags}). The clearest feature of the coherence function between optical and X-rays is a broad peak around a Fourier frequency of 1 Hz, and this is associated with a time lag that is approximately constant at $\approx$150 ms, the peak CCF delay. Changes in coherence values appear to be associated with changeovers in the dominant optical (or X-ray) PSD Lorentzian components with Fourier frequency (Fig.~\ref{fig:cohlag}).

  \item We find a large optical--to--X-ray flux ratio (\S~\ref{sec:sed}) which is apparently higher than on any previous occasion when simultaneous broad-band data was observed. This cannot be easily explained by a disc unless it is reprocessing an atypically-high fraction of incident X-rays. Furthermore, it requires that any jet model of the source in this state produce a high fraction of pre-shock optical synchrotron in the base as compared to the contribution from non-thermal optically-thin X-rays (\S~\ref{sec:sedmodels}).

  \item Several lines of reasoning are presented against reprocessing of X-rays dominating the optical power and the rapid variability, at least according to canonical linear transfer function scenarios in the outer disc. These include: the low source X-ray Eddington fraction; the sharpness of the fast CCF peak and its small time lag; the anti-correlation seen in the CCF on times of $\ltsim$few seconds; and, the narrowness of the optical ACFs with respect to X-rays. Any other complex reprocessing scenario, if true, must incorporate atypically strong reprocessing ratios, and must also be able to isolate the fast X-ray flares from the slower ones (\S~\ref{sec:reprocessedcomponent_fast}). 

  \item Standard heating of the outer disc predicts a weak reprocessed flux about 10 times lower than the total observed power. The slower ($\sim$10 s) variability component highlighted in the CCFs computed from the heavily binned light curves, and also in the large coherence values and lags at low Fourier frequencies, increases at blue ($u'$) wavelengths, and is consistent with disc reprocessing  (\S~\ref{sec:reprocessedcomponent_slow}). 

  \item The rms energy spectrum of fast fluctuations is best described by an X-ray variable component matching the overall energy spectrum, and a separate optical component redder than the total optical energy spectrum (\S~\ref{sec:rmsspectrum}). The contribution of a steady component such as a disc is constrained to be $<$20 per cent in $g'$, and a typical \diskir\ model extending from soft X-rays can easily satisfy this. A model in which perturbations related to accretion instabilities in the outer disc propagate to the inner flow and then to the jet, triggering hard X-ray flares via Comptonisation and optical cyclosynchrotron emission, can qualitatively explain the main results of our analysis, and can also produce linear rms--flux relations in both bands. The low optical:X-ray coherence peak value of $\approx$0.1 is a result of interaction between these independent components. The $B$ field necessary to produce the fast and strong optical flares that we observe must be $\gtsim$10$^4$ G. Furthermore, in such a scenario, the 150~ms time lag is related to the propagation delay between the X-ray corona and the jet optical emission regions.

\end{enumerate}

\section{Conclusions}
Our wide optical/X-ray spectral and timing simultaneous coverage gives new insight on the processes occurring on fast timescales in the inner accretion regions of this important source. Fast optical timing is beginning to probe interactions of the jet, corona and disc. This provides good starting material for detailed modelling of the accretion environment. But many questions remain to be answered, including the origin of the optical QPO, the correlation of the observed lags with the various variability components found in the PSDs, the origin of the optical emission lines, and the underlying driver of the fast fluctuations and observed power in \gx339, and also other sources. The present inferences are based on relatively short datasets, with the best simultaneous dataset having a total duration of less than one hour. Longer follow-up observations in various states of the sources should enhance the constraints presented herein. Indeed, several systems are now seen to show interesting variability characteristics on fast times, and such work ought to continue with new instruments being mounted on large telescopes and offering such modes on an increasing basis: e.g. FORS/VLT high time resolution mode, SALT/SALTICAM and Ultraspec, among others. Further insight could come from improving the time resolution of the observations, or through detailed dynamical cross-correlation analyses utilising each segment of the light curves separately (cf. Durant et al. 2010, in prep.). Coordination with the infrared offers another excellent opportunity, as this may be able to probe optically-thick jet emission. Indeed, \citet{casella10} report the detection of fast infrared flaring in GX~339--4 with the infrared also delayed by $\sim$100 ms with respect to X-rays. Finally, instruments such as the LAXPC on board \astrosat\ will provide increased hard X-ray sensitivity as compared to \rxte/PCA\ \citep{astrosat} and thus also help to break some of the model degeneracies discussed herein. 

\section*{Acknowledgments}
PG acknowledges a JAXA International Top Young Fellowship and a RIKEN Foreign Postdoctoral Research fellowship during parts of this work. ACF acknowledges the Royal Society for support. TS and MD acknowledge Spanish Ministry grants AYA2004 02646 \& AYA2007 66887. TRM acknowledges support from the Science and Technology Facilities Council (ST/F002599/1). JM acknowledges financial support from CNRS, ANR and GDR PCHE in France. PC acknowledges funding via a EU Marie Curie Intra-European Fellowship under contract no. 2009-237722. 

PG thanks C.B. Markwardt for making the \idlextract\ code public. Patrick Wallace is gratefully acknowledged for use of his SLA C library. PG thanks Paul Vreeswijk for his enthusiasms, and Dipankar Maitra and David Russell for comments on an initial draft. The ESO DDT time allocation committee is thanked for the granted time on the optical spectroscopy presented herein, and the efforts of the \rxte\ Guest Observer Facility in helping with the scheduling of the simultaneous observations are appreciated. The anonymous referee's comments helped to make the discussion and presentation of results more rigorous.

\begin{table*}
  \begin{center}
  \begin{tabular}{lccccccr}
    \hline
Band     & Lorentzian  &      $r$             &   $\nu_0$  & $\Delta$   &    \numax   &  $Q$      &      $\chi^2$/dof \\
         &  component  &                      &    (Hz)    &   (Hz)     &     (Hz)    &           &                   \\
    \hline				    		              					
         &  1          &       0.16(3)        & 0.012(2)   & 0.011(4)   &    0.017(3) & 0.6(2)    &                  \\
Full PCA &  2          &       0.49(1)        & 0          & 0.113(9)   &    0.113(9) &   0       &         37/48    \\
         &  3          &       0.34(3)        & 0.6(5)     & 2.1(2)     &     2.1(2)  & 0.14(11)  &                   \\
\\
         &  1          &       0.18(10)       & 0.010(5)   & 0.015(12)  &    0.018(10)& 0.3(3)    &                   \\
2--5 keV &  2          &       0.41(11)       & 0.015(51)  & 0.10(1)    &    0.10(2)  & 0.1(2)    &         53/48     \\
         &  3          &       0.37(2)        & 0          & 1.7(4)     &     1.7(4)  &  0        &                   \\
\\
         &  1          &       0.14(3)        & 0.012(2)   & 0.009(4)   &    0.015(3) & 0.7(3)    &                   \\
5--20 keV&  2          &       0.51(1)        & 0          & 0.114(9)   &    0.114(9) &   0       &         46/48     \\
         &  3          &       0.38(6)        & 0.2(8)     & 2.4(3)     &     2.4(3)  & 0.04(17)  &                   \\
        \hline
  \end{tabular}
  \caption{Multiple Lorentzian fits to the X-ray PCA PSD averaged for all three nights, listed in order of increasing \numax. The Lorentzian functional form is $P(\nu)=r^2\Delta/\pi/[\Delta^2+(\nu-\nu_0)^2]$, as defined by \citet{belloni02}. $\nu_{\rm max}=\sqrt{\nu_0^2+\Delta^2}$ and $Q=\nu_0/2/\Delta$, where $r$, \numax\ and $Q$ refer to the integrated rms (over the full range of --$\infty$ to +$\infty$), characteristic frequency, and quality factor respectively. $n$ numbers in brackets denote the 1$\sigma$ uncertainties on the last $n$ digits. \label{tab:xpsdfits}}
\end{center}
\end{table*}

\begin{table*}
  \begin{center}
  \begin{tabular}{lccccccr}
    \hline
Band  & Lorentzian &      $r$             &      $\nu_0$   & $\Delta$  &   \numax   &  $Q$      &      $\chi^2$/dof \\
      & component  &                      &       (Hz)     &   (Hz)    &    (Hz)    &           &                   \\
    \hline				  					
             \multicolumn{8}{c}{{\underline{\textsl{Night 1}}}}\\
      &   1        &       0.029(4)       &     0.049(1)   & 0.004(2)  &  0.049(1)   & 6.2(2.9) &                   \\
$r'$  &   2        &       0.074(7)       &     0          &  0.06(1)  &  0.06(1)    &  0       &         95/82     \\
      &   3        &       0.192(3)       &     0          &  0.90(6)  &  0.91(6)    &  0       &                   \\
      &   4        &       0.055(8)       &     2.9(3)     &  1.8(2)   &  3.4(3)     &  0.8(1)  &                   \\
\\
      &   1        &       0.024(4)       &     0.049(1)   & 0.004(2)  &  0.049(1)   & 5.9(2.6) &                   \\
$g'$  &   2        &       0.058(6)       &     0          &  0.06(1)  &  0.06(1)    &  0       &         94/82     \\
      &   3        &       0.154(3)       &     0          &  0.89(6)  &  0.89(6)    &  0       &                   \\
      &   4        &       0.057(6)       &     2.9(3)     & 2.2(2)    &  3.6(3)     & 0.7(1)   &                   \\
\\
      &   1          &       0.05(1)        &    0         &  0.02(1)  &  0.02(1)    &    0     &                  \\
$u'$  &   2          &       0.020(4)       &    0.050(2)  & 0.005(2)  &   0.050(3)  & 5.5(2.7) &         22/19    \\
      &   3          &       *              &          *   &        *  &      $>$0.3 &        * &                  \\
\\
             \multicolumn{8}{c}{{\underline{\textsl{Night 2}}}}\\
      &   1        &       0.055(6)       &     0          & 0.011(6)  &  0.011(6)   &  0       &                   \\
$r'$  &   2        &       0.032(3)       &     0.053(1)   & 0.004(1)  &  0.053(1)   & 6.1(1.8) &         76/67     \\
      &   3        &       0.132(10)      &     0          &  0.31(5)  &  0.31(5)    &  0       &                   \\
      &   4        &       0.177(7)       &     0          &  1.31(10) &  1.3(1)     &  0       &                   \\
\\
      &   1        &       0.052(6)       &     0          & 0.016(7)  &  0.016(7)   &  0       &                   \\
$g'$  &   2        &       0.026(3)       &     0.053(1)   & 0.005(2)  &  0.053(1)   & 5.1(1.6) &         62/67     \\
      &   3        &       0.076(22)      &     0.08(9)    &  0.22(7)  &  0.23(7)    & 0.2(2)   &                   \\
      &   4        &       0.155(6)       &     0          &  1.23(9)  &  1.23(9)    &  0       &                   \\
\\
             \multicolumn{8}{c}{{\underline{\textsl{Night 3}}}}\\
      &   1        &       0.054(4)       &     0.052(2)   & 0.011(3)  &  0.053(2)   &  2.3(5)  &                   \\
$r'$  &   2        &       0.178(21)      &     0          & 0.47(8)   &  0.47(8)    & 0        &         43/38     \\
      &   3        &       0.141(26)      &     0          & 1.5(4)    &  1.5(4)     &  0       &                   \\
\\
      &   1        &       0.036(5)       &     0.052(1)   & 0.006(3)  &  0.053(2)   & 4.2(2.1) &                   \\
$g'$  &   2        &       0.077(9)       &     0          & 0.13(4)   &  0.13(4)    & 0        &         46/38     \\
      &   3        &       0.160(4)       &     0          & 0.86(6)   &  0.86(6)    & 0        &                   \\
    \hline
  \end{tabular}
  \caption{Multiple Lorentzian fits to the optical PSDs from Night 1. See caption of Table~\ref{tab:xpsdfits} for details. Components marked with an asterisk are largely unconstrained. 
\label{tab:opsdfits}}
\end{center}
\end{table*}

\bibliographystyle{mnras}
\bibliography{gandhi10}

\begin{thebibliography}{}

\bibitem[\protect\citeauthoryear{{Agrawal}}{{Agrawal}}{2006}]{astrosat}
{Agrawal} P.~C., 2006, Advances in Space Research, 38, 2989

\bibitem[\protect\citeauthoryear{{Appenzeller} et~al.}{{Appenzeller}
  et~al.}{1998}]{fors}
{Appenzeller} I. et~al., 1998, The Messenger, 94, 1

\bibitem[\protect\citeauthoryear{{Arai} et~al.}{{Arai} et~al.}{2009}]{arai09}
{Arai} A. et~al., 2009, \pasj, 61, L1

\bibitem[\protect\citeauthoryear{{Belloni} \& {Hasinger}}{{Belloni} \&
  {Hasinger}}{1990}]{bellonihasinger90}
{Belloni} T.,  {Hasinger} G., 1990, \aap, 230, 103

\bibitem[\protect\citeauthoryear{{Belloni}, {Psaltis}, \& {van der
  Klis}}{{Belloni} et~al.}{2002}]{belloni02}
{Belloni} T., {Psaltis} D.,  {van der Klis} M., 2002, \apj, 572, 392

\bibitem[\protect\citeauthoryear{{Bendat} \& {Piersol}}{{Bendat} \&
  {Piersol}}{1986}]{bendatpiersol86}
{Bendat} J.~S.,  {Piersol} A.~G., 1986, {Random Data: Analysis and Measurment
  Procedures}.
\newblock New York, Wiley-Interscience, 1986

\bibitem[\protect\citeauthoryear{{Blackburn}}{{Blackburn}}{1995}]{ftools}
{Blackburn} J.~K., 1995, in Astronomical Society of the Pacific Conference
  Series, Vol.~77, {R.~A.~Shaw, H.~E.~Payne, \& J.~J.~E.~Hayes} , ed,
  Astronomical Data Analysis Software and Systems IV, p. 367

\bibitem[\protect\citeauthoryear{{Bradt}, {Rothschild}, \& {Swank}}{{Bradt}
  et~al.}{1993}]{rxte}
{Bradt} H.~V., {Rothschild} R.~E.,  {Swank} J.~H., 1993, \aaps, 97, 355

\bibitem[\protect\citeauthoryear{{Buxton} \& {Bailyn}}{{Buxton} \&
  {Bailyn}}{2007}]{buxtonbailyn07}
{Buxton} M.,  {Bailyn} C., 2007, The Astronomer's Telegram, 1109, 1

\bibitem[\protect\citeauthoryear{{Buxton} \& {Vennes}}{{Buxton} \&
  {Vennes}}{2003}]{buxton03}
{Buxton} M.,  {Vennes} S., 2003, \mnras, 342, 105

\bibitem[\protect\citeauthoryear{{Cardelli}, {Clayton}, \& {Mathis}}{{Cardelli}
  et~al.}{1989}]{cardelli89}
{Cardelli} J.~A., {Clayton} G.~C.,  {Mathis} J.~S., 1989, \apj, 345, 245

\bibitem[\protect\citeauthoryear{{Casella}, {Belloni}, \& {Stella}}{{Casella}
  et~al.}{2005}]{casella05}
{Casella} P., {Belloni} T.,  {Stella} L., 2005, \apj, 629, 403

\bibitem[\protect\citeauthoryear{{Casella} et~al.}{{Casella}
  et~al.}{2010}]{casella10}
{Casella} P. et~al., 2010, \mnras, 404, L21

\bibitem[\protect\citeauthoryear{{Casella} \& {Pe'er}}{{Casella} \&
  {Pe'er}}{2009}]{casellapeer09}
{Casella} P.,  {Pe'er} A., 2009, \apjl, 703, L63

\bibitem[\protect\citeauthoryear{{Chiang} et~al.}{{Chiang}
  et~al.}{2010}]{chiang10}
{Chiang} C.~Y., {Done} C., {Still} M.,  {Godet} O., 2010, \mnras, 403, 1102

\bibitem[\protect\citeauthoryear{{Corbel} \& {Fender}}{{Corbel} \&
  {Fender}}{2002}]{corbel02}
{Corbel} S.,  {Fender} R.~P., 2002, \apjl, 573, L35

\bibitem[\protect\citeauthoryear{{Corbel} et~al.}{{Corbel}
  et~al.}{2000}]{corbel00}
{Corbel} S., {Fender} R.~P., {Tzioumis} A.~K., {Nowak} M., {McIntyre} V.,
  {Durouchoux} P.,  {Sood} R., 2000, \aap, 359, 251

\bibitem[\protect\citeauthoryear{{Corbel} et~al.}{{Corbel}
  et~al.}{2003}]{corbel03}
{Corbel} S., {Nowak} M.~A., {Fender} R.~P., {Tzioumis} A.~K.,  {Markoff} S.,
  2003, \aap, 400, 1007

\bibitem[\protect\citeauthoryear{{Dhillon} et~al.}{{Dhillon}
  et~al.}{2007}]{ultracam}
{Dhillon} V.~S. et~al., 2007, \mnras, 378, 825

\bibitem[\protect\citeauthoryear{{di Matteo}, {Celotti}, \& {Fabian}}{{di
  Matteo} et~al.}{1997}]{dimatteo97}
{di Matteo} T., {Celotti} A.,  {Fabian} A.~C., 1997, \mnras, 291, 805

\bibitem[\protect\citeauthoryear{{di Matteo}, {Celotti}, \& {Fabian}}{{di
  Matteo} et~al.}{1999}]{dimatteo99_gx339}
{di Matteo} T., {Celotti} A.,  {Fabian} A.~C., 1999, \mnras, 304, 809

\bibitem[\protect\citeauthoryear{{Dickey} \& {Lockman}}{{Dickey} \&
  {Lockman}}{1990}]{dickeylongman90}
{Dickey} J.~M.,  {Lockman} F.~J., 1990, \araa, 28, 215

\bibitem[\protect\citeauthoryear{{Done}, {Gierli{\'n}ski}, \& {Kubota}}{{Done}
  et~al.}{2007}]{done07}
{Done} C., {Gierli{\'n}ski} M.,  {Kubota} A., 2007, \aapr, 15, 1

\bibitem[\protect\citeauthoryear{{Dunn} et~al.}{{Dunn} et~al.}{2008}]{dunn08}
{Dunn} R.~J.~H., {Fender} R.~P., {K{\"o}rding} E.~G., {Cabanac} C.,  {Belloni}
  T., 2008, \mnras, 387, 545

\bibitem[\protect\citeauthoryear{{Durant} et~al.}{{Durant}
  et~al.}{2008}]{durant08}
{Durant} M., {Gandhi} P., {Shahbaz} T., {Fabian} A.~P., {Miller} J., {Dhillon}
  V.~S.,  {Marsh} T.~R., 2008, \apjl, 682, L45

\bibitem[\protect\citeauthoryear{{Durant} et~al.}{{Durant}
  et~al.}{2009}]{durant09}
{Durant} M., {Gandhi} P., {Shahbaz} T., {Peralta} H.~H.,  {Dhillon} V.~S.,
  2009, \mnras, 392, 309

\bibitem[\protect\citeauthoryear{{Edelson} \& {Krolik}}{{Edelson} \&
  {Krolik}}{1988}]{edelsonkrolik88}
{Edelson} R.~A.,  {Krolik} J.~H., 1988, \apj, 333, 646

\bibitem[\protect\citeauthoryear{{Eikenberry} et~al.}{{Eikenberry}
  et~al.}{1998}]{eikenberry98}
{Eikenberry} S.~S., {Matthews} K., {Morgan} E.~H., {Remillard} R.~A.,  {Nelson}
  R.~W., 1998, \apjl, 494, L61

\bibitem[\protect\citeauthoryear{{Esin} et~al.}{{Esin} et~al.}{2001}]{esin01}
{Esin} A.~A., {McClintock} J.~E., {Drake} J.~J., {Garcia} M.~R., {Haswell}
  C.~A., {Hynes} R.~I.,  {Muno} M.~P., 2001, \apj, 555, 483

\bibitem[\protect\citeauthoryear{{Fabian} et~al.}{{Fabian}
  et~al.}{1982}]{fabian82}
{Fabian} A.~C., {Guilbert} P.~W., {Motch} C., {Ricketts} M., {Ilovaisky} S.~A.,
   {Chevalier} C., 1982, \aap, 111, L9

\bibitem[\protect\citeauthoryear{{Fender} et~al.}{{Fender}
  et~al.}{1997}]{fender97}
{Fender} R.~P., {Pooley} G.~G., {Brocksopp} C.,  {Newell} S.~J., 1997, \mnras,
  290, L65

\bibitem[\protect\citeauthoryear{{Gallo}, {Fender}, \& {Pooley}}{{Gallo}
  et~al.}{2003}]{gallo03}
{Gallo} E., {Fender} R.~P.,  {Pooley} G.~G., 2003, \mnras, 344, 60

\bibitem[\protect\citeauthoryear{{Gandhi}}{{Gandhi}}{2009}]{g09_rmsflux}
{Gandhi} P., 2009, \apjl, 697, L167

\bibitem[\protect\citeauthoryear{{Gandhi} et~al.}{{Gandhi} et~al.}{2008}]{g08}
{Gandhi} P. et~al., 2008, \mnras, 390, L29 {\bf (Paper I)}

\bibitem[\protect\citeauthoryear{{Gaskell} \& {Peterson}}{{Gaskell} \&
  {Peterson}}{1987}]{gaskellpeterson87}
{Gaskell} C.~M.,  {Peterson} B.~M., 1987, \apjs, 65, 1

\bibitem[\protect\citeauthoryear{{Gierli{\'n}ski}, {Done}, \&
  {Page}}{{Gierli{\'n}ski} et~al.}{2009}]{diskir}
{Gierli{\'n}ski} M., {Done} C.,  {Page} K., 2009, \mnras, 392, 1106

\bibitem[\protect\citeauthoryear{{Hamuy} et~al.}{{Hamuy}
  et~al.}{1994}]{hamuy94}
{Hamuy} M., {Suntzeff} N.~B., {Heathcote} S.~R., {Walker} A.~R., {Gigoux} P.,
  {Phillips} M.~M., 1994, \pasp, 106, 566

\bibitem[\protect\citeauthoryear{{Hasinger} \& {van der Klis}}{{Hasinger} \&
  {van der Klis}}{1989}]{hasingervanderklis89}
{Hasinger} G.,  {van der Klis} M., 1989, \aap, 225, 79

\bibitem[\protect\citeauthoryear{{Homan} et~al.}{{Homan}
  et~al.}{2005}]{homan05}
{Homan} J., {Buxton} M., {Markoff} S., {Bailyn} C.~D., {Nespoli} E.,  {Belloni}
  T., 2005, \apj, 624, 295

\bibitem[\protect\citeauthoryear{{Horne}}{{Horne}}{1985}]{horne85}
{Horne} K., 1985, \mnras, 213, 129

\bibitem[\protect\citeauthoryear{{Hynes} et~al.}{{Hynes}
  et~al.}{2003a}]{hynes03_xtej1118}
{Hynes} R.~I. et~al., 2003a, \mnras, 345, 292

\bibitem[\protect\citeauthoryear{{Hynes} et~al.}{{Hynes}
  et~al.}{2006}]{hynes06}
{Hynes} R.~I. et~al., 2006, \apj, 651, 401

\bibitem[\protect\citeauthoryear{{Hynes} et~al.}{{Hynes}
  et~al.}{2003b}]{hynes03_gx339}
{Hynes} R.~I., {Steeghs} D., {Casares} J., {Charles} P.~A.,  {O'Brien} K.,
  2003b, \apjl, 583, L95

\bibitem[\protect\citeauthoryear{{Hynes} et~al.}{{Hynes}
  et~al.}{2004}]{hynes04}
{Hynes} R.~I., {Steeghs} D., {Casares} J., {Charles} P.~A.,  {O'Brien} K.,
  2004, \apj, 609, 317

\bibitem[\protect\citeauthoryear{{Imamura} et~al.}{{Imamura}
  et~al.}{1990}]{imamura90}
{Imamura} J.~N., {Kristian} J., {Middleditch} J.,  {Steiman-Cameron} T.~Y.,
  1990, \apj, 365, 312

\bibitem[\protect\citeauthoryear{{Ingram}, {Done}, \& {Fragile}}{{Ingram}
  et~al.}{2009}]{ingram09}
{Ingram} A., {Done} C.,  {Fragile} P.~C., 2009, \mnras, 397, L101

\bibitem[\protect\citeauthoryear{{Isobe} et~al.}{{Isobe}
  et~al.}{1990}]{isobe90}
{Isobe} T., {Feigelson} E.~D., {Akritas} M.~G.,  {Babu} G.~J., 1990, \apj, 364,
  104

\bibitem[\protect\citeauthoryear{{Jahoda} et~al.}{{Jahoda}
  et~al.}{1996}]{rxtepca}
{Jahoda} K., {Swank} J.~H., {Giles} A.~B., {Stark} M.~J., {Strohmayer} T.,
  {Zhang} W.,  {Morgan} E.~H., 1996, in Presented at the Society of
  Photo-Optical Instrumentation Engineers (SPIE) Conference, Vol. 2808,
  {O.~H.~Siegmund \& M.~A.~Gummin} , ed, Society of Photo-Optical
  Instrumentation Engineers (SPIE) Conference Series, p.~59

\bibitem[\protect\citeauthoryear{{Jenkins} \& {Watts}}{{Jenkins} \&
  {Watts}}{1969}]{jenkinswatts69}
{Jenkins} G.~M.,  {Watts} D.~G., 1969, {Spectral analysis and its
  applications}.
\newblock Holden-Day Series in Time Series Analysis, London: Holden-Day, 1969

\bibitem[\protect\citeauthoryear{{Jester} et~al.}{{Jester}
  et~al.}{2005}]{jester05}
{Jester} S. et~al., 2005, \aj, 130, 873

\bibitem[\protect\citeauthoryear{{Kalemci} et~al.}{{Kalemci}
  et~al.}{2007}]{kalemci07}
{Kalemci} E. et~al., 2007, The Astronomer's Telegram, 1074, 1

\bibitem[\protect\citeauthoryear{{Kanbach} et~al.}{{Kanbach}
  et~al.}{2001}]{kanbach01}
{Kanbach} G., {Straubmeier} C., {Spruit} H.~C.,  {Belloni} T., 2001, \nat, 414,
  180

\bibitem[\protect\citeauthoryear{{Kotov}, {Churazov}, \& {Gilfanov}}{{Kotov}
  et~al.}{2001}]{kotov01}
{Kotov} O., {Churazov} E.,  {Gilfanov} M., 2001, \mnras, 327, 799

\bibitem[\protect\citeauthoryear{{Maitra} et~al.}{{Maitra}
  et~al.}{2009}]{maitra09}
{Maitra} D., {Markoff} S., {Brocksopp} C., {Noble} M., {Nowak} M.,  {Wilms} J.,
  2009, \mnras, 398, 1638

\bibitem[\protect\citeauthoryear{{Makishima} et~al.}{{Makishima}
  et~al.}{1986}]{makishima86}
{Makishima} K., {Maejima} Y., {Mitsuda} K., {Bradt} H.~V., {Remillard} R.~A.,
  {Tuohy} I.~R., {Hoshi} R.,  {Nakagawa} M., 1986, \apj, 308, 635

\bibitem[\protect\citeauthoryear{{Makishima} et~al.}{{Makishima}
  et~al.}{2008}]{makishima08}
{Makishima} K. et~al., 2008, \pasj, 60, 585

\bibitem[\protect\citeauthoryear{{Malzac} et~al.}{{Malzac}
  et~al.}{2003}]{malzac03}
{Malzac} J., {Belloni} T., {Spruit} H.~C.,  {Kanbach} G., 2003, \aap, 407, 335

\bibitem[\protect\citeauthoryear{{Malzac} \& {Belmont}}{{Malzac} \&
  {Belmont}}{2009}]{malzacbelmont09}
{Malzac} J.,  {Belmont} R., 2009, \mnras, 392, 570

\bibitem[\protect\citeauthoryear{{Malzac}, {Merloni}, \& {Fabian}}{{Malzac}
  et~al.}{2004}]{malzac04}
{Malzac} J., {Merloni} A.,  {Fabian} A.~C., 2004, \mnras, 351, 253

\bibitem[\protect\citeauthoryear{{Markoff}, {Nowak}, \& {Wilms}}{{Markoff}
  et~al.}{2005}]{markoff05}
{Markoff} S., {Nowak} M.~A.,  {Wilms} J., 2005, \apj, 635, 1203

\bibitem[\protect\citeauthoryear{{Marsh} \& {Horne}}{{Marsh} \&
  {Horne}}{1988}]{marsh88}
{Marsh} T.~R.,  {Horne} K., 1988, \mnras, 235, 269

\bibitem[\protect\citeauthoryear{{McClintock} et~al.}{{McClintock}
  et~al.}{2001}]{mcclintock01}
{McClintock} J.~E., {Garcia} M.~R., {Caldwell} N., {Falco} E.~E., {Garnavich}
  P.~M.,  {Zhao} P., 2001, \apjl, 551, L147

\bibitem[\protect\citeauthoryear{{McHardy} et~al.}{{McHardy}
  et~al.}{2004}]{mchardy04}
{McHardy} I.~M., {Papadakis} I.~E., {Uttley} P., {Page} M.~J.,  {Mason} K.~O.,
  2004, \mnras, 348, 783

\bibitem[\protect\citeauthoryear{{Merloni}, {Di Matteo}, \& {Fabian}}{{Merloni}
  et~al.}{2000}]{merloni00}
{Merloni} A., {Di Matteo} T.,  {Fabian} A.~C., 2000, \mnras, 318, L15

\bibitem[\protect\citeauthoryear{{Miller} et~al.}{{Miller}
  et~al.}{2006}]{miller06}
{Miller} J.~M., {Homan} J., {Steeghs} D., {Rupen} M., {Hunstead} R.~W.,
  {Wijnands} R., {Charles} P.~A.,  {Fabian} A.~C., 2006, \apj, 653, 525

\bibitem[\protect\citeauthoryear{{Miyamoto} et~al.}{{Miyamoto}
  et~al.}{1992}]{miyamoto92}
{Miyamoto} S., {Kitamoto} S., {Iga} S., {Negoro} H.,  {Terada} K., 1992, \apjl,
  391, L21

\bibitem[\protect\citeauthoryear{{Motch}, {Ilovaisky}, \& {Chevalier}}{{Motch}
  et~al.}{1982}]{motch82}
{Motch} C., {Ilovaisky} S.~A.,  {Chevalier} C., 1982, \aap, 109, L1

\bibitem[\protect\citeauthoryear{{Motch} et~al.}{{Motch}
  et~al.}{1985}]{motch85}
{Motch} C., {Ilovaisky} S.~A., {Chevalier} C.,  {Angebault} P., 1985, Space
  Science Reviews, 40, 219

\bibitem[\protect\citeauthoryear{{Motch} et~al.}{{Motch}
  et~al.}{1983}]{motch83}
{Motch} C., {Ricketts} M.~J., {Page} C.~G., {Ilovaisky} S.~A.,  {Chevalier} C.,
  1983, \aap, 119, 171

\bibitem[\protect\citeauthoryear{{Mu{\~n}oz-Darias} et~al.}{{Mu{\~n}oz-Darias}
  et~al.}{2007}]{munoz-darias07}
{Mu{\~n}oz-Darias} T., {Mart{\'{\i}}nez-Pais} I.~G., {Casares} J., {Dhillon}
  V.~S., {Marsh} T.~R., {Cornelisse} R., {Steeghs} D.,  {Charles} P.~A., 2007,
  \mnras, 379, 1637

\bibitem[\protect\citeauthoryear{{Negoro}, {Kitamoto}, \& {Mineshige}}{{Negoro}
  et~al.}{2001}]{negoro01}
{Negoro} H., {Kitamoto} S.,  {Mineshige} S., 2001, \apj, 554, 528

\bibitem[\protect\citeauthoryear{{Nowak} et~al.}{{Nowak}
  et~al.}{1999a}]{nowak99_cygx1_ii}
{Nowak} M.~A., {Vaughan} B.~A., {Wilms} J., {Dove} J.~B.,  {Begelman} M.~C.,
  1999a, \apj, 510, 874

\bibitem[\protect\citeauthoryear{{Nowak}, {Wilms}, \& {Dove}}{{Nowak}
  et~al.}{1999}]{nowak99_gx339}
{Nowak} M.~A., {Wilms} J.,  {Dove} J.~B., 1999, \apj, 517, 355

\bibitem[\protect\citeauthoryear{{Nowak} et~al.}{{Nowak}
  et~al.}{1999b}]{nowak99_cygx1_iii}
{Nowak} M.~A., {Wilms} J., {Vaughan} B.~A., {Dove} J.~B.,  {Begelman} M.~C.,
  1999b, \apj, 515, 726

\bibitem[\protect\citeauthoryear{{O'Brien} et~al.}{{O'Brien}
  et~al.}{2002}]{obrien02}
{O'Brien} K., {Horne} K., {Hynes} R.~I., {Chen} W., {Haswell} C.~A.,  {Still}
  M.~D., 2002, \mnras, 334, 426

\bibitem[\protect\citeauthoryear{{Pooley} \& {Fender}}{{Pooley} \&
  {Fender}}{1997}]{pooleyfender97}
{Pooley} G.~G.,  {Fender} R.~P., 1997, \mnras, 292, 925

\bibitem[\protect\citeauthoryear{{Reis} et~al.}{{Reis} et~al.}{2008}]{reis08}
{Reis} R.~C., {Fabian} A.~C., {Ross} R.~R., {Miniutti} G., {Miller} J.~M.,
  {Reynolds} C., 2008, \mnras, 387, 1489

\bibitem[\protect\citeauthoryear{{Remillard} \& {McClintock}}{{Remillard} \&
  {McClintock}}{2006}]{remillardmcclintock06}
{Remillard} R.~A.,  {McClintock} J.~E., 2006, \araa, 44, 49

\bibitem[\protect\citeauthoryear{{Rothschild} et~al.}{{Rothschild}
  et~al.}{1998}]{rxtehexte}
{Rothschild} R.~E. et~al., 1998, \apj, 496, 538

\bibitem[\protect\citeauthoryear{{Russell} et~al.}{{Russell}
  et~al.}{2006}]{russell06}
{Russell} D.~M., {Fender} R.~P., {Hynes} R.~I., {Brocksopp} C., {Homan} J.,
  {Jonker} P.~G.,  {Buxton} M.~M., 2006, \mnras, 371, 1334

\bibitem[\protect\citeauthoryear{{Schachter}, {Filippenko}, \&
  {Kahn}}{{Schachter} et~al.}{1989}]{schachter89}
{Schachter} J., {Filippenko} A.~V.,  {Kahn} S.~M., 1989, \apj, 340, 1049

\bibitem[\protect\citeauthoryear{{Shahbaz}, {Fender}, \& {Charles}}{{Shahbaz}
  et~al.}{2001}]{shahbaz01}
{Shahbaz} T., {Fender} R.,  {Charles} P.~A., 2001, \aap, 376, L17

\bibitem[\protect\citeauthoryear{{Shahbaz} et~al.}{{Shahbaz}
  et~al.}{1996}]{shahbaz96}
{Shahbaz} T., {Smale} A.~P., {Naylor} T., {Charles} P.~A., {van Paradijs} J.,
  {Hassall} B.~J.~M.,  {Callanan} P., 1996, \mnras, 282, 1437

\bibitem[\protect\citeauthoryear{{Soria}, {Wu}, \& {Johnston}}{{Soria}
  et~al.}{1999}]{soria99}
{Soria} R., {Wu} K.,  {Johnston} H.~M., 1999, \mnras, 310, 71

\bibitem[\protect\citeauthoryear{{Spruit} \& {Kanbach}}{{Spruit} \&
  {Kanbach}}{2002}]{spruitkanbach02}
{Spruit} H.~C.,  {Kanbach} G., 2002, \aap, 391, 225

\bibitem[\protect\citeauthoryear{{Steiman-Cameron} et~al.}{{Steiman-Cameron}
  et~al.}{1990}]{steiman-cameron90}
{Steiman-Cameron} T., {Imamura} J., {Middleditch} J.,  {Kristian} J., 1990,
  \apj, 359, 197

\bibitem[\protect\citeauthoryear{{Steiman-Cameron} et~al.}{{Steiman-Cameron}
  et~al.}{1997}]{steiman-cameron97}
{Steiman-Cameron} T.~Y., {Scargle} J.~D., {Imamura} J.~N.,  {Middleditch} J.,
  1997, \apj, 487, 396

\bibitem[\protect\citeauthoryear{{Tomsick} et~al.}{{Tomsick}
  et~al.}{2008}]{tomsick08}
{Tomsick} J.~A. et~al., 2008, \apj, 635, 593

\bibitem[\protect\citeauthoryear{{Uemura} et~al.}{{Uemura}
  et~al.}{2002}]{uemura02_v404cyg}
{Uemura} M. et~al., 2002, \pasj, 54, L79

\bibitem[\protect\citeauthoryear{{Uttley}, {McHardy}, \& {Vaughan}}{{Uttley}
  et~al.}{2005}]{uttley05}
{Uttley} P., {McHardy} I.~M.,  {Vaughan} S., 2005, \mnras, 359, 345

\bibitem[\protect\citeauthoryear{{van der Klis}}{{van der
  Klis}}{1989}]{vanderklis89}
{van der Klis} M., 1989, in {{\"O}gelman} H.,  {van den Heuvel} E.~P.~J., ed,
  Timing Neutron Stars, p.~27

\bibitem[\protect\citeauthoryear{{van der Klis}}{{van der
  Klis}}{1995}]{vanderklis95}
{van der Klis} M., 1995, in {W.~H.~G.~Lewin, J.~van Paradijs, \& E.~P.~J.~van
  den Heuvel} , ed, X-ray binaries, p. 252 - 307, p. 252

\bibitem[\protect\citeauthoryear{{van der Klis}}{{van der
  Klis}}{1997}]{vanderklis97}
{van der Klis} M., 1997, in {G.~J.~Babu \& E.~D.~Feigelson} , ed, Statistical
  Challenges in Modern Astronomy II, p. 321

\bibitem[\protect\citeauthoryear{{van Paradijs} \& {McClintock}}{{van Paradijs}
  \& {McClintock}}{1994}]{vanparadijsmcclintock94}
{van Paradijs} J.,  {McClintock} J.~E., 1994, \aap, 290, 133

\bibitem[\protect\citeauthoryear{{Vaughan} \& {Nowak}}{{Vaughan} \&
  {Nowak}}{1997}]{vaughannowak97}
{Vaughan} B.~A.,  {Nowak} M.~A., 1997, \apjl, 474, L43

\bibitem[\protect\citeauthoryear{{Vaughan} et~al.}{{Vaughan}
  et~al.}{2003}]{vaughan03}
{Vaughan} S., {Edelson} R., {Warwick} R.~S.,  {Uttley} P., 2003, \mnras, 345,
  1271

\bibitem[\protect\citeauthoryear{{Wardzi{\'n}ski} \&
  {Zdziarski}}{{Wardzi{\'n}ski} \& {Zdziarski}}{2000}]{wardzinski00}
{Wardzi{\'n}ski} G.,  {Zdziarski} A.~A., 2000, \mnras, 314, 183

\bibitem[\protect\citeauthoryear{{White} \& {Peterson}}{{White} \&
  {Peterson}}{1994}]{whitepeterson94}
{White} R.~J.,  {Peterson} B.~M., 1994, \pasp, 106, 879

\bibitem[\protect\citeauthoryear{{Wijnands} \& {van der Klis}}{{Wijnands} \&
  {van der Klis}}{1999}]{wijnandsvanderklis99}
{Wijnands} R.,  {van der Klis} M., 1999, \apj, 514, 939

\bibitem[\protect\citeauthoryear{{Wilkinson} \& {Uttley}}{{Wilkinson} \&
  {Uttley}}{2009}]{wilkinson09}
{Wilkinson} T.,  {Uttley} P., 2009, \mnras, 397, 666

\bibitem[\protect\citeauthoryear{{Wilms} et~al.}{{Wilms}
  et~al.}{1999}]{wilms99}
{Wilms} J., {Nowak} M.~A., {Dove} J.~B., {Fender} R.~P.,  {di Matteo} T., 1999,
  \apj, 522, 460

\bibitem[\protect\citeauthoryear{{Wu} et~al.}{{Wu} et~al.}{2001}]{wu01}
{Wu} K., {Soria} R., {Hunstead} R.~W.,  {Johnston} H.~M., 2001, \mnras, 320,
  177

\bibitem[\protect\citeauthoryear{{Yuan}, {Cui}, \& {Narayan}}{{Yuan}
  et~al.}{2005}]{yuan05}
{Yuan} F., {Cui} W.,  {Narayan} R., 2005, \apj, 620, 905

\bibitem[\protect\citeauthoryear{{Zdziarski} et~al.}{{Zdziarski}
  et~al.}{2004}]{zdziarski04}
{Zdziarski} A.~A., {Gierli{\'n}ski} M., {Miko{\l}ajewska} J., {Wardzi{\'n}ski}
  G., {Smith} D.~M., {Harmon} B.~A.,  {Kitamoto} S., 2004, \mnras, 351, 791

\bibitem[\protect\citeauthoryear{{Zdziarski} et~al.}{{Zdziarski}
  et~al.}{1998}]{zdziarski98}
{Zdziarski} A.~A., {Poutanen} J., {Mikolajewska} J., {Gierlinski} M., {Ebisawa}
  K.,  {Johnson} W.~N., 1998, \mnras, 301, 435

\bibitem[\protect\citeauthoryear{{Zurita} et~al.}{{Zurita}
  et~al.}{2008}]{zurita08}
{Zurita} C., {Durant} M., {Torres} M.~A.~P., {Shahbaz} T., {Casares} J.,
  {Steeghs} D., 2008, \apj, 681, 1458

\end{thebibliography}

\label{lastpage}

\end{document}